\documentclass[sigconf]{acmart} %, anonymous

\copyrightyear{2020} 
\acmYear{2020} 
\setcopyright{acmlicensed}\acmConference[CCS '20]{Proceedings of the 2020 ACM SIGSAC Conference on Computer and Communications Security}{November 9--13, 2020}{Virtual Event, USA}
\acmBooktitle{Proceedings of the 2020 ACM SIGSAC Conference on Computer and Communications Security (CCS '20), November 9--13, 2020, Virtual Event, USA}
\acmPrice{15.00}
\acmDOI{10.1145/3372297.3423347}
\acmISBN{978-1-4503-7089-9/20/11}

\usepackage{graphicx}
\usepackage{amsmath,amssymb,latexsym}
\usepackage{pifont}
\usepackage{xspace}
\usepackage{import}
\usepackage{graphicx}
\usepackage{multirow}
\usepackage{booktabs}
\usepackage{textgreek}
\newcommand{\const}[1]{\ensuremath{\mathsf{#1}\xspace}}
  
\usepackage{versions}
\excludeversion{RedundantContent}

\begin{document}
%\fancyhead{}
\title{Usage Patterns of Privacy-Enhancing Technologies} % TODO: replace with your title

\author{Kovila P.L. Coopamootoo}
\affiliation{%
  \institution{Newcastle University}
  \city{Newcastle-Upon-Tyne}
  \country{UK} 
}
\email{kovila.coopamootoo@newcastle.ac.uk}

\begin{abstract}
The steady reports of privacy invasions online paints a picture of the Internet growing into a more dangerous place.
This is supported by reports of the potential scale for online harms facilitated by the mass deployment of online technology and the data-intensive web. 
While Internet users often express concern about privacy, 
some report taking actions to protect their privacy online.
We investigate the methods and technologies that individuals employ to protect their privacy online.
We conduct two studies, of N=180 and N=907, to elicit individuals' use of privacy methods online, within the US, the UK and Germany.
We find that non-technology methods are among the most used methods in the three countries.
We identify distinct groupings of privacy methods usage in a cluster map.
The map shows that together with non-technology methods of privacy protection, simple PETs that are integrated in services, 
form the most used cluster, whereas more advanced PETs form a different, least used cluster.
We further investigate user perception and reasoning for mostly using one set of PETs in a third study with N=183 participants.
We do not find a difference in perceived competency in protecting privacy online between advanced and simpler PETs users.
We compare use perceptions between advanced and simpler PETs and report on user reasoning for not using advanced PETs, as well as support needed for potential use.
This paper contributes to privacy research by eliciting use and perception of use across $43$ privacy methods, including $26$ PETs across three countries and provides a map of PETs usage.
The cluster map provides a systematic and reliable point of reference for future user-centric investigations across PETs.
Overall, this research provides a broad understanding of use and perceptions across a collection of PETs, and can lead to future research for scaling use of PETs.
\end{abstract}

% TODO: replace this section with code generated by the tool at https://dl.acm.org/ccs.cfm


%\ccsdesc{Security and privacy~Use https://dl.acm.org/ccs.cfm to generate actual concepts section for your paper}
% -- end of section to replace with generated code

 % TODO: replace with your keywords

\maketitle

\section{Introduction}
The deployment of technology at massive scale is thought to have enabled the rapid emergence of a set of `online harms' such as privacy abuses, data breaches, cyber-attacks, inappropriate uses of personal data, hate crimes, and self-harm/suicide among others~\cite{wright2019online}.
%\paragraph{Pace of data collection vs human ability}
Today's data-intensive web is characterized with mass sharing, collection and aggregation
%, mining and selling 
of individuals' data, %data capitalist world
that enables the provisioning of customised services but unfortunately also engenders targeted advertising~\cite{chandramouli2013real}, digital discrimination~\cite{edelman2014digital}, privacy invasive algorithmic computations~\cite{forbes2019}, and a general fuzzyness about privacy rights online.
The recent years have indeed seen reports of 
various high profile privacy infringement cases involving mass unauthorised transfer and use of sensitive data~\cite{garrett2019social,murgia2019how,privacyint2019nobodys}.
% have also been reported in the recent years~\cite{garrett2019social,murgia2019how,privacyint2019nobodys}.
% including influencing voters via social media~\cite{garrett2019social}, top health websites' sharing of sensitive data with advertisers, including Google and Facebook~\cite{murgia2019how},
%as investigated by the Financial Times, 
%and period apps' (such as Maya and MIA) extensive sharing of sensitive personal data pertaining to women's health with third parties, including Facebook~\cite{privacyint2019nobodys}. % as found by Privacy International

\begin{RedundantContent}
Internet users often express discomfort with the data collection that enables personalization, and a large portion takes some kind of action such as clearing cookies and browsing history~\cite{rainie2013anonymity}.
However, the methods employed by individuals may not be enough to protect one's privacy, because, for example a particular web browser on a specific machine comprises a unique fingerprint that can be traced by web servers across the web~\cite{nikiforakis2013cookieless}. %, and this information in conveyed through headers that are automatically exchanged by every web browser and web server behind the scenes~\cite{nikiforakis2013cookieless}.
When individuals share online, they are also willing to take a number of steps towards social or interpersonal privacy for data shared as part of social interactions. %they are at the mercy of other individuals' or companies' behavior. 
These include the use of privacy settings for boundary regulation, as well as coping strategies to reduce emotional distress from privacy loss~\cite{wisniewski2012fighting}, such as blocking others, creating multiple profiles, censoring themselves, untagging and offline negotiations~\cite{stutzman2012boundary}.

%Yet, individuals may not perceive the connection between social and informational privacy, where strategies such as specifying privacy settings or having multiple profiles do not support better control over informational privacy, because 

Yet, the architectures and algorithms that collect data and make inferences are mostly invisible to users~\cite{de2005two} who may not imagine the extent of privacy loss. %It is therefore difficult for them to manage informational privacy, when they are unaware of disclosures~\cite{Paula Ding Dourish - Two experiences designing for effective security}.
Privacy protection is therefore perceived to be a `losing game' for individuals as more and more data about them is being generated faster and faster from more and more devices~\cite{kerry2019why}. 
\end{RedundantContent}

%***Review_D on justification, state-of-the-art.
%We will reshape introduction and discuss (1) the impact of large-scale, evidence-based studies on generalisations, (2) scaling of PETs investigations based on usage similarities rather than individual PETs as previous research, (3) how classification pattern can enable meta-investigations of PETs and advance methodology, (4) pace of PETs usage versus online harms, (5) learning from interaction of similarly-used PETs.

%\paragraph{Online Harms \& Usage of PETs.}
%demonstrate high pace/spread of online harms vs limited usage of PETs.
At the forefront of the narrative that ``the Internet is a dangerous place", is the UK, where the %, a perspective that emphasises the risks and potential for significant harm.
British Government has released an Online Harms white paper that elaborates on the scale and extent of harms faced by individuals online. % perpetrated by other individuals or organisations. 
It proposes a mission of mitigating online harms without inhibiting online innovation~\cite{wright2019online}. One of the strands of this mission is to empower citizens via technology, in particular via awareness, usage and integration of Privacy-Enhancing Technologies (PETs) in real life situations. %***************
%\paragraph{Privacy-Enhancing Technologies}
%define: privacy-enhancing technology
%The EU Agency for CyberSecurity defines privacy-enhancing technologies (
PETs are defined by the EU Agency for CyberSecurity, as technologies shaped according to privacy principles, where PETs ``covers the broader range of technologies that are designed for supporting privacy and data protection"~\cite{ENISA2020}.
Integrating PETs in daily life and scaling use of PETs, first requires a broad understanding of current use of PETs.

%%%% REVIEW %%%
%\textbf{\textcolor{red}{Introduce and discuss oxIS as motivation}}
%In the context of the need to address online harms, 
For a user perspective of online harms, the Oxford Internet Institute looked into 
%the trends of online threats from the perspective of Internet users, in particular 
how UK individuals' experience of problems online influence their behavior, in an Oxford Internet Survey (OxIS)~\cite{blank2019perceived}.
%, as well as (2) compared users and non-users of the Internet}.
%OxIS found that $10\%$ of their survey respondents are non-users of the Internet with worry about privacy being the most important reason they are not online. 
They found that non-users of the Internet (who made up $10\%$ of the respondents), were more concerned than users about privacy violations, consequently keeping themselves offline. %, and are presumed to listen to the narrative of the internet being a dangerous place.
Surprisingly, Internet users were not more concerned about the possibility of being a victim online in 2019 compared to 2013.
%However, most Internet users in 2019 were not strongly concerned about the possibility of being a victim online.
In 2019, general privacy concern increased only by $4\%$, and $26$-$31\%$ of respondents have taken action to protect their purchases, age, marital status or medical details and $40\%$ their contact details.
%In addition they observe an apparent divide in the UK separating Internet users and non-users, where non-users are more concerned about the dangers of the internet than users, and less likely to believe that technology make things better.
%\textcolor{red}{Say something about user concern of privacy/online harm - in particular of a few years ago} --- generally people claim to be worried, anxious and fearful when it comes to privacy
According to Pew Research Center of the US, although a majority of US consumers think that they have little or no control over how their personal information is collected and used by companies ($81\%$ in 2019 compared to $91\%$ in 2013)~\cite{auxier2019americans,madden2014public}, between $10\%$ to $19\%$ feel that they have a lot of control over different forms of personal information, such as physical location, social media posts, private conversations, purchases, or websites visited~\cite{auxier2019americans}.
%Pew Research Centre reports a drop from $91\%$ in 2013 to $81\%$ in 2019....
The European Commission reports a Eurobarometer survey in 2019 with findings that 
%for example $65\%$ have heard of the right to access their data
$51\%$ of those who provide personal information online felt that they have partial control over this information, while $14\%$ felt that they have complete control. $62\%$ of those feeling partial or no control were concerned about not having complete control.
In addition, a majority of the respondents in 2019 have heard of the rights guaranteed by the GDPR, and some have exercised these rights~\cite{eurobarometer2019european}. 
%have heard of most of the rights guaranteed by the GDPR,

%Consequently, the following questions arise: 
Given (1) the above reports of nuances in privacy concerns over the years and that some individuals take actions towards privacy, as well as (2) the well known \emph{privacy paradox} phenomenon~\cite{SpiGro2001,AcqGro2005,gerber2018explaining,kokolakis2017privacy},  
%While we wonder about the privacy protection ways of Internet users, 
this paper seeks to investigate the following questions: \emph{`what privacy methods or PETs do Internet users employ to protect their privacy? How do Internet users perceive the use of PETs?'}
%How do Internet users perceive use of PETs?'
%Does the stabilisation in concern reflect a perception of using (more) effective PETs?'
%Concern about viruses and malware has declined by $17\%$ and $43\%$ less respondents have taken taken action to prevent a virus in 2019.
%**********************
\begin{RedundantContent}
%\paragraph{Investigating Individual PETs vs. Patterns of PETs}
User-centric privacy research has widely covered privacy concerns~\cite{} or privacy behavior in general~\cite{1,2,3}~\cite{}, why individuals use individual PETs or privacy methods~\cite{}, including user perceptions~\cite{}.....\textcolor{red}{Include vastness of area}........
A few studies have in particular investigated the adoption of individual PETs, such as secure encrypted communication~\cite{abu2017obstacles}, anonymous credentials~\cite{benenson2015user}, anonymity service~\cite{harborth2018examining}, and VPN~\cite{namara2020emotional}. %E2EE
While Abu-Salma~\cite{abu2017obstacles} looked into secure communication, Harboth and Pape~\cite{harborth2018examining} and Benenson~\cite{benenson2015user} investigated users' perceived usefulness and effectiveness, %finding that they do not match the anonymous technology's offering. 
and Namara et al.~\cite{namara2020emotional} investigated the emotional and practical considerations of the adoption or abandonment of VPNs as PETs.

While previous research already advances our understanding of the adoption and usage of individual PETs, %they also address individual PETs.
%......this understanding is still limited because: 
%Problems with individual investigations: 
(1) a lot of investigations look into individual PETs~\cite{1,2,3},
(2) we consequently end up with a relatively small-scale targeted outcome, 
(2) we learn about why individuals use PETs, but we do not have classifications based on patterns of use of PETs, 
(3) little is known about inter-PETs usage or PETs usage interactions, and
(4) we have not developed methodology to better understand clusters of PETs. %while we better understand particular PETs, 

%Small scale studies into individual PETs to understand adoption of the particular PETs -- say which individual PETs have been investigated for adoption: E2EE, anonymous communication, VPN, Tor...
%In particular, perceived usefulness and effectiveness do not match the technology's offering, and users exhibit poor trust in the technology~\cite{abu2017obstacles,benenson2015user,harborth2018examining}, and in-correct mental models~\cite{abu2017obstacles}.
\end{RedundantContent}
%In this paper, %we explore the privacy methods Internet users employ to protect their privacy online.
We contribute to the rich landscape of user-centric privacy research, %different approach %build on previous
%Our investigation is different from previous research because 
that has expansively addressed privacy behavior, including via the (extent of) disclosure of personal information~\cite{AcqGro2006,SpiGro2001,norberg2007privacy,AcqGro2005,barnes2006privacy,dienlin2015privacy}, privacy strategies~\cite{oomen2008privacy,coles2011practice,abu2017obstacles,wang2011regretted} or the use of privacy controls and individual PETs~\cite{gerber2019johnny,dienlin2015privacy,utz2009privacy,abu2017obstacles,renaud2014doesn,benenson2015user,harborth2018examining,namara2020emotional}, with a large-scale and cross-national study of the use of a collection of PETs. % across three countries.
%the adoption and usage of an individual PET~\cite{1,2,3} or privacy protection method~\cite{4,5,6}, privacy concern~\cite{} or privacy behavior in general~\cite{1,2,3}. 

We first investigate the use/non-use of a range of privacy methods and PETs across three countries, %as an aspect of behavior, 
rather than taking an in-depth look at how users interact with individual PETs~\cite{renaud2014doesn,benenson2015user,harborth2018examining,namara2020emotional} or engage with controls~\cite{miltgen2015exploring,gerber2019johnny}. 
%XXX.......say why use/non-use as an aspect of behavior is important......XXX
This provides a broad view of what privacy methods and PETs individuals use, and enables us to uncover patterns of use and preferences. %differences in 

Second, while various factors may influence the use/non-use of particular privacy controls and PETs, such as perceived risks~\cite{gerber2019johnny,garg2014privacy}, perceived usefulness~\cite{harborth2018examining,benenson2015user} or demographics~\cite{park2015men,oomen2008privacy},
we focus on what happens following individuals' concern about their privacy, in particular, whether they are aware of PETs and how they perceive use of PETs (c.f. the staircase/steps towards using a particular PET by Renaud et al.~\cite{renaud2014doesn}).
We postulate that PETs use patterns may be evidence of differences in use perceptions, in rationale for use or support needed.
Therefore, rather than investigating perception of use of individual PETs~\cite{benenson2015user,harborth2018examining}, % and why users choose particular PETs, 
we evaluate perceptions across clusters of PETs. % in Study 3. 
This enables us to investigate whether (1) use perceptions and rationale are clearly demarcated across the patterns, %as use is, %that is patterns of use is a reflection of differing use perceptions
as well as (2) to simultaneously understand why individuals prefer a particular collection of PETs.

\begin{RedundantContent}
We first classify patterns of use of privacy methods online (including PETs and non-technology methods), in Studies 1 \& 2 via a clustering approach. 
The classification enables better understanding of PETs usage via patterns of use.
This provides a systematic, evidence-based and reliable point of reference for PETs usage categorisation and future research.
It provides a methodological addition to the state-of-the-art on PETs usage, that can further drive investigations and understanding of PETs use via comparison of clusters, transition between clusters or similarity in usage of PETs from particular clusters.
Our classification can therefore enable investigations of PETs use at a larger scale than previous individual PETs research, and fast-track PETs usage understanding, while also providing a categorisation of PETs based on usage rather than design or protection provided. %(2) via a different classification methodology, 
This provides a pathway to investigate adoption of PETs from a user-centric perspective. %, bottom-up from usage to perception of use, design features and privacy protection, rather than top-down.

Second, we demonstrate the use of the classification to investigate use perceptions across PETs usage categories, as well as the reasoning for use of a particular type of PETs and support required, in Study 3. 

%\paragraph{Methodology}
%Why is similarity in usage important? Similar usage can inform identification of common features that work, trust in HCI and features that work for a particular culture. 

%--- on the need for large scale investigations/combined PETs investigations, PATTERN investigations versus individual investigation of PETs usage, to scale up use

%However, the field yet to have a meta-investigation of privacy methods and PETs usage, as well as patterns of PETs usage that emerge from similarity of use. Such patterns could provide a classification to drive PETs investigations (1) on a larger scale than individual PETs investigations, and fast-track PETs adoption and usage research and (2) via a different classification methodology, that is usage rather than design or protection provided. 
%--- on the need for other methods of investigations, bottom up, from the user/usage rather than design - which is what the cluster map allows\\
%and possibly in the future personalised privacy solutions.

%--- what can we learn from similar usage of PETs --- in cluster, between countries - legal cultural impact on usage\\

A small number of recent research has specifically investigated the challenges or barriers to adoption of privacy technologies.
For example, a first investigation looked into the user acceptance of anonymous credentials, differentiating between primary and secondary goal property of PETs~\cite{benenson2015user}. They found that perceived usefulness of PETs for the primary goal of interaction was the most important factor to adoption of PETs, even outweighing usability and perceived usefulness of the PET (as secondary task).

A second investigation extending the technology acceptance model, found that perceived anonymity (perceived privacy as primary goal) and trust were strong determinants of behavioural intentions and actual use behavior. % and for the extended model, the explained variance is $25.64\%$ more than the one with the core TAM predictors (perceived usefulness and perceived ease of use)~\cite{harborth2018examining}. 

A third investigation into users' understanding of the protection of secure communication tools as well as whether they value that protection~\cite{abu2017obstacles}. %(end-to-end encryption) 
The researchers found that to be adopted, secure tools have to offer their users utility (that is the ability to reach communication partners, the primary goal) and that users do not understand the essential concept of end-to-end encryption or had incorrect mental models, thereby limiting their motivation to adopt secure tools.
\end{RedundantContent}

% The cluster map (i) is in itself a systematic, evidence-based and reliable point of reference, useful for future research decisions, (ii) raises important questions with regards to cluster preference, cluster transition and conditions for transition, (iii) offers a classification which advances methodology enabling further meta-investigations into PETs (based on their usage aggregation).
% Study1&2: Non-technology methods are among most-used methods. 
%Impact: future research much needed on why.

\emph{Contributions.}
 %say a little bit on mixed methodology
%\paragraph{Statistics Argument}
We provide both the quantitative grounding of classification, as well as the rich qualitative investigation of why individuals prefer one type of PETs over another.
We provide a large-scale, cross-national, mixed-methods (quantitative and qualitative) investigation of PETs usage.
%--- large studies allow for generalisations? more likely results are representative of population?\\
We classify patterns of use of privacy methods online (including PETs and non-technology methods), in Studies 1 \& 2 via a clustering approach. 
%The classification enables better understanding of PETs usage via patterns of use.
This provides a systematic, evidence-based and reliable point of reference for PETs usage categorisation and future research, and
%This approach provides
a classification based on patterns of use of PETs (rather than design type or protection provided).
%then enables us to investigate the use of a collection of PETs via 
We therefore also provide re-usable methodology to better understand clusters of PETs, that can further drive investigations and understanding of PETs use via comparison of clusters, transition between clusters or inter-PETs usage.
%similarity in usage of PETs from particular clusters.
%, and raise questions about inter-PETs usage and transition from a PET cluster to another. 

%**Our classification can therefore enable investigations of PETs use at a larger scale than previous individual PETs research, while also providing a categorisation of PETs based on usage rather than design or protection provided.

We find that the cluster map distinguishes usage of PETs into Advanced and Other PETs, and that non-technology privacy methods are among the most used methods online.
We demonstrate use of the PETs usage classification within an investigation of why certain PETs are preferred over others, in Study 3.
%We find that information, social support and ease of use are important potential contributions to enhancing usage of Advanced PETs.
While we find support for themes identified in previous privacy controls and PETs use research, in particular with regards to awareness of privacy protection~\cite{renaud2014doesn,shirazi2014deters}, perceived usefulness~\cite{harborth2018examining,benenson2015user}, usability~\cite{renaud2014doesn,abu2017obstacles}, 
or trust~\cite{harborth2018examining,benenson2015user},
we expand on the themes of information about PETs, privacy needed and usability
as well as report other rationales for not using advanced PETs, such as perceived monetary cost and social support.
We also find that there is no statistical difference in individuals' perceived competency to protect their privacy online, whether they are Advanced or Other PETs users.

\emph{Outline.}
The rest of the paper is organised as follows: we first review background research, and %into privacy protection behavior, 
%user-centric cross-national investigations of privacy, 
%patterns of technology use and privacy perceptions.
then provide a section describing Study 1 and 2, via their aim, methodology and results.
We proceed into presenting Study 3, also via aim, methodology and results.
We complete the paper with an overall discussion and conclusion, and provide an Appendix with additional support, including the questionnaires used in the studies.

\section{Background}
Given our paper computes patterns of privacy methods usage and investigates user perceptions across clusters, 
we provide a review of literature addressing 
(1) privacy behavior, of which use of PETs is one aspect, and
(2) technology adoption and classification of user responses in privacy research. %, as well as, 
%how perception of use influence adoption.

\subsection{Privacy Behavior} %notable strands
\label{sec:privacy_behavior}
%While privacy behavior research is rich,
We review a few strands of privacy behavior related research, in particular,  %of privacy behavior research
%(1) behavior as a response to attitude and the privacy paradox, 
(1) different protection practices, %views on privacy behavior, % (including the use of privacy controls and PETs),  
(2) factors impacting use of privacy controls and PETs, and
(3) cross-national investigations of privacy concerns and behavior.

\begin{RedundantContent}
\subsubsection{Privacy Paradox}
%1- Paradox - privacy concern vs behavior - 3 SLR - concern measured in xyz ways Preibusch \\
Social psychology views behavior as a response to attitude, with literature debating the linkage between the two~\cite{fazio1986attitudes,fazio1981direct,eagly1993psychology}.
With regards to privacy, a number of studies observed the dichotomy between privacy attitude and behavior---a phenomenon which researchers coined the \emph{privacy paradox}---where on the one hand users express concerns about the handling of their personal data and desire protection, while on the other hand, they voluntarily disclose via social networks or rarely make an effort to protect their data actively
~\cite{SpiGro2001,AcqGro2005,AcqGro2006,chellappa2005personalization,norberg2007privacy,dienlin2015privacy}.

%*** explain better + say something about the privacy calculus***
The paradox has been explained via (1) the privacy calculus, where disclosing intentions and behaviors result from an anticipatory, joint assessment of perceived risks and benefits connected to the disclosure of private information~\cite{culnan1999information, dinev2006extended} and (2) individuals' psychological limitations leading to cognitive biases and heuristics~\cite{acquisti2007can,knijnenburg2013dimensionality}.
%individuals' psychological limitations, such as bounded rationality~\cite{AcqGro2005} or attempt for immediate gratification~\cite{acquisti2004}.
Observations of the privacy paradox have also been discussed~\cite{preibusch2013guide,dienlin2015privacy}
and systematic reviews of privacy paradox research 
conducted~\cite{gerber2018explaining,barth2017privacy,kokolakis2017privacy}.
% The privacy paradox, as accurate representation of phenomena,
The resulting observation is that (1) research have conceptualised both privacy attitude and behavior in different ways~\cite{gerber2018explaining,barth2017privacy,kokolakis2017privacy}, and (2) a number of theoretical considerations and factors may explain the privacy attitude-behavior link~\cite{gerber2018explaining}.
%Various instruments have been devised to measure privacy concern~\cite{preibusch2013guide}, 
%For example, attitude has been matched to concern for data collection~\cite{AcqGro2005}, concern about data use~\cite{SpiGro2001}, concern about identifying information~\cite{chellappa2005personalization}, concern about what others know~\cite{barnes2006privacy} and concern about identity theft or access by others~\cite{dienlin2015privacy}.
\end{RedundantContent}

\subsubsection{Privacy Protection Practices}
\label{sec:protection_practices}
%2- privacy behavior investigated in many different ways - from my excel disclosure vs privacy controls \\
The privacy research community is well acquainted with the  \emph{privacy paradox} phenomenon---where on the one hand users express concerns about the handling of their personal data and desire protection, while on the other hand, they voluntarily disclose via social networks or rarely make an effort to protect their data actively
~\cite{SpiGro2001,AcqGro2005,AcqGro2006,chellappa2005personalization,norberg2007privacy,dienlin2015privacy}.
Observations of the privacy paradox have also been discussed~\cite{preibusch2013guide,dienlin2015privacy}
and systematic reviews of privacy paradox research 
conducted~\cite{gerber2018explaining,barth2017privacy,kokolakis2017privacy}.
One of the resulting observations is that both privacy attitude and behavior can be conceptualised in different ways~\cite{gerber2018explaining,kokolakis2017privacy,coopamootoo2017whyprivacy}.

Research has operationalised privacy behavior in different ways, %~\cite{gerber2018explaining,kokolakis2017privacy}, 
including via (the extent of) disclosure, via use of privacy controls and PETs, or via the adoption of protection strategies.
First, privacy behavior as disclosure include observations such as  revelations to an online bot~\cite{SpiGro2001}, Facebook membership~\cite{AcqGro2006} and revelations in a bank and pharmaceutical scenario~\cite{norberg2007privacy} or self-reports such as usage of Facebook or information disclosed~\cite{AcqGro2005,barnes2006privacy,dienlin2015privacy}.
Second, research has also operationalised privacy behavior as engaging in protective control actions and using privacy technology including the general use of privacy controls~\cite{gerber2019johnny}, use of controls in the context of social networks~\cite{dienlin2015privacy,utz2009privacy}, %,joinson2010privacy} - check what protection behavior
and use of secure encrypted communication~\cite{abu2017obstacles,renaud2014doesn}, anonymous credentials~\cite{benenson2015user}, anonymity service~\cite{harborth2018examining}, and VPN~\cite{namara2020emotional}, or reading the privacy policy~\cite{tsai2011effect}.
%We refer to active use of PETs as privacy-protective behavior.
%How is behavior linked to use of tech?
Third, protection strategies have been investigated such as providing incorrect~\cite{oomen2008privacy} or false~\cite{coles2011practice} information, delivering sensitive information in-person or using a ``code" to deliver sensitive information to others~\cite{abu2017obstacles}, and deleting friends or rejecting friend requests, self-censoring or deleting content on social network~\cite{wang2011regretted}.

\subsubsection{Factors Influencing Protective Privacy Behavior}
\label{sec:factors_on_behavior}
%3 - predictors of behavior? (from meta-analysis of most important attitude-behavior links Gerber) \\
Various research have pointed to factors affecting protective control actions 
%use of privacy controls and PETs 
including a systematic review with comparison of predictor effect sizes~\cite{gerber2018explaining}, and obstacles to adoption of PETs~\cite{gerber2019johnny,renaud2014doesn,shirazi2014deters} and secure communication~\cite{abu2017obstacles}.
We provide a summary of factors in Table~\ref{tab:behavior_review} (focusing on use of controls and PETs as behavior, rather than disclosure or protection strategies described above in Section~\ref{sec:protection_practices}).
We also name the context of investigation, such as social networks for inbuilt PETs or name the standalone PET investigated. % \textbf{\textcolor{red}{***make this into a table***}}
We note that some factors were found to be more important than others, for example the risk of sharing~\cite{garg2014privacy}, awareness of consequences of privacy violations~\cite{gerber2019johnny} and contextual aspects of the messaging tool such as fragmented user bases~\cite{abu2017obstacles}, were found to be more important than usability.

%Usability was also reported but found not to be the major obstacle for adoption~\cite{abu2017obstacles,renaud2014doesn} compared aspects contextual to the messaging tool are important such as fragmented user bases with regards to secure communication tools~\cite{abu2017obstacles}.

\begin{table} %[h]
\centering
\footnotesize
\caption{Factors Influencing Protective Privacy Behavior}
\label{tab:behavior_review}
\resizebox{\columnwidth}{!}{
\begin{tabular}{ll} %p{3cm}
\toprule
\textbf{Factor} & \textbf{PETs/Context of Use} \\
\midrule
Years of internet experience & social/technical protection~\cite{park2015men}\\
Internet skill & Facebook privacy settings~\cite{hargittai2010facebook} \\
Perceived rewards in disclosure & general/technical protection~\cite{miltgen2015exploring} \\
Privacy risks concerns &general/technical protection~\cite{miltgen2015exploring} \\
Unawareness of risks & social networks~\cite{gerber2019johnny}\\ %\multirow{2}{*}{social networks}
Risks of sharing &social networks~\cite{garg2014privacy} \\
Perceived usefulness \& ease of use & anonymizing technology~\cite{harborth2018examining,benenson2015user} \\
Emotional considerations  & VPN use~\cite{namara2020emotional}\\
Knowledge about how to protect oneself & E2EE~\cite{renaud2014doesn} \\
Awareness of protection tools & E2EE \& tracking~\cite{renaud2014doesn,shirazi2014deters}\\
No perceived need to act& E2EE \& tracking~\cite{renaud2014doesn,shirazi2014deters}\\
Inability to use protection & tracking~\cite{shirazi2014deters}\\
Becoming side-tracked & tracking~\cite{shirazi2014deters}\\
Usability & E2EE \& secure msg~\cite{abu2017obstacles,renaud2014doesn}\\
Social Influence & secure msg \& social networks~\cite{abu2017obstacles,gerber2019johnny}\\
Education &Facebook sharing~\cite{garg2014privacy}\\
Gender& technical protection~\cite{park2015men,oomen2008privacy}, \\
& social network control~\cite{tifferet2019gender}\\
%&\\
\bottomrule
\end{tabular}
}
\footnotesize{\emph{Note: E2EE refers to end-to-end encryption}}
\vspace{-.4cm}
\end{table}

\begin{RedundantContent}
\begin{enumerate}
\item years of internet experience~\cite{park2015men}, 
\item perceived rewards in disclosure~\cite{miltgen2015exploring}, 
\item privacy risks concerns~\cite{miltgen2015exploring}, 
\item unawareness of risks~\cite{gerber2019johnny} or risks of sharing~\cite{garg2014privacy} in the context of social networks, 
\item perceived usefulness and ease of use of anonymizing technology~\cite{harborth2018examining,benenson2015user}, 
\item emotional and practical considerations with regards to VPN use~\cite{namara2020emotional},
%demographic variables such as 
\item lack of concern and knowledge about how to protect oneself/awareness of protection tools, no perceived need to act, inability to use protection tool or becoming side-tracked In the context of end-to-end encryption and tracking protection~\cite{renaud2014doesn,shirazi2014deters},
\item too much effort in reading complex policies, lack of knowledge of protection possibilities, sharing behavior of others and social pressure, convenience or utility of social web for expression in the context of social media and tools~\cite{gerber2019johnny},
\item education in the context of Facebook sharing~\cite{garg2014privacy},
\item gender influence for more technical protection~\cite{park2015men,oomen2008privacy}. % technical protection behavior
\end{enumerate}
\end{RedundantContent}

\begin{RedundantContent}
In the context of end-to-end encryption and tracking protection, obstacles to the adoption are lack of concern and knowledge about how to protect oneself/awareness of protection tools, no perceived need to act, inability to use protection tool or becoming side-tracked~\cite{renaud2014doesn,shirazi2014deters}.
%Obstacles to using of tracking protection tools: worry about other privacy issues, not aware of consequences, not concerned/nothing to hide, not aware of protection tools, not able to use PETs properly, or become side-tracked~\cite{shirazi2014deters}.
With regards to secure communication tools~\cite{abu2017obstacles,renaud2014doesn}, usability was found not to be the major obstacle for adoption but aspects contextual to the messaging tool are important such as fragmented user bases~\cite{abu2017obstacles}.
In the context of social media and tools such as WhatsApp, Telegram or Facebook, obstacles named are too much effort in reading complex policies, lack of knowledge of protection possibilities, sharing behavior of others and social pressure, convenience or utility of social web for expression~\cite{gerber2019johnny}.
%Garg - usability not as important as perceived risk of sharing
\end{RedundantContent}

%Usability not the major obstacle for adoption of secure communication tools~\cite{abu2017obstacles,renaud2014doesn} but aspect contextual to the messaging tool are important  such as fragmented user bases.

%3.1----- why people choose certain controls --- perception of risk most important determinant of privacy behavior (Garg 2014) \\
%3.2----- still in 2019 (Gerber et al 2019), people are unaware of risks albeit personalised ads \& finance \\
%Harboth and Pape~\cite{harborth2018examining} and Benenson~\cite{benenson2015user} investigated users' perceived usefulness and effectiveness, %finding that they do not match the anonymous technology's offering. 
%and Namara et al.~\cite{namara2020emotional} investigated the emotional and practical considerations of the adoption or abandonment of VPNs as PETs.

%\subsubsection{Supporting Privacy Behavior}
%4 - how privacy behavior could be encouraged - Tsai Egelman 2011

\subsubsection{Cross-National Influence}
%Privacy is a human right~\cite{assembly1948universal}, yet with no universally agreed definition of privacy. 
%In the online context, however, a common understanding of privacy is the right to determine when, how, and to what extent personal data can be shared with others.
While there is no universal privacy or data protection law that applies across the whole Internet, a number of international and national privacy frameworks have largely converged to form a set of core, baseline privacy principles, 
%These frameworks 
that establish fair information practices for consumers, businesses and organisations. % in the age of data-centricity.
Whereas the General Data Protection Regulation (GDPR) provides the legal backbone for data protection and privacy in Europe~\cite{voigt2017eu}, the US does not have an all-encompassing law like the GDPR but has a variety of federal and state laws that aim to protect a citizen’s privacy and online data~\cite{piper2019data} and the UK established the Data Protection Act 2018 as its implementation of the GDPR~\cite{parliament2018data}.
%Cross-country differences in privacy regulation have been thought to impact trends on the Internet~\cite{Erbschloe2001}.
With regards to Internet users, differences have been observed in privacy concern, behavior or valuation of information across countries~\cite{cho2009multinational,marshall2008social,reed2016thumbs,cvrcek2006study,bellman2004international,thomson2015socio}, with some influence of national culture (where national culture is viewed as the collective mindset distinguishing members of one nation from another~\cite{hofstede1984culture}, and is embedded in the way members think, feel and act~\cite{hofstede2001culture}),
%that influences a person's actions~\cite{triandis1994culture}) 
in particular via the collectivist versus individualist distinctions~\cite{cho2009multinational,marshall2008social,reed2016thumbs} or relational mobility~\cite{thomson2015socio}.
Others have looked into the privacy calculus across countries in contexts such as e-commerce~\cite{dinev2006privacy}, social network~\cite{krasnova2010privacy}, health records~\cite{dinev2016individuals} or driving behavior~\cite{kehr2015blissfully}. %driving behavior data
These studies leveraged population samples worldwide~\cite{reed2016thumbs,bellman2004international}, or compared the US versus Asian countries~\cite{thomson2015socio,marshall2008social}, the US versus Asia-Pacific countries~\cite{cho2009multinational}, the US versus European countries~\cite{dinev2006privacy,krasnova2010privacy,dinev2016individuals,kehr2015blissfully}, or across European countries~\cite{cvrcek2006study,hallinan2012citizens}.
%Individuals across European countries value the privacy of their information differently~\cite{cvrcek2006study}.

\subsection{Technology Adoption \& Patterns}
%XXX....Say why patterns are important + say which areas of technology they have been applied to + how patterns can be derived via cluster analysis + how perceptions have been clusterised in privacy research ....XXX

%\subsubsection{Why are patterns/pattern analysis important (to tech use)?}
%Gives an overview of trends in usage among a collection of objects/techs, helps us find characteristics of usage that manifest beyond single tech use but across a pool of them.

%Detecting patterns of technology use is important because

\subsubsection{Model of Technology Adoption}
\label{sec:models_of_tech_adoption}
%This viewpoint is supported by the Technology Acceptance Model (TAM), whose key purpose is to provide a basis for tracing the impact of external variables on internal beliefs, attitudes and intentions~\cite{legris2003people}.
The Technology Acceptance Model (TAM) provides the theoretical background for understanding why users accept or reject technology~\cite{davis1989perceived}.
It has been empirically demonstrated to successfully predict $40\%$ of system use~\cite{legris2003people}.
TAM suggests that perceived ease of use (PEOU) and perceived usefulness (PU) are two most important factors explaining technology use.
%TAM has been empirically demonstrated to be successful in predicting $40\%$ of system use~\cite{legris2003people} and has been used in various information system contexts, including security~\cite{serge paper, other paper cited b4} and privacy behavior~\cite{benenson2015user}.
Privacy research employing TAM as psychological construct ranged from (1) understanding adoption of specific PETs, such as user acceptance of anonymous credentials~\cite{benenson2015user}, anonymity technology~\cite{harborth2018examining}, and VPNs~\cite{namara2020emotional}; 
to (2) perceived privacy in adoption of technologies such as social media~\cite{rauniar2014technology}, 
e-commerce~\cite{lallmahamood1970examination,vijayasarathy2004predicting}, biometrics~\cite{miltgen2013determinants},
Snapchat~\cite{lemay2017passion} or trading systems~\cite{roca2009importance}.

%\subsubsection{Technology Patterns \& Cluster Analysis}
\subsubsection{Cluster Analysis for User-Centric Privacy}
Investigations into patterns of use of technology may enable identification of broad usage trends across single or a collection of technologies. 
This can be conducted via cluster analysis, which supports the identification of patterns and provides an objective methodology for quantifying structural characteristics of observations, 
%is useful for taxonomy description, data simplification, 
and relationship identification~\cite{balijepally2011we,macqueen1967some}.
%A few investigations of patterns of use of technology have been conducted in different contexts, including young adults' use of technology for academic purposes~\cite{thompson2013digital,sharpe2019exploring} and families' use of technology in the home~\cite{chesley2006families,turk2016households}, including the use of cluster analysis as methodology to investigate technology usage patterns~\cite{ebbert2020patterns}.
Cluster analysis has also been employed in user-centric studies in the privacy context to better understand perceptions or behaviors. Example investigations using cluster analysis include
SNS engagement and privacy habits in relation to transport~\cite{ebbert2020patterns}, 
privacy management strategies in Facebook~\cite{lankton2017facebook},
privacy perceptions based on geographical regions~\cite{huang2016privacy},
perceptions of information sensitivity between countries~\cite{schomakers2019internet}, %germany vs brazil?
cross-country comparison of adolescents' privacy perceptions~\cite{soffer2014privacy},
privacy versus sharing perceptions~\cite{olson2005study}, and
privacy and risks perceptions to better understand behavior~\cite{coventry2014perceptions}.
%privacy and security perceptions between age groups~\cite{kaiser2016privacy}, and
%trust perceptions in E-Commerce~\cite{Atchariyachanvanich2008 - Cluster Analysis of E-Commerce Customer Profiles Based on Trust Perception}.

\begin{RedundantContent}
They mainly focused on usage for young adults in the context of education context and usage patterns in the home. These included qualitative investigations~\cite{thompson2013digital,sharpe2019exploring,chesley2006families} and cluster analyses~\cite{ebbert2020patterns,mccarthy2016technology}.

\paragraph{In the context of learning \& for young adults}
Thompson~\cite{thompson2013digital} investigated how digital natives' use of technology is associated with their approaches to learning.
Gell et al~\cite{gell2015patterns} investigated technology use (use of e-mail/text messages and the internet) by sociodemographic and health characteristics and prevalence ratios for technology usage by disability status, while
Sharpe~\cite{sharpe2019exploring} looked into UK college students' engagement with institutional virtual learning environments to Web 2.0 tools in both personal/social and educational contexts. Some support was found for a connection between personal/social and academic uses of technology. The majority of learners were using educational technologies in fairly simplistic ways that satisfy the demands of their course. Significant relationships were found between digital practices in college and subject studied and gender. 

\paragraph{In the home context}
Investigations included (1) longitudinal usage patterns for families, which suggest a trend toward adoption and use of e-mail, the Internet, cell phones, and pagers over time, this trend toward continuing use is stronger for some technologies (e-mail, the Internet) than for others (cell phones, pagers)~\cite{chesley2006families}, and
and technology adoption and use patterns of households of specific technology, such as the digital terrestrial television~\cite{turk2016households}. % in Italy.
%These dimensions were further investigated through their relations with demographic variables of primary decision-makers in households.

%Crabtree et al. provide a methodological pattern language framework that grounds design in the socially organized patterns of technology usage that exist in the home~\cite{crabtree2001patterns}.

\subsubsection{Cluster Analysis Computations of Technology Use}
Ebbert and Dutke~\cite{ebbert2020patterns} used cluster analysis to describe German students' distinct usage patterns of a lecture recording service, in particular characterised by usage frequency, repetitiveness and selectivity in watching, lecture attendance, and social context and location.

McCarthy et al.~\cite{mccarthy2016technology} explores the possibility of grouping respondents to a transport survey based on their answers to questions on their social networking (SNS) use via cluster analysis. Their aim was to contribute to an understanding of respondents' technology engagement and privacy habits.

%\subsubsection{Privacy Methods Usage}
%To-date, there are no published research studies on patterns of use of privacy methods and technologies.
%Such research could depict methods used (as in this paper), as well as how privacy methods are used and for what purposes
% patterns could be what methods (as I did), but can also be how privacy methods are used, for what purposes
\end{RedundantContent}

\begin{RedundantContent}
\label{sec:models_of_tech_adoption}
\subsubsection{Psychological Model}
%Two main: TAM + UTAUT\\
The core Technology Acceptance Model (TAM) was proposed by Davis to address why users accept or reject technology~\cite{davis1989perceived, davis1989user}. 
%explain users' behavior intentions to use technology
A key purpose of TAM is to provide a basis for tracing the impact of external variables on internal beliefs, attitudes and intentions~\cite{legris2003people}.
It suggests that perceived ease of use (PEOU) and perceived usefulness (PU) are two most important factors explaining technology use.

TAM has been empirically demonstrated to be successful in predicting $40\%$ of system use~\cite{legris2003people} and has been used in a number of information system contexts, including ecommerce, telemedicine software and employee adoption of information security among others. It has also been used to investigate adoption of specific PETs~\cite{benenson2015user}. % ---well there are privacy examples too ----
%The core model depicts the mediating role of perceived ease of use and perceived usefulness on the probability of system use.
%effects of perceived usefulness and perceived ease of use as predictors on behavioural intention, where perceived ease of use also influences perceived usefulness. 
%In particular, perceived usefulness and effectiveness do not match the technology's offering, and users exhibit poor trust in the technology~\cite{abu2017obstacles,benenson2015user,harborth2018examining}, and in-correct mental models~\cite{abu2017obstacles}.

TAM has been extended to include categories of variables during investigations including %\begin{itemize}
(1) external precursors such as prior usage, self-efficacy/perceived competency;
(2) factors from other theories such as subjective norm, expectation, risk and trust;
(3) contextual factors such as gender, culture and technology characteristics; and
(4) consequence measures such as perceptual and actual usage, and attitude~\cite{king2006meta}.
%\end{itemize}

\subsubsection{Privacy Technology Adoption}
While TAM provides a foundation for better understanding technology adoption, 
we are aware of only three recent research strands that have specifically investigated the challenges or barriers to adoption of privacy technologies, with the help of constructs from the TAM.
First, Benenson et al. looked into the user acceptance of anonymous credentials, differentiating between primary and secondary goal property of PETs~\cite{benenson2015user}. The authors found that perceived usefulness of PETs for the primary goal of interaction was the most important factor to adoption of PETs, even outweighing usability and perceived usefulness of the PET (as secondary task).

Second, Harboth and Pape extended the TAM, and found that perceived anonymity (perceived privacy as primary goal) and trust were strong determinants of behavioural intentions and actual use behavior and for the extended model, the explained variance was $25.64\%$ more than the one with the core TAM predictors~\cite{harborth2018examining}.  %core predictors

In a third recent study, Namara et al. used the TAM constructs to guide understanding of the factors that motivate individuals to adopt or abandon using VPNs as PETs. They find that individuals who adopt VPNs are motivated by emotions, whereas those who abandon them are motivated by practical motivations
~\cite{namara2020emotional}. %look into both the practical and emotional considerations for adoption, determining 
\end{RedundantContent}

%\section{Theoretical Foundation}
%Although privacy breaches are commonly advertised and the scale of online harms are reportedly increased, Internet users are not more strongly concerned about their privacy online. 
%While the status of user concern may be explained via their choice for economic incentives, cognitive biases, habituation, lack of personal experience of privacy issues, lack of protection knowledge, social influence or an illusion of control, 
%it may also be explained by users' current privacy practices (or lack thereof) on the Internet, use/non-use of PETs and 
%their perception of use and experience of particular privacy methods and PETs.

%However, the stabilisation in users concern also depicts a level of comfort with current to privacy practices (or lack thereof) on the Internet.
%This comfort, in turn, may stem from perception of use and experience in using particular privacy methods and PETs.

\section{Study 1 \& 2}
\label{sec:study1-2}
\subsection{Aim}
We explore usage of privacy methods online in visual maps.
%and investigate patterns and interactions among PETs and non-technology methods  %different groups.
We sample participants from the US, UK and German populations.

\subsubsection{Privacy Method Patterns}
We posit that individuals likely engage with privacy via a habitual collection of methods or pattern of actions, % or that become a habit. 
where habits are actions that have become automatically triggered by situational cues~\cite{lally2010habits}.
%and habit development follows a period of successful self-regulation of a new behaviour. 
%**Consciously controlled behaviors become automatic over time with repetition, where high levels of cognitive processing or reflection is usually not necessary when activities become habitual. 
We investigate as \textbf{RQ1}, 
%\begin{question}[\textsf{RQ-E}]
%How do users' tasks assessment of maintaining their privacy versus sharing online differ across the four cognitions of psychological empowerment?\\
``What patterns emerge in individuals' privacy methods usage and preference?
How similar are usage patterns across privacy methods?" %preference
Because we do not have a-priori expectations of the nature of relationships between privacy method preferences, we employ 
a Cluster Analysis~\cite{macqueen1967some} to naturally distinguish different patterns of use, and further look into usage patterns in relation to country via a Correspondence Analysis~\cite{greenacre1988clustering,greenacre2017correspondence}.
%Correspondence Analysis as an exploratory method to tease out similarities and patterns, and further distinguish between different patterns via Cluster Analysis.

\subsubsection{Cross-National Privacy Method Similarities} % \& Differences} %& Sharing 
%Accessibility and use of privacy and sharing methods may have cross-cultural influences.
%\subsubsection{Methods}
%We define privacy/sharing habits as the list of individual privacy/sharing methods used by the participant.
%We compare the list of individual privacy and sharing methods.
%\begin{question}[\textsf{RQ-T1}]
The diffusion and adoption of technology do not necessarily follow a common pattern in terms of rates or timing across countries. %*While some countries are receptive to new technologies or those for a particular purpose, others are not.
Adoption of technology is often influenced by cross-national factors~\cite{bagchi2003influence,erumban2006cross}, where in particular, use of technology for privacy may be influenced by regulation.
We investigate as \textbf{RQ2}, 
``What methods are mostly used to protect one's privacy online? what similarities emerge between countries?"

\begin{RedundantContent}
\subsubsection{Interaction between Privacy Methods Usage}
We investigate as \textbf{RQ3},
``How does the extent of use of different patterns of privacy methods (including PETs and non-technology methods) impact the likelihood of using particular PETs?"
\end{RedundantContent}

\subsection{Method}
We conduct two survey studies online. %within subject studies where participants answer both privacy and sharing questions that are then used in comparison.
The first study (Study 1) is aimed at identifying a preferred list of privacy methods.
The second study (Study 2) employs the compiled list of methods from Study 1 to query participants about their use of the range of privacy methods identified.

%\subsubsection{Assignment}
%In the second study, we compared preferred privacy methods between countries, thereby including a between-subject analysis. 

%We compared privacy and sharing empowerment for each participant. 
%We randomly assigned participants to answer either the privacy or sharing empowerment questions first.

\begin{RedundantContent}
We conduct a first study with $N=180$ participants, querying participants about empowerment perception and their preferred list of privacy and sharing methods. %with participants from the USA, UK and Germany 
We compile lists of preferred privacy and sharing methods. 
Using these lists, we design and run a second study with a representative sample with $N=606$ participants. We ask participants to select their preferences from the lists provided. %from the USA and the UK
\end{RedundantContent}

\subsubsection{Participants}
\label{sec:study1_participants}
%[XXXTODOXXX say why we sampled from the UK, US \& Germany]
With their advanced digital economies, Europe and the US may be considered as the drivers of protection technology around the globe.
However, Europe versus the US differ in privacy regulation~\cite{pwc2016data} and 
%consequently 
potentially also in individuals' privacy protection patterns.
%individuals' privacy protection culture.
%end-user salience wrt privacy protection.
%Similarities: advanced digital economies, Europe and US may my the drivers of using technology for protection around the globe \\
%Differences: 

%\subsubsection{Recruitment}
For Study 1, we sampled $N=180$ participants, comprising $N=58$ US participants, $N=62$ UK participants and $N=60$ German (DE) participants. We sampled participants from Prolific Academic.  
%The US sample was recruited from population of Amazon Mechanical Turk workers, while the UK and DE sample were from Prolific Academic. 
%Prolific Academic's participant pool is more naive and less dishonest, and 
The data quality of Prolific Academic has good reproducibility~\cite{peer2017beyond}, and 
 is comparable to Amazon Mechanical Turk's which is widely used in security and privacy user studies.
%We chose these platforms because \dots [a bit on the pros and cons of these sampling services]~\cite{Peer et al. 2017 - Beyond the Turk - Alternative platforms for crowdsourcing behavioral research}.

For Study 2, we recruited an $N=907$ sample from the US, UK and DE via Prolific Academic. 
The samples from the US and the UK were representative of age and gender of the respective countries, as provided by Prolific Academic. %Note that, US individuals with hispanic ethnicity may describe themselves as mixed, other or white. %and ethnicity demographics 
For the DE sample, we did not achieve a representative sample in terms of gender and age.
%While we use that sample to investigate our research questions, we foresee extending to representative samples of other countries in the future.
%focused the second and main study US and UK representative samples

Table~\ref{tab:demographics} provides a summary of the demographic details for the two studies.
The studies lasted between $10$ to $20$ minutes. Participants were compensated at a rate of \pounds$7.5$ per hour, slightly above the minimum rate of \pounds$5$ per hour suggested by Prolific Academic.

\begin{RedundantContent}
%\subsubsection{Demographics}
Table~\ref{tab:demographics} provides a summary of the demographic details for the two studies, with sample size $N$, mean age, gender, education level and ethnicity.
$5\%$ of the German sample had an education level lower than high school for the first study and $1\%$ for the second study.
For the second study, $6$ UK participants reported to have a PhD, $4$ for the US and $9$ for DE.
% $N = 68$, mean age of $36$ years old, $32$ female and $36$ male and the UK population via Prolific Academic, $N = 62$, mean age of $30.65$ years old, $43$ female and $19$ male.
%[data on other demographics data perhaps a table for ethnicity etc]
\end{RedundantContent}

\begin{table}[h]
\centering
\caption{Participant Characteristics}
\label{tab:demographics}
\footnotesize
%\resizebox{\textwidth}{!}{
\begin{tabular}{lcrcrr} %crrrrlrrrrr}
\toprule
&\textbf{Country}& \textbf{$N$} & \textbf{Mean Age} & \multicolumn{2}{c}{\textbf{Gender}} \\ %&&\multicolumn{4}{c}{\textbf{\%Education Level}} &&\multicolumn{5}{c}{\textbf{\% Ethnicity}} \\
%ethnicity - asian, black, mixed, other, white
\cline{5-6}
%\cline{8-11}
%\cline{13-17}
&&&& \#Female & \#Male \\ %&&HighSchool&College&Undergrad&Masters/PhD && White & Black & Asian & Mixed & Other\\
\midrule
\multirow{3}{*}{\textbf{Study 1}}&US & $58$ & $35.53$ & $29$ & $29$ \\%&& 24.1&31.0&36.2&8.6 && 82.8 & 5.2 & 5.1&5.2&1.7\\
&UK & $62$ & $30.65$ & $43$ & $19$ \\%&& 22.6&19.4&41.9&16.1&& 88.7 &3.2&3.2&4.8&-\\
&DE & $60$ & $30.68$ & $27$ & $33$ \\%&& 30.0 & 13.3 & 28.3 & 21.7&&96.7&-&-&3.3&-\\
\midrule 
\multirow{3}{*}{\textbf{Study 2}}& US &303 & 43.72&155&148 \\%&&39.9 &22.1 &20.1 &14.2&&69.3&14.9&8.9&4.3&2.6\\
&UK & 303 & 44.21 &154 &149 \\%&&26.7&17.5&32.0&18.5 && 77.6 &5.3 & 10.9 & 4.3 & 2.0\\
&DE & 301 & 28.91 & 115 & 186\\% && 31.2 & 15.6 & 28.6 & 23.6&& 93.0&0.7&1.9&3.7&0.7\\
%\midrule 
%\multirow{3}{*}{\textbf{Third Study}}& US &40 & 31.50 &25&15 &&35.0&20.0&32.5&12.5 &&70.0&2.5&7.5&7.5&12.5\\
%& UK & 101 &37.23 & 58& 43  &&18.8&25.7&38.6&14.9 &&76.2&4.0& 13.9&5.0& 0\\
%& DE  & 42 & 28.98 & 16&26&&26.2&19.1&28.6&23.8 && 100.0&-&-&-&-\\

\bottomrule
\end{tabular}
%}\\
%\vspace{-.4cm}
\end{table}

\subsubsection{Procedure}
\label{sec:procedure}
Study 1 aims to identify and compile a list of privacy methods preference. %and sharing 
We do so via an open-ended question, across three countries.
Study 1 consisted of %\begin{inparaenum}[(a)]
a questionnaire on demographics, and
%\item a description of privacy online, and the four psychological empowerment questions, %purpose defined as achieving privacy online 
an open-ended query to list three to five tools most often employed to achieve the purpose of privacy online.
%\item a description of sharing online, and the four psychological empowerment questions. %, and %purpose defined as achieving privacy online
%\item an open-ended query to list three to five tools most often employed to achieve the purpose of sharing online.
%\end{inparaenum}

\begin{figure}
\centering
\includegraphics[keepaspectratio,width=1\columnwidth]{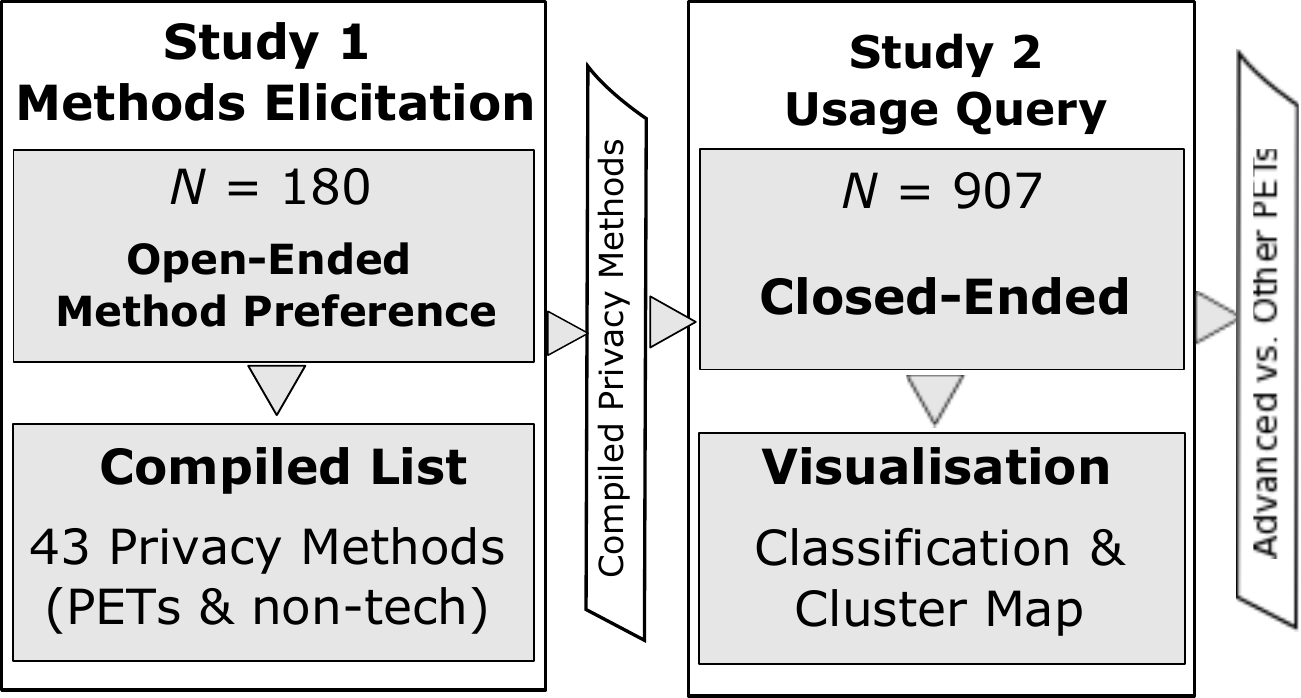}
%[scale=0.7{\textwidth}]
%\vspace{-.2cm}
\caption{Studies 1 \& 2 design.}
\label{fig:structure}
\vspace{-.4cm}
\end{figure}

Study 2 followed the same format as Study 1, except that we changed the open-ended queries of the first study to close-ended privacy methods questions, for participants to select the methods they mostly use from the whole list provided.
We also shifted to a larger sample for the three countries. % and a representative pool for the US and the UK as provide by Prolific Academic.
%\end{inparaenum}
\begin{RedundantContent}
We defined privacy and sharing for the two studies, thereby focusing participants to a specific meaning.
We developed the definition of [privacy/sharing] online with inputs from Coopamootoo \& Gro{\ss}'s findings of the cognitive content of individuals' [privacy/sharing] attitude~\cite{coopamootoo2017whyprivacy}. 
In particular, privacy attitude has contents of `others as individuals or organisations who pose a threat, %They experience a feeling of fear. %such as by asking individuals ``What does privacy online mean to you?", i
while sharing attitude includes `others as connections including friends, family'. % and experience happiness~\cite{coopamootoo2017whyprivacy}.
%For the description of [privacy/sharing] online, we developed the definition with inputs from the content analysis of individuals' [privacy/sharing] attitude, conducted as part of an empirical study~\cite{coopamootoo2017whyprivacy}.
%\begin{tcolorbox}
We defined privacy online as \\
\emph{``to control access to information that are sensitive or personal, to be informed of other individual and business practices such as collection, processing and use of personal information disclosed, and to have the choice on disclosure and how one's information is dealt with."}\\
We defined Sharing online as \\
\emph{``to create content and share with other web users (such as sharing one's opinion or expertise) and also to share personal information or life events with close connections, friends and family."}
 %\end{tcolorbox}
\end{RedundantContent}
We provide a summary of the procedure in Figure~\ref{fig:structure}.

\subsubsection{Measurement Apparatus}
\label{sec:measurement_study1-2}
%\subsubsection{Perception of Privacy protection}
\emph{Study 1:} %Privacy Methods Usage}
We queried participants on the individual privacy methods they most often use, eliciting their own methods via the instruction 
%an open-ended question in the first study. 
%In particular, we ask 
``List between 3 to 5 tools that you have most often employed before to achieve the purpose of privacy online."

\emph{Study 2:}
%We request preference report from the compiled list of methods from the first study.
We asked participants to rate the list of privacy methods provided with whether they use them `very often' or `very rarely/not at all'. 
While we opted for binary data, we found that querying `If you have used them before: Yes/No' would not fit our inquiry. `Yes' would include `once and never again' usage type. We wanted to capture methods that are popular with users (and include protections that are often but not always used by choice) via `very often' versus those participants have not thought of or have used in very rare circumstances (not the go-to protection). 

We provide the full question and wording as provided to participants in Table~\ref{tab:study2_questionnaire} in the Appendix~\ref{sec:study2_questionnaire}. 
We pre-tested this list for comprehension with $N=7$ colleagues and acquaintances in the computing and social science departments, and believe that participants in Study 2 understood the wordings, as they were named by participants in a similar sample in Study 1.
We also provide an open-ended option for `other privacy method'.

We encouraged participants to respond truthfully by stating that the surveys are anonymous in the consent form, and also asking for truthful answers in the questionnaires.
We included attention check questions throughout the survey and the list of privacy methods presented to participants in Study 2 was randomised. %$43$

%...xxxx TODO xxxx ..... ALSO FULL WORDING OF FOR STUDY 2 TO REFER TO HOW THEY WERE NAMED BY PARTICIPANTS THEMSELVES IN STUDY1....argue that participants understand as we speak their language -How about OTHER OPTION ....

%In addition, we compute privacy behavior as the total number of different methods participants employ to protect their privacy online.
%Similarly, we compute sharing behavior as the total number of different methods participants employ to to share information online.

\begin{table*}[h]
\centering
\footnotesize
\caption{Privacy Methods Categorised by Design Type and Privacy Protection (as elicited in Study 1).\\
We provide the shorthand shown in the Cluster Map in brackets, where applicable.}
\label{tab:all_privacy_methods}
\resizebox{\textwidth}{!}{
\begin{tabular}{lllll} %p{3cm}
\toprule
\multirow{28}{*}{\textbf{Protection}} && \multicolumn{3}{c}{\textbf{Design Type}} \\
\cline{3-5}
&&\textbf{Built-in} & \textbf{Standalone} & \textbf{Non-Technology}\\
&\\
%\hline

%\midrule
%\multirow{10}{*}{\textbf{Anonymity}} 
&\textbf{Anonymity}&Encryption&Erasery&Not Store Info (notstore)\\
&&Clear/Delete info/history (clearhistory) &Tor&Anonymous profile names (anonprofile)\\
&&Pseudonyms/Onion (pseudonyms)&Proxy&NotGivePI / LimitSharing / MinimalInfo (limitshare)\\
&&&IPHider&Several/Bogus / LimitedUse Emails (afewemail)\\
&&&Virtual machine (virtualma)&Fake Info\\
&&&&Limit Use of SNS Accounts (limitsns)\\ %//No SNS\\ Del FB\\
&&&&SwitchOffCamera/Devices/PortableHD (switchoffcam)\\
&&&&No Access Acc In Public Place/Networks (nopubac)\\  %/No Shared Computers\\
&&&&Not use FB (noFB)\\
&&&&Not Engaging Online/Careful/Not Signing Up (not-engage)\\

&\\
%\midrule
%\multirow{4}{*}{Browsing History \& Tracking Prevention} 
&\textbf{Browsing History \&}&Private Browsing/incognito (privbrowsing)&DuckDuckGo&\\
&\textbf{Tracking Prevention} &Anti-tracking addon (antitrack) &Ghostery&\\
& &No location tracking (nolocation) &NoScript&\\
&&Clear/Limit cookies (clearcookies)&&\\

&\\
%\midrule
%\multirow{2}{*}{Communication \& Filtering} 
&\textbf{Communication \&} &Adblock&Firewall&\\
&\textbf{Filtering} &HTTPS&VPN&\\

%\midrule
&\\
&\textbf{Prevent Leaking \&}&Privacy settings (privset) &Password manager (pwd mger)&Not save (notsavepwd) or reuse password (notreusepwd)\\
&\textbf{Stealing of Data} &Opt out &Paypal&Read terms of service (readterms)\\ %Antivirus
&&Private profiles (anonprofile)&Anti-spyware (antispy)& Request data collected, GDPR (reqdatacol)\\ %Read business practice\\
&&&Anti-malware (antimal)& no newsletter, think twice (nonewslet)\\
& &&Kapersky& Website care/No suspicious sites (suspiciousweb)\\

\bottomrule
\end{tabular}
}
\end{table*}

\subsubsection{Limitations}
\label{sec:study2_limits}
\emph{Representative Sample:}
While in Study 2 our sample is representative of the US and UK populations, with respect to age and gender, we did not achieve a representative sample for DE (Germany). This explains the difference in mean age for DE compared to the US and UK, in Table~\ref{tab:demographics}.
From the 2019 demographics record, the German population has a median age of $47.4$, with $37.3\%$ of the population aged over $54$~\cite{indexmundi}.
The crowd sourcing platform could not cater for the representative sample for the sample size we required.
Our relatively large sample size however decreases the probability of assuming as true a false premise.
In addition, our sample characteristic is likely transferable to the larger German population, where participants in our sample exhibit behaviors of the general population~\cite{polit2010generalization}.
However, in future research, it will be valuable to strive towards representativeness across the sample, including via attributes such as ethnicity and education level. % or computer skills. %as well as investigate 
Alternative crowdsourcing platforms for participant recruitment may also need to be investigated.

%We however believe, ...LARGE SAMPLE SIZE - COULD WE GENERALISE, OVERRIDE BIAS?

\emph{Self-Report Bias:}
Our studies rely mainly on self-reports rather than system observations. 
Self-report is a valuable and widely used method of querying users in security and privacy user studies.
While self-reports can be argued to induce bias, research investigating response bias in security user studies has found that self-report insights can translate to real-world environments~\cite{redmiles2018asking}. %Our findings therefore contribute to the field on usage of privacy methods.
%+ this study is different from crawler/automated observation of use studies 

The list of privacy methods presented to participants in Study 2 were named from a pool of participants of $N=180$ across the three countries in Study 1.
We pre-tested the list for comprehension for Study 2 with researchers who may be thought to bias the test outcomes. However, the list was sourced from participants themselves in Study 1, therefore came from the `field'.

\subsubsection{Ethics}
We obtained approval from the University's Faculty Ethics Committee before starting the research, for all the studies (Studies 1 \& 2, as well as Study 3).
We sought participants' consent for data collection prior to their responding to the questionnaires, and we did not collect identifying information. 
Participants were free to leave the survey at anytime and were compensated slightly above the advised rate as described in Section~\ref{sec:study1_participants}.

\subsection{Results}
\subsubsection{Compiled Privacy Methods from Study 1}
\label{sec:privacy_methods}
We collect participants' responses of $3$ to $5$ most used privacy tools or methods in Study 1, with the $N=180$.
We end up with $43$ privacy methods coded across the three countries and $11$ participants stating they `don't know'.

%From the list of methods, privacy design type and privacy protection categories emerged, as depicted in Table~\ref{tab:all_privacy_methods}. 
%\subsubsection{Design Type}
%The lists of methods emerging from the first study are provided in `Compiled Privacy Methods' section, and 
Participants reported privacy methods that may be designed as %\begin{inparaenum}[(a)]
(a) a standalone privacy technology, or as %where privacy is the primary goal for the user
(b) privacy controls or settings integrated (built-in) within other tools such as browsers or messaging tools.
Participants also reported strategies that do not involve using privacy controls, such as `give fake info', which we name as `non-technology' methods.
%where privacy is the secondary goal for the user.
%\end{inparaenum} 
%There are also debates in literature about the need for privacy by design~\cite{}. %, as well as arguments and evidence~\cite{} that the design type influence adoption of technologies [privacy and/or general].
%**We categorize participants' privacy method preference into whether they were standalone or builtin technology/methods or whether they were non-technology ways of ensuring privacy protection, which we categorize as user-defined methods.
%\subsubsection{Privacy Protection}
We also loosely name four possible protection categories, namely %\begin{inparaenum}[(1)]
(1) anonymity (ANO), 
(2) browsing history and tracking prevention (BHP), 
(3) communication privacy and filtering (COP), and
(4) preventing leaking and stealing of data (PLS).
%\end{inparaenum}
%Table~\ref{tab:all_privacy_methods} summarizes the privacy methods and categories.
We present the privacy methods elicited according to its design type and privacy protection type provided, as depicted in Table~\ref{tab:all_privacy_methods}. 
(Note that this loose categorisation only serves to better present the types of privacy methods named, and do not affect the rest of the analysis.)

\subsubsection{Privacy Method Patterns from Study 2}
\label{sec:cluster_CA}
%Talk about the dimensionality of privacy methods usage data - that there could be a large number of variables explaining why people who use a certain privacy method also use/not use the $42$ other methods. TODO DIMENSIONALITY OF DATA We want to find out if the data can be explained by a couple of dimensions only.
%We identify associations and patterns of behavior in the reported use of privacy methods, and 
We investigate \textsf{RQ1}, 
``What patterns emerge in individuals' privacy methods usage and preference?
How similar are usage patterns across privacy methods?"
%How similar are individuals' privacy methods usage and preference?
%What patterns of use emerge?"
%What are the patterns of privacy methods preference, based on gender and sample country? What are the components of the different patterns?"
%**We consider our dataset of privacy methods preference across the three countries to have high dimensionality, where a number of variables can explain why individuals who use a certain privacy method also use or not use the $42$ other methods, and which combination of usage of privacy methods are connected.
Our dataset is a contingency table of 43 rows (privacy methods) and 3 columns (count of participants in the US, UK and DE who reported to `very often’ use the privacy method). We provide the dataset in Table~\ref{tab:dataset} in Appendix~\ref{sec:study2_dataset}.

\begin{figure*}
\centering
\includegraphics[keepaspectratio,width=.8\textwidth]{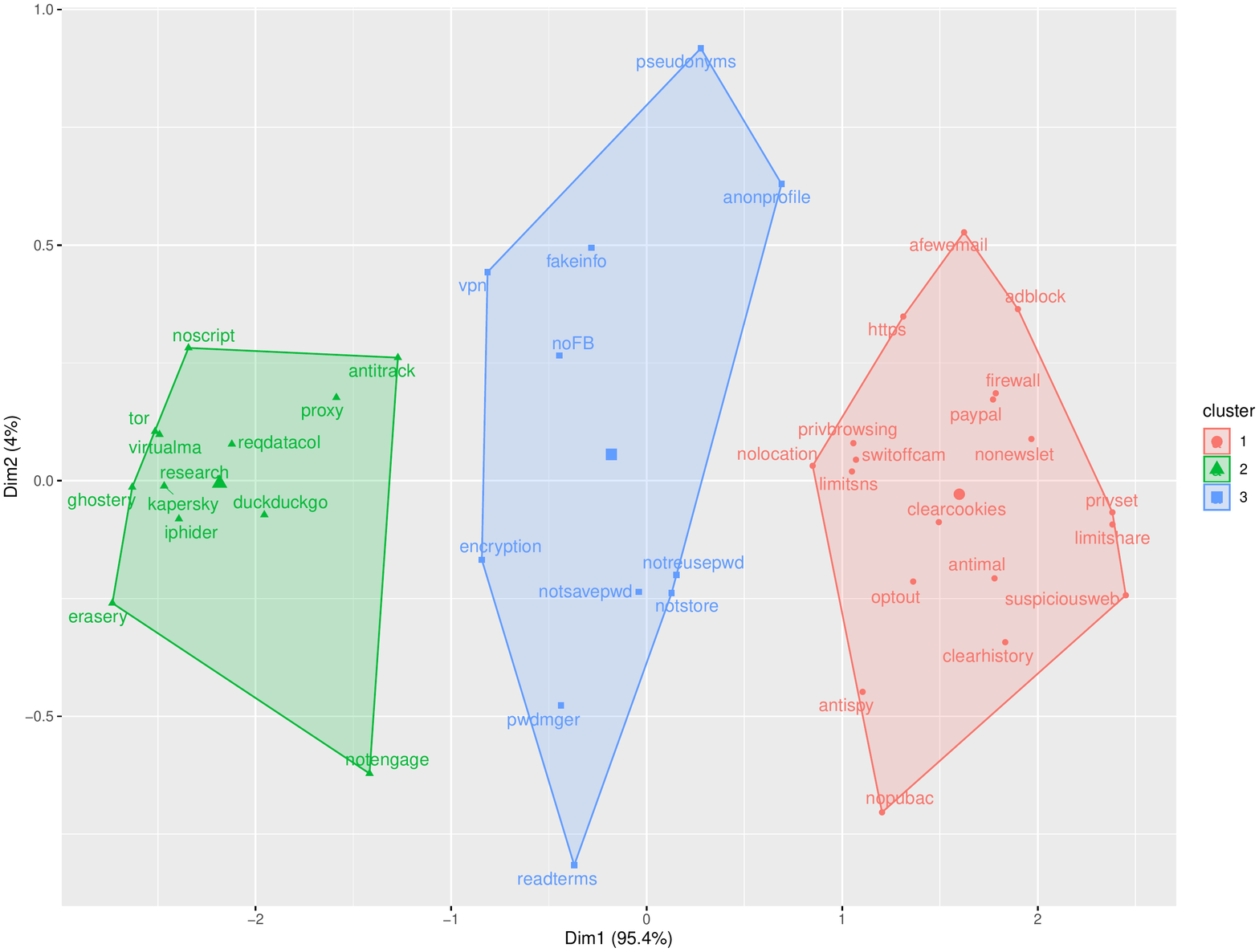}
\vspace{-.2cm}
\caption{Privacy Methods Clusters along the Popularity of Privacy Method Dimension; left to right = least to most used.}
\label{fig:clusters}
\vspace{-.3cm}
\end{figure*}

\begin{figure*}
\centering
%\begin{subfigure}{.5\textwidth}%
%\centering
\includegraphics[height = 10cm, width=\textwidth]{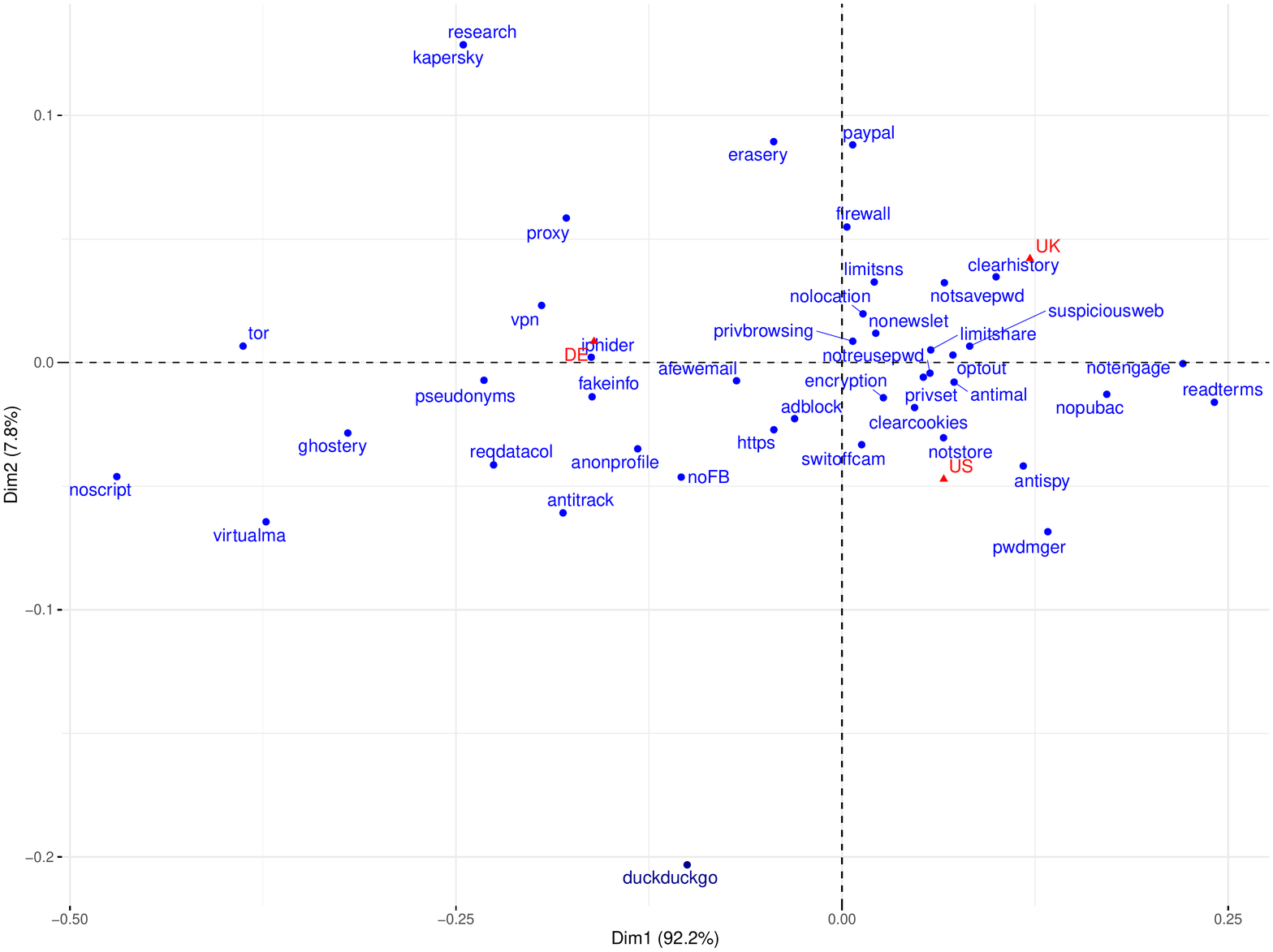}
\caption{Spatial Plot of Privacy Methods \& Country.}
\label{fig:symmetric_plot}
\vspace{-.3cm}
\end{figure*}

We conduct two multivariate analyses and visualise our dataset: 
(1) via a Cluster Analysis~\cite{kaufman2009finding,romesburg2004cluster} that classifies the set of $43$ privacy methods as suggested by natural groupings in the data themselves, 
%*Clusters are created based on characteristics the data possesses, where the data is classified into a small number of manageable and meaningful groups.
%The visualisation, in the form of a 
and produce a cluster map as a simplified depiction of the relationships between privacy methods;
and (2) via a Correspondence Analysis~\cite{hair1998multivariate,greenacre2017correspondence} that investigates and visualises the relationship between the $43$ privacy methods across $3$ countries, as well as the similarities between privacy methods given their country profiles.
%Our data set consists of $43$ rows for privacy methods and $3$ columns for countries. % 43 instead of 42 because 'not save' & 'not reuse' passwords are two items in CA and Cluster Analysis.
The main results for this subsection are Figures~\ref{fig:clusters} and~\ref{fig:symmetric_plot}.

\emph{\textbf{Cluster Analysis.}}
%*We aim to identify groupings or clusters of privacy methods with similar characteristics, where the privacy methods in the same cluster are as similar to each other as possible, and privacy methods in different clusters are as different to each other as possible.
In the next few paragraphs, we report on the cluster analysis and results via the following steps: \begin{enumerate}
\item we conduct a dimensionality reduction, 
\item we determine the optimum number of clusters and compute the cluster analysis, 
\item we assess the internal validity for the cluster analysis, and
\item we visualise the cluster map and interpret the dimensions.
\end{enumerate}
We use the R package Factoextra~\cite{kassambara2017package} for the cluster computation and visualisation. %, in particular via the \emph{fviz\_cluster} function.

%The function transforms the initial set of variables into a new set of variables using Principal Component Analysis (PCA)~\cite{wold1987principal} as dimensionality reduction technique.
We first compute a Principal Component Analysis (PCA)~\cite{wold1987principal} as dimensionality reduction technique, using the \emph{prcomp} function in R. 
The PCA operates on the 3 country variables to output 3 factors or principal components (PC), namely PC1, PC2 and PC3.
%factors, namely principle component 1 (PC1), principle component 2 (PC2) and principle component 3 (PC3).
PC1 accounts for 95.4\% of the variation in the data, while PC2 accounts for 4\% of variation and PC3 accounts for 0.5\%.
We describe the importance of the principal components in Table~\ref{tab:PC_importance}.
Together, PC1 and PC2 account for $99.5\%$ of the variation in the data, which is a large amount of variance.
We therefore focus on PC1 and PC2.

\begin{table}[H]
\centering
\caption{Importance of Principal Components.}
\label{tab:PC_importance}
\footnotesize
%\resizebox{\textwidth}{!}{
\begin{tabular}{lrrr}
\toprule
 &PC1&PC2 & PC3\\
%\midrule
\cline{2-4}
Standard deviation & 1.692& 0.348 & 0.128 \\
Proportion of Variance &  0.954& 0.040 & 0.005  \\
Cumulative Proportion&  0.954 & 0.995 &1.000  \\
\bottomrule
\end{tabular}
%}
\vspace{-.1cm}
\end{table}

We report the factor loadings for PC1 and PC2 in Table~\ref{tab:factor_loadings}, %and PC3 
where the country variables load similarly to PC1, while the US and the UK load in opposite direction to DE for PC2.

\begin{table}[H]
\centering
\caption{Factor Loadings.}
\label{tab:factor_loadings}
\footnotesize
%\resizebox{\textwidth}{!}{
\begin{tabular}{lrr}
\toprule
Variables &PC1&PC2\\
\midrule
US& 0.585& -0.303 \\%&-0.752 \\
UK&  0.580&-0.492 \\ %& 0.649  \\
DE&  0.567& 0.816 \\%& 0.111  \\
\bottomrule
\end{tabular}
%}
\vspace{-.1cm}
\end{table}

Second, we employ $k$-means as cluster analysis method on the principal components. % and visualise a cluster map, 
$k$-means clustering is the most commonly used unsupervised machine learning algorithm for partitioning a given data set into a set of $k$ groups or clusters~\cite{macqueen1967some}, and groups observations by minimizing Euclidean distances between them.
Our cluster analysis focuses on the first two principal components that explain $99.5\%$ variance, and visualise them as Dimension 1 (Dim1) and Dimension 2 (Dim2), where in general, dimensions are variables or \emph{features} of data objects~\cite{hair1998multivariate}. %, to represent the original variables. %, and is a projection of the original dataset.
Dim1 refers to PC1, while Dim2 refers to PC2.

\begin{figure}
%\vspace{15cm}
\centering
\includegraphics[keepaspectratio,width=.95\columnwidth]{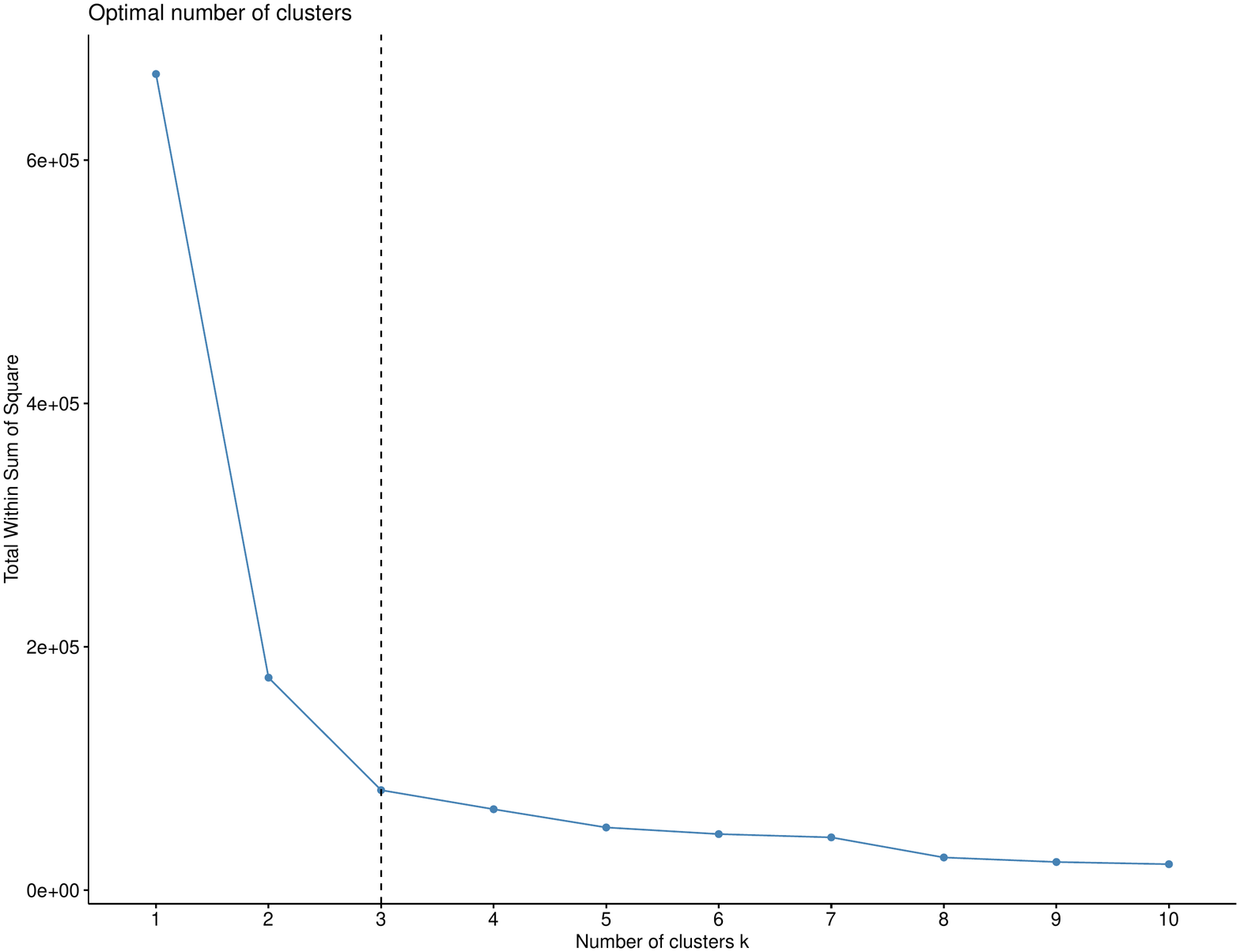}
\vspace{-.3cm}
\caption{Determine number of clusters via \emph{Elbow Method}}
\label{fig:elbow}
\vspace{-.4cm}
\end{figure}

To determine the optimal number of clusters, 
we use the \emph{elbow method}, which consists of optimising the within-cluster sum of squares (WSS)~\cite{hair1998multivariate}. % and looks at the total WSS as a function of the number of clusters.
%\emph{k}-means clustering defines clusters such that the total intra-class cluster variation or total WSS is minimised. The total WSS measures the compactness of the clustering. We therefore want total WSS to be as small as possible~\cite{hair1998multivariate}.
%The \emph{elbow method} looks at the total WSS as a function of the number of clusters.
We choose a number of clusters so that adding another cluster does not improve the total WSS much better.
From the plot in Figure~\ref{fig:elbow}, the total WSS does not show a sharp improvement after $3$ clusters. 
We subsequently conduct a \emph{k}-means cluster analysis with $3$ clusters, and 
%The classification of observations into groups requires computation of \emph{distances} or \emph{similarity} between each pair of observation, resulting in a distance or dissimilarity matrix.
%*We use the \emph{Euclidean} distance as a classical method for distance measure and 
summarise distances between pairs of clusters in Table~\ref{tab:dist_matrix}.
The $3$ clusters correspond to $3$ distinct groups of privacy methods preference, of sizes $19$, $13$, and $11$.
Privacy methods in the same cluster are as similar as possible in terms of user preference, with low intra-class variation of $11.7\%$, %this was the total within cluster variation                high-intra class similarity -  lowest within cluster variation 
whereas privacy methods in different clusters are as dissimilar as possible, with high inter-class variation of $88.3\%$. 
% (i.e with low inter-class similarity) or high inter-class variation.
We provide further cluster statistics in Table~\ref{tab:cluster_stats}.
\begin{table}[h]
\centering
\caption{Separation Matrix, that is a matrix of separation values between all pairs of clusters.}
\label{tab:dist_matrix}
\footnotesize
%\resizebox{\textwidth}{!}{
\begin{tabular}{lrrr}
\toprule
Cluster &1&2&3\\
\midrule
1&  0.0000 \\ %&160.70159 &48.91830 \\
2&    160.7016  & 0.00000 \\ %& 38.48376  \\
3&     48.9183  &38.48376 & 0.00000  \\
\bottomrule
\end{tabular}
%}
\vspace{-.3cm}
\end{table}

\begin{table}[h]
\centering
\caption{Cluster Statistics}
\label{tab:cluster_stats}
\footnotesize
\resizebox{\columnwidth}{!}{
\begin{tabular}{lrrr}
\toprule
Value Description & Cluster 1 & Cluster 2 & Cluster 3 \\
\midrule
Cluster size & 19& 13& 11\\
Diameter & 122.69& 115.90& 134.38\\
%Avg. within cluster distance & 57.28 &49.30& 68.53\\
Cluster separation & 48.92& 38.48& 38.48\\
Cluster avg. silhouette width & 0.58& 0.64& 0.43\\
\midrule
Avg. distance b/w clusters &\multicolumn{3}{c}{204.53}\\
Avg. distance within clusters & \multicolumn{3}{c}{57.74}\\
Within cluster sum of squares &\multicolumn{3}{c}{82868.96}\\
Between cluster sum of squares & \multicolumn{3}{c}{623240.80}\\
\bottomrule
\end{tabular}
}
Diameter = maximum within cluster distance \\
Cluster separation = vector of clusterwise minimum distances of a point in the cluster to a point of another cluster.
\end{table}

%Computing correlational and euclidean distance of similarity
%*\paragraph{Cluster Validation}
%*Cluster validation evaluates the goodness of clustering algorithm results, where internal cluster validation is a method that uses the internal information of the clustering process to evaluate the goodness of the clustering structure without reference to external information.
%*Since we do not have external ``true" cluster values in advance, we use the internal method.

\begin{figure}
%\vspace{15cm}
\centering
\includegraphics[keepaspectratio,width=.95\columnwidth]{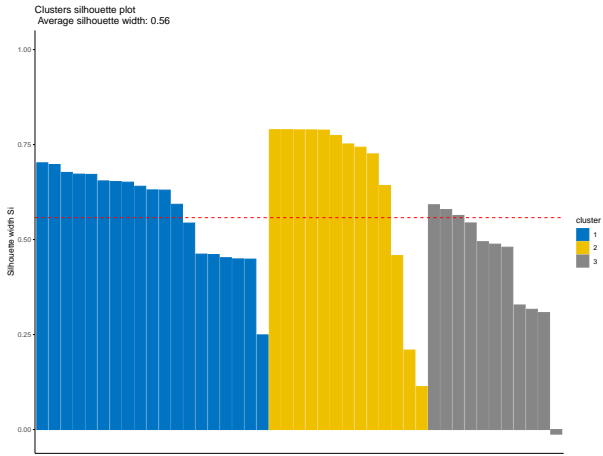}
\vspace{-.3cm}
\caption{Silhouette Plot for Cluster Validity}
\label{fig:silh}
\vspace{-.5cm}
\end{figure}

%Internal validation measures often reflect the compactness, connectedness and the separation of the cluster partitions~\cite{liu2010understanding}.
Third, for internal validation of the cluster partitions~\cite{liu2010understanding}, we conduct a \emph{Silhouette Analysis} which measures how well an observation is clustered and estimates the average distance between clusters~\cite{rousseeuw1987silhouettes}.
The silhouette plot in Figure~\ref{fig:silh} displays a measure of how close each point in one cluster is to points in the neighbouring cluster, with an average silhouette width $S_i$ of $0.56$ across the $3$ clusters.
$S_i$ values can range from $-1$ to $1$, where negative values signify that observations are placed in the wrong cluster and values close to $1$ signify that observations are well clustered.
Our $S_i$ value of $+0.56$ shows that our clustering is okay.

%For cluster validation, we compute a \emph{Silhouette Plot} and the \emph{Dunn Index}.

%\paragraph{Cluster Map \& Dimensions}
Fourth, we visualise the cluster analysis on a two-dimensional Cluster Map, as provided in Figure~\ref{fig:clusters} % 
(using the \emph{fviz\_cluster} function in R).
%patterns. K-means clustering with 3 clusters of sizes 19, 12, 11
%The first dimension accounts for $95.2\%$ of variance in the data.
Dim1/PC1 explains $95.4\%$ of variation in the data and have similar contribution from the three countries as shown by PC1 in Table~\ref{tab:factor_loadings} and the $x$-axis of Figure~\ref{fig:clusters}.
The high variance explained by this factor likely represent the main characteristic of our dataset, that is the usage counts.
In particular, Figure~\ref{fig:clusters} shows that `erasery' and `suspiciousweb' are at the extreme ends of the $x$-axis and from Table~\ref{tab:dataset}, they are the least and most used method, respectively.
%(the $x$-axis of Figure~\ref{fig:clusters}, and PC1 in Table~\ref{tab:factor_loadings}) 
We therefore interpret Dim1 to characterise `The Popularity of Privacy Method' (both visually and confirming with the dataset).
The red, right-most cluster pertains to most used privacy methods while the green and left-most cluster pertains to the least used privacy methods.

Dim2 (the $y$-axis of Figure~\ref{fig:clusters}, from PC2 in Table~\ref{tab:factor_loadings}) depicts the country of use, where the positive end of the $y$-axis refers privacy methods usage more associated to DE, while negative end refers to usage more associated to the US and UK.
For example, `pseudonyms' and `readterms' are both in the same cluster but are at the opposite ends of the $y$-axis.
From Table~\ref{tab:dataset}, we find that `pseudonyms' is most used in DE compared to US/UK while `readterms' is least used in DE compared to US/UK.

\begin{RedundantContent}
%\paragraph{Cluster Dendogram}
We further illustrate how closely the privacy methods in the clusters are associated via a dendogram in Figure~\ref{fig:dendo} of the Appendix.
When interpreting the dendrogram we focus on the height at which any two privacy methods are joined together. In Figure~\ref{fig:dendo}, we can see that `limit sharing' and `privacy settings' are most similar, as the height of the link that joins them together is the smallest. The next two most similar methods are `TOR' and `virtual machines'.
\end{RedundantContent}

\emph{\textbf{Correspondence Analysis.}}
The Correspondence Analysis (CA) is an exploratory technique that graphically represents the relations among rows and columns of a contingency table, in a spatial map, and is 
%CA is an 
increasingly popular for dimensional reduction and perceptual mapping~\cite{hair1998multivariate}.
%ref in Hair et al 7th edition book - multivariate data analysis pg 595}.
%*CA has been beneficial in areas ranging from health and medicine to social sciences, archaeology, ecology, software development and market research~\cite{van1994correspondence,sourial2010correspondence,loslever2009using}.
%We then run a Cluster Analysis~\cite{macqueen1967some} to clarify similarity and differences between groups of privacy methods.
\begin{RedundantContent}
The steps in CA are similar to traditional decompositional Multi Dimensional Scaling (MDS) methods in that it includes  %\begin{inparaenum}[(1)]
(1) gathering measures of similarity or preference,
(2) using dimension scaling/reduction technique to estimate relative position of each object in multidimensional space, and
(3) identitying and interpreting the axes of the dimensional space in terms of perceptual and/or objective attributes.
%\end{inparaenum}
\end{RedundantContent}
\begin{RedundantContent}
However, compared to decompositional MDS and compositional factor and discriminant analyses, CA is %\begin{inparaenum}[(1)]
(4) compositional rather than decompositional because the spatial map is based on association between objects and a set of descriptive characteristics specified by researchers,
(5) its most direct application is in portraying the correspondence of categories of variables, particularly those measured in nominal scale. The correspondence is the basis for the spatial map.
(6) CA is uniquely beneficial in representing rows and columns in joint space.
%\end{inparaenum}
\end{RedundantContent}
%\paragraph{Computing Correspondence Analysis}
%\emph{\textbf{Contingency Table:}}
%We build a table of privacy methods preference across the three countries.
%We build a contingency table with $3$ columns for countries US, UK, DE and $43$ rows for each of the privacy methods.
%Each cell in the table contains the number of participants choosing that privacy method in a particular country.
We use the dataset of privacy methods count across each country as in Table~\ref{tab:dataset}.
%*\emph{Measure of Association or Similarity: } %association or similarity
%*From the cross-tabulated data, we can view the relative number of privacy methods for each country in the contingency table. %To identify any pattern in countries, for example that a country has more or less the same number of particular privacy methods or that certain privacy methods are used similarly in US, UK and DE, we need to define what the `expected' preference count would be so we can say whether the actual preference count is more or less.
%To identify these patterns of `more or less', 
%*We standardize cell counts to make them comparable and the standardized measure simultaneously considers the differences in row and total columns.
%*CA uses \emph{Chi-Square} statistics ($X^2$) to standardize cell frequency values of the contingency table and form the basis for association or similarity.
%The $X^2$ value can be easily transformed into a similarity measure. 
%The process of calculating $X^2$ (squaring the difference) removes the directionality of the similarity. to restore the directionality, we use the sign of the original difference, but reversed to make it more intuitive.
%Like MDS, positive/larger differences are greater association and negative/smaller values are less association.
%\emph{\textbf{CA Result:}} % \& Assessing Overall Fit: 
In the next few paragraphs, we report on the CA and results via the following steps: %\begin{enumerate}
(1) we compute the CA, 
(2) we visualise the spatial plot, and 
(3) we interpret the dimensions.
%\end{enumerate}
%We use the R package Factoextra~\cite{kassambara2017package}

First, we identify `DuckDuckGo' as an outlier, and treat it as a supplementary row, a practice used to treat outliers in correspondence analysis~\cite{bendixen1996practical}.
We compute the CA via the \emph{CA} function from the Factoextra package~\cite{kassambara2017package} in R. We find significant independence of variables with $X^2 = 317.016$, $p<.001$, and 
%We consider the number of axes or dimensions that are relevant for the interpretation and good representation of the data.
that the first dimension accounts for $92.16\%$ of the variance in the data while the second dimension accounts for $7.84\%$. Together these two dimensions account for $100\%$ of variability. %, which is the maximum variance to be explained. %The more variance explained, the fewer insights will be missed.
 %an acceptably large percentage to represent our data.
\begin{RedundantContent}
Figure~\ref{fig:screeplot} provides a \emph{scree plot} depicting the contribution of each dimension in the variability, as well as a red dash line showing the average eigenvalue of approximately $11\%$. Dimensions $3$ and $4$ explain too little variance to be further considered in the analysis.
/Users/nkpc1/Dropbox/Human in SP/Coop2019a_PrivacyvSharing Technologies/figures
\begin{figure}
\centering
\includegraphics[keepaspectratio,width=1\columnwidth]{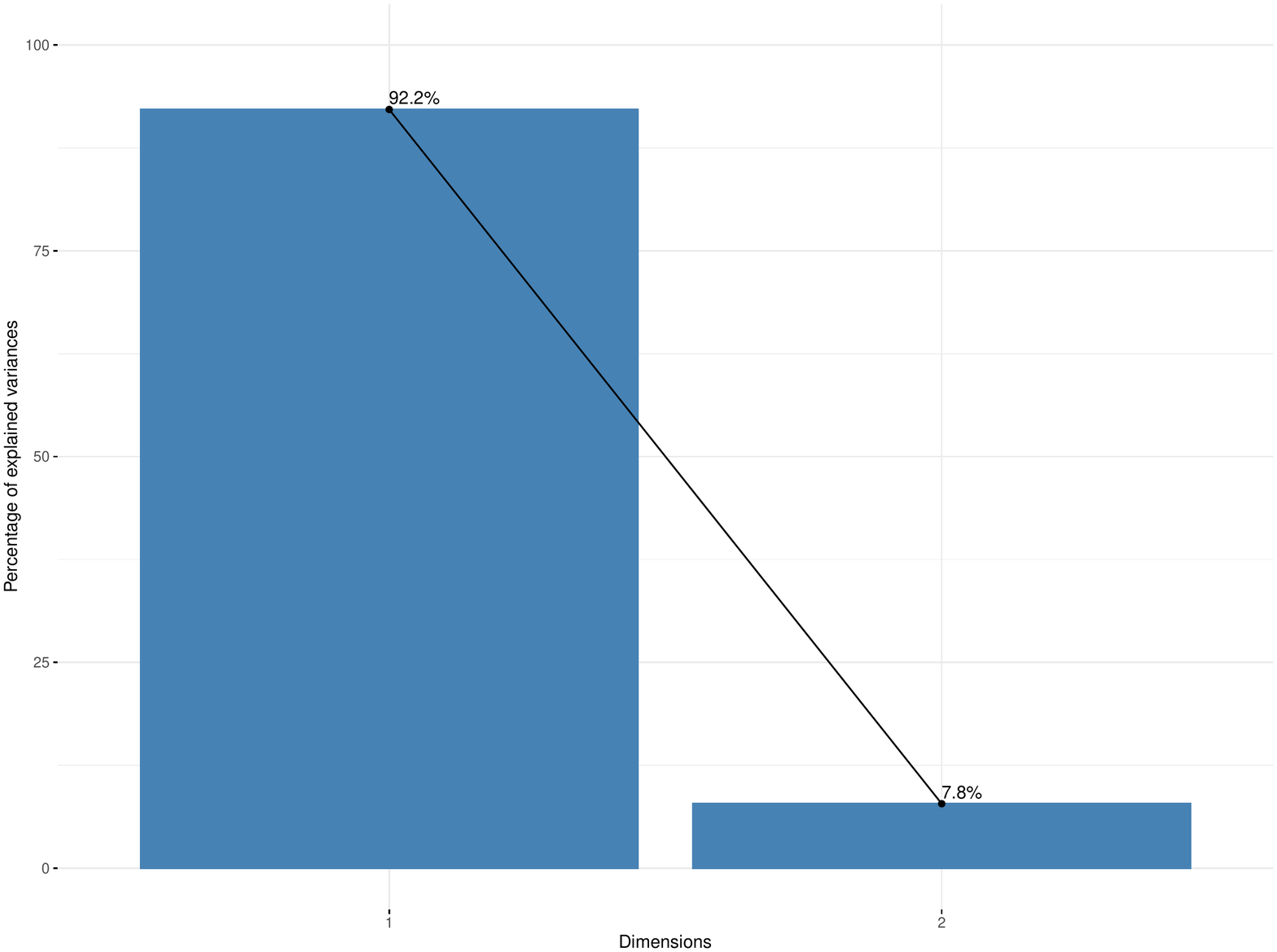}
\caption{Scree plot showing the \% of variance explained by dimensions.}
\label{fig:screeplot}
\end{figure}
%keepaspectratio,width=.8\textwidth
\end{RedundantContent}

%\paragraph{Spatial Map}
%The spatial map shows the relative positioning of objects.
%***The similarity (signed $X^2$) values provide a standardized measure of association, with which CA creates a spatial map.
%**With this association measure, %/similarity 
%**CA creates a spatial map by using the standardized measure to estimate orthogonal dimensions upon which the categories can be placed to best account for the strength of association represented by the $X^2$ distances.
%*Similar pairs of objects are positioned so that the distance between them in multidimensional space is smaller than the distance between any other two pairs of objects. 
%*CA displays both rows and columns in a reduced-dimensional space by decomposing the variability, of the data table and identifying the factors that best synthesize the data variability.
%by decomposing the total inertia, that is variability, of the data 
Second, we visualise a spatial plot, using the \emph{fviz\_ca\_biplot} function in R, as provided %(also known as a symmetric plot) %showing a global pattern in within the data
%in particular a scatterplot 
in Figure~\ref{fig:symmetric_plot}. The plot shows the row and column profiles simultaneously in common space. 
%Figure~\ref{fig:symmetric_plot} summarises the associations between privacy methods across country (and gender) profiles, methods that are close in proximity are more similar in country profiles than those that are far apart.
%to detect and represent underlying structures in the privacy methods preference dataset. CA allows exploration of relationships between privacy methods to reveal patterns, and a global picture of salient relationships in the dataset.
%CA describes the patterns geometrically by locating each variable of analysis as a point in low-dimensional space. The results are interpreted on the basis of the relative positions of the points and their distribution along the dimensions. As categories become more similar in distribution, the closer (distance between points) they are represented in space.
%Although it is used as an exploratory technique, it is particularly powerful as it ``uncovers" groupings of variable categories in dimensional space, providing key insights on relationships between categories. %without needing to meet assumptions requirments of other technisues to analyse categorical data.
%\emph{Proximity of Privacy Methods:}
%*The graphical output of CA is a set of two-dimensional scatterplots where rows/columns are represented as points.
The distance between data points of the same type (row to row, that is privacy method to privacy method) is related to the degree to which the rows have similar profiles in the columns, that is relative frequencies in country variable.
The more points belonging to the same set (pattern) are close to each other and the more similar their profiles are.
As example, individuals who use IPHider and those who give fake information as privacy methods have similar country profiles, %(likely from DE) 
whereas those who use Tor and `read terms and conditions' as privacy methods have very dissimilar country profiles. % (DE vs UK or US).

%\paragraph{Interpreting the Dimension 1}
%The visual display of data helps the interpretation and allows patterns to emerge. 
Third, to interpret the dimensions, we look into the privacy methods and country that contributes most to defining and characterising each dimension.
We provide a plot of the contribution of the privacy methods to the first dimension in the Appendix~\ref{sec:2nd_correspondence_plot} as Figure~\ref{fig:contribplot}.
%degree of association between the privacy methods and the first dimension (measured via cos2) in the Appendix as Figure~\ref{fig:qualityplot}.
%\emph{\textbf{Dimension 1:}}
The top $10$ privacy methods contributing to the definition of Dimension 1 (the x-axis of Figure~\ref{fig:symmetric_plot}) in decreasing rank, with positive (+ve) or negative (-ve) contributions are: 
%\begin{inparaenum}[(1)]
(1) NoScript ($-ve$),
(2) pseudonyms (-ve), 
(3) read terms and conditions ($+ve$),
(4) Tor (-ve),
(5) no public access (+ve),
(6) virtual machine (-ve),
(7) VPN (-ve), 
(8) not engage (+ve),
(9) Ghostery (-ve), 
(10) give fake info (-ve).
%\end{inparaenum}
In addition, with regards to country contribution, DE contributes $61\%$ to the definition of Dimension 1.
We interpret Dimension 1 to characterise `Radicalness of Protective Method' ranging from the extreme positive end of the x-axis with `Not Engage', `No Public Access' and `Read terms and conditions' as non-technological solutions to the extreme negative end of the x-axis with NoScript, pseudonyms, Tor, virtual machine, VPN, Ghostery as PETs and not engaging with `Give Fake Information'.
By our interpretation, the `Most Radical Protective Methods' are at the extreme ends of the x-axis thereby contributing most positively and negatively, and the `Least Radical Protective Methods' are nearer to origin ($x=zero$).

\begin{RedundantContent}
The top $5$ privacy methods most associated with Dimension 1 are: 
\begin{figure}
\centering
\includegraphics[keepaspectratio,width=1\columnwidth]{./figures/contrib_dim1_top5}
\caption{Contribution of Methods to Dimension 1.}
\label{fig:contrib_dim1}
\end{figure}
\end{RedundantContent}

\begin{RedundantContent}
\emph{\textbf{Dimension 2:}}
The top $10$ privacy methods contributing to the definition of Dimension 2 (the y-axis of Figure~\ref{fig:symmetric_plot}) in decreasing rank, with positive or negative contributions are: 
\begin{inparaenum}[1)]
\item Paypal (+ve),
\item research before engaging (+ve),
\item Kaspersky (+ve),
\item firewall (+ve),
\item password manager (-ve), 
\item anti-tracking extension (-ve), 
\item anti-spyware (-ve), 
\item not using Facebook (-ve), 
\item proxy (+ve), and 
\item clear browser history (+ve).
\end{inparaenum}
The US contributes $57\%$ to the definition of Dimension 2.
We interpret Dimension 2 to characterise `Visibility of Privacy Method - hidden' / `conscious choice'??? TODO ......
\end{RedundantContent}

\begin{RedundantContent}
\subsubsection{Representation of first dimension}
Since the first dimension explains $83.9\%$ of the variance, we provide a depiction of the (1) quality (degree) of association between the privacy methods and the first dimension in Figure~\ref{fig:quality_plot}, and (2) the privacy methods with the largest contributions to the first dimension in Figure~\ref{fig:contrib_plot}.

\begin{figure*}%
\centering
\begin{subfigure}{.5\textwidth}%
\centering
\includegraphics[height = 6cm, width= 7cm]{./figures/viz_ca_barrplot_cos2_top15_dim1}
\caption{Association}
\label{fig:quality_plot}
\end{subfigure}%
\begin{subfigure}{.5\textwidth}%
\centering
\includegraphics[height = 6cm, width= 7cm]{./figures/viz_ca_barrplot_contrib_dim1_top15}
\caption{Contribution}
\label{fig:contrib_plot}
\end{subfigure}%
\caption{Quality of Association \& Contribution of Privacy Methods to Dimension 1}
\end{figure*}
\end{RedundantContent}

We note the difference between the cluster and spatial maps.
While the cluster map shows the natural grouping of privacy methods usage, the spatial map shows similarities in privacy methods usage based on their relationship to the different countries. 
In addition, the cluster analysis uses euclidean distance, whereas the correspondent analysis uses chi-square statistic.
Therefore, Dimension 1 of both maps do not refer to the same characteristic of privacy method usage.

\subsubsection{Cross-National Method Use Similarities}
%\subsubsection{Method Similarities}
From Study 2, we investigate \textsf{RQ2}
``What methods are mostly used to protect one's privacy online? What similarities emerge between countries?"
%``Is there an association between [privacy/sharing] method used and country of residence of the participant?"
%\subsubsection{Privacy}
%\subsubsection{Tools Preference}
%Figure~\ref{fig:pr_all} provides a descriptive view of the Top10 most commonly employed privacy method across all countries ($N=180$), and 
Table~\ref{tab:top10} shows a depiction of the top $10$ privacy methods preferences across the three countries, where we observe
that $4$ of the privacy methods appear in the top $10$ most reported methods in all three countries. These methods are 
%\begin{inparaenum}[(1)]
(1) privacy settings, 
(2) limit sharing, 
(3) website care, and 
(4) no newsletter.
%\end{inparaenum}
In addition, we find $8$ privacy methods similarities in the top $10$ most reported methods for both the UK and US, $6$ methods similarities between the UK and DE, and $5$ methods similarities between the US and DE.
\vspace{-.24cm}

\begin{table*}[h]
\centering
\caption{Top 10 Privacy Methods by Country starting with most frequently mentioned}
\label{tab:top10}
\footnotesize
\resizebox{\textwidth}{!}{
\begin{tabular}{l@{}lll c l@{}lll c l@{}lll}
\toprule
\multicolumn{4}{c}{\textbf{United States}}&&\multicolumn{4}{c}{\textbf{United Kingdom}}&&\multicolumn{4}{c}{\textbf{Germany}}\\
\cline{1-4}
\cline{6-9}
\cline{11-14}
\multicolumn{2}{l}{\textbf{Method}}&\textbf{Design}&\textbf{CAT}&&\multicolumn{2}{l}{\textbf{Method}}&\textbf{Design}&\textbf{CAT}&&\multicolumn{2}{l}{\textbf{Method}}&\textbf{Design}&\textbf{CAT}\\
\midrule
%1&Incognito\_PrivBrowsing\_DuckDuckGo&1& NotGivePI\_LimitSharing\_MinimalInfo&1&AdBlock\\
%1&Private Browsing&BI &BHP &&1& Limit Sharing&UD&ANO&&1&AdBlock&BI&COP\\
1 &Website care &NT&PLS&&1&Website care& NT & PLS &&1&AdBlock&BI&COP\\
%2 &AdBlock&FIL&&2 &PrivacySettings\_Controls\_RouterSettings&2&VPN\\
%2 &AdBlock&BI&COP&&2 &Privacy Settings&BI&PLS&&2&VPN&ST&COP\\
2 & Privacy settings &BI&PLS&&2 & Limit Sharing &NT&ANO && 2& Bogus Emails &NT&ANO\\
%3 &VPN&COP&&3 &PrivateProfiles\_LimitPublicAccess&3&Incognito\_PrivBrowsing\_DuckDuckGo\\
%3 &VPN&ST&COP&&3 &Private Profiles&BI&PLS&&3&Private Browsing&ST&BHP\\
3 & Limit Sharing &NT&ANO&&3 & Privacy settings&BI&PLS&&3 & Privacy settings&BI&PLS\\
%4 &Clear\_DeleteInfo\_History\_NotStoreInfo\_Erasery&4& Incognito\_PrivBrowsing\_DuckDuckGo&4&NotGivePI\_LimitSharing\_MinimalInfo\\
%4 &Clear Info/History&BI&ANO&&4& Private Browsing&ST&BHP&&4&Limit Sharing&UD&ANO\\
4 & Research before engaging &NT&ANO&&4 & Clear Info/History &BI&ANO && 4&  Limit Sharing &NT&ANO\\
%5 &Clear\_DisAllow\_LimitCookies&5& Clear\_DisAllow\_LimitCookies&5&Clear\_DisAllow\_LimitCookies\\
%5 &Clear/Limit Cookies&BI&BHP&&5& Clear/Limit Cookies&BI&BHP&&5&Clear/Limit Cookies&BI&BHP\\
5 & Anti-Malware &ST&PLS&&5 & Paypal&ST&PLS &&5& No Newsletter &NT&PLS\\
%6 &Antivirus&DLP-S&&6 &VPN&COP&&6&NotEngagingOnline_ResearchBefore_Careful_NotSigningUp_NoFBSignup
%6 &Antivirus&ST&PLS&&6 &VPN&ST&COP&&6&Not Engage&UD&ANO\\
6 & No Newsletter &NT&PLS&&6 & Research before engaging&NT&ANO && 5&Paypal&ST&PLS \\
%7 &Ghostery&TRP&&7 &WebsiteCare\_NoSuspiciousSites\_Reputable&7&SeveralEmails\_BogusEmails\\
%7 &Ghostery&ST&BHP&&7 &Website Care&UD&PLS&&7&Several Emails&UD&ANO\\
7 & AdBlock &BI&COP&&7 &No Newsletter &NT&PLS&&5 &Website care& NT & PLS \\
%8 &WebsiteCare\_NoSuspiciousSites\_Reputable&8 &Antivirus&8&Antivirus\\
%8 &Website Care&UD &PLS&&8 &Antivirus&ST&PLS&&8&Antivirus&ST&PLS\\
8 & Clear Info/History &BI&ANO&&8 & Firewall &ST&COP&&5& Firewall &ST&COP\\
%9 &Password Manager&DLP-P&&9 &Clear\_DeleteInfo\_History\_NotStoreInfo\_Erasery&9&AnonymousProfileNames\\
%9 &Password Manager&ST&PLS&&9 &Clear Info/History&BI&ANO&&9&Ano,-Pseudo Names&UD&ANO\\
9 & Clear/Limit Cookies &BI&BHP&&9 & Anti-Malware&ST&PLS &&9&HTTPS&BI&COP\\
%10 & AntiSpy\_Malware&10 &SeveralEmails\_BogusEmailsForUnimportantUs\_LimitedUse&10&FakeInfo\\
%10 & Anti Spy-,Mal-ware&ST&PLS&&10 &Several Emails&UD&ANO&&10&Fake Info&UD&ANO\\
10 & Not Access Accts in Public Place &NT&ANO&&10 & Not Access Accts in Public Place &NT &ANO &&10& Pseudonyms &BI&ANO\\
\bottomrule 
\end{tabular}
}\\
\emph{BI}, \emph{ST} \& \emph{NT} refer to design type of built-in, standalone and non-technology respectively.\\
\emph{ANO}, \emph{BHP}, \emph{COP} \& \emph{PLS} refer to privacy protection categories of anonymity, browsing history and tracking prevention, communication privacy \& filtering, and preventing leaking \& stealing of data respectively.
\end{table*}

\section{Study 3}
\label{sec:study3}
\subsection{Aim}
%\subsection{User Cognition}
While individuals are generally reported to be concerned about their privacy~\cite{blank2019perceived,auxier2019americans,madden2014public,eurobarometer2019european}, they are not necessarily observed to actively use PETs, as a consequence of their concern. %~\cite{}.
In particular, a number of intervening steps may influence the \emph{`privacy concern --- use of PETs'} link, such as whether individuals perceive a need to act, are aware of the usefulness PETs or are able to use PETs~\cite{renaud2014doesn}, amongst the various other factors impacting behavior listed in Table~\ref{tab:behavior_review}.

We aim to understand use perceptions and reasons for choosing a collection of PETs.
Using the clusters visualised in the cluster map of Figure~\ref{fig:clusters} (Study 2) as classification of privacy methods usage patterns, 
we examine individuals' perceptions of the PETs in the different clusters, and their rationale for a particular choice.
%What do these patterns say about individuals' perceptions? What do the patterns reveal? do the patterns actually have psychological grounding?
\label{sec:list contents}
We create two lists of PETs from the cluster map, that we name Advanced PETs (Adv.PETs) and Other PETs (Oth.PETs): % and investigate individuals' perception and rationale for mostly using one list. 
%We build two lists of PETs,  %. , from the Cluster Map of Figure~\ref{fig:clusters}.
%These lists are used in Study 3 to query participants on their perception of use.

\emph{Advanced PETs List:}
Adv.PETs is populated with the PETs in the leftmost cluster, which contains the least used and more advanced solution to privacy protection. We also add VPN and encryption, which fall on the left of the centroid of the middle cluster, to the list.
In particular, Adv.PETs refers to a list of the following PETs: Erasery, Ghostery, virtual machine, Tor, NoScript, IPHider, Kaspersky, DuckDuckGo, proxy, anti-tracking extension, VPN, and encryption.

\emph{Other PETs List:}
Oth.PETs is populated with the PETs in the rightmost cluster, which contains the most used privacy methods. We also add pseudonyms and `anonymous profile', which fall on the right of the centroid of the middle cluster, to the list.
Oth.PETs refers to the following list of PETs: switch off location tracking, private browsing, HTTPS, anti-spyware, opt-out (of data collection), clear cookies, anti-malware, clear history, Paypal, firewall, Adblock, privacy settings, pseudonyms and anonymous profile.

\emph{Research Questions.}
The Technology Acceptance Model (TAM), as discussed in Section~\ref{sec:models_of_tech_adoption}
depicts how perceptions explain use of technology~\cite{davis1989perceived}.
% has been shown to explain usage of technology, including use of PETs~\cite{harborth2018examining,benenson2015user}.
%as investigated in studies with individual PETs.
We focus on the main components of the TAM, namely perceived usefulness (PU), perceived ease of use (PEOU) as well as behavior intention (BI), to investigate perceptions of use of Adv.PETs and Oth.PETs.
%Figure~\ref{fig:study3_design} provides an overview of the procedure. 

We first ask ``Do individuals prefer to use Adv.PETs (Adv.Users) or Oth.PETs (Oth.Users)?"
We then investigate as \textbf{RQ3},
%``How do individuals perceive use of PETs in both lists?
``How do Adv.Users' perceptions of use differ between Adv.PETs and Oth.PETs?"
and ``How do Oth.Users' perceptions of use differ between Adv.PETs and Oth.PETs?" via the hypotheses: \\
%How does perception of use of Adv.PETs and Oth.PETs differ between Adv.Users and Oth.Users?" 
\const{H_{3a},_0}: Adv.Users show no difference in perceptions of use between Adv.PETs and Oth.PETs. \\
\const{H_{3o},_0}: Oth.Users show no difference in perceptions of use between Adv.PETs and Oth.PETs. \\
\const{H_{3a},_1}: Adv.Users show a significant difference in perceptions of use between Adv.PETs and Oth.PETs. \\
\const{H_{3o},_1}: Oth.Users show a significant difference in perceptions of use between Adv.PETs and Oth.PETs.

We also investigate other factors such as \emph{competency}, as an antecedent to technology acceptance, as well as factors
contributing to the use of privacy technology, such as  \emph{awareness} and \emph{social influence}. Social influence is a factor well investigated in relation to technology adoption and usage~\cite{vannoy2010social,talukder2011impact}.

We investigate as \textbf{RQ4},
``Does perceived competency at managing one's privacy online differ between Adv.Users and Oth.Users?"\\
\const{H_4,_0}: There is no difference in perceived privacy competency between Adv.Users and Oth.Users.\\
\const{H_4,_1}: There is a significant difference in perceived privacy competency between Adv.Users and Oth.Users.

We investigate as \textbf{RQ5},
``Do Adv.Users' and Oth.Users' [awareneness of/perceived social influence to use] differ between Adv.PETs and Oth.PETs?" via the hypotheses \\
\const{H_{5a},_0}: Adv.Users show no difference in [awareneness of/perceived social influence to use] between Adv.PETs and Oth.PETs. \\
\const{H_{5o},_0}: Oth.Users show no difference in [awareneness of/perceived social influence to use] between Adv.PETs and Oth.PETs. \\
\const{H_{5a},_1}: Adv.Users show a significant difference in awareneness of/perceived social influence to use] between Adv.PETs and Oth.PETs.\\
\const{H_{5o},_1}: Oth.Users show a significant difference in awareneness of/ perceived social influence to use] between Adv.PETs and Oth.PETs.

%**********Reason for preference
We further query participants on why they chose Adv.PETs or Oth.PETs, as well as the support they would need to use the other type of PETs, as \textbf{RQ6}.

\subsection{Method}
\subsubsection{Participants}
For the third study, we recruited an $N=183$ from the US, UK and DE via Prolific Academic.
The demographics across the 3 countries is provided in Table~\ref{tab:demo_study3}.

\begin{table}[h]
\centering
\caption{Participant Characteristics}
\label{tab:demo_study3}
\footnotesize
%\resizebox{\textwidth}{!}{
\begin{tabular}{lcrcrr} %crrrrlrrrrr}
\toprule
&\textbf{Country}& \textbf{$N$} & \textbf{Mean Age} & \multicolumn{2}{c}{\textbf{Gender}} \\ %&&\multicolumn{4}{c}{\textbf{\%Education Level}} &&\multicolumn{5}{c}{\textbf{\% Ethnicity}} \\
%ethnicity - asian, black, mixed, other, white
\cline{5-6}
%\cline{8-11}
%\cline{13-17}
&&&& \#Female & \#Male \\ %&&HighSchool&College&Undergrad&Masters/PhD && White & Black & Asian & Mixed & Other\\
\midrule
\multirow{3}{*}{\textbf{Study 3}}& US &40 & 31.50 &25&15 \\ %&&35.0&20.0&32.5&12.5 &&70.0&2.5&7.5&7.5&12.5\\
& UK & 101 &37.23 & 58& 43 \\ %&&18.8&25.7&38.6&14.9 &&76.2&4.0& 13.9&5.0& 0\\
& DE  & 42 & 28.98 & 16&26 \\ %&&26.2&19.1&28.6&23.8 && 100.0&-&-&-&-\\

\bottomrule
\end{tabular}
%}\\
\vspace{-.3cm}
\end{table}

\subsubsection{Procedure}
\label{sec:procedure_study3}
\begin{figure}
\centering
\includegraphics[keepaspectratio,width=1\columnwidth]{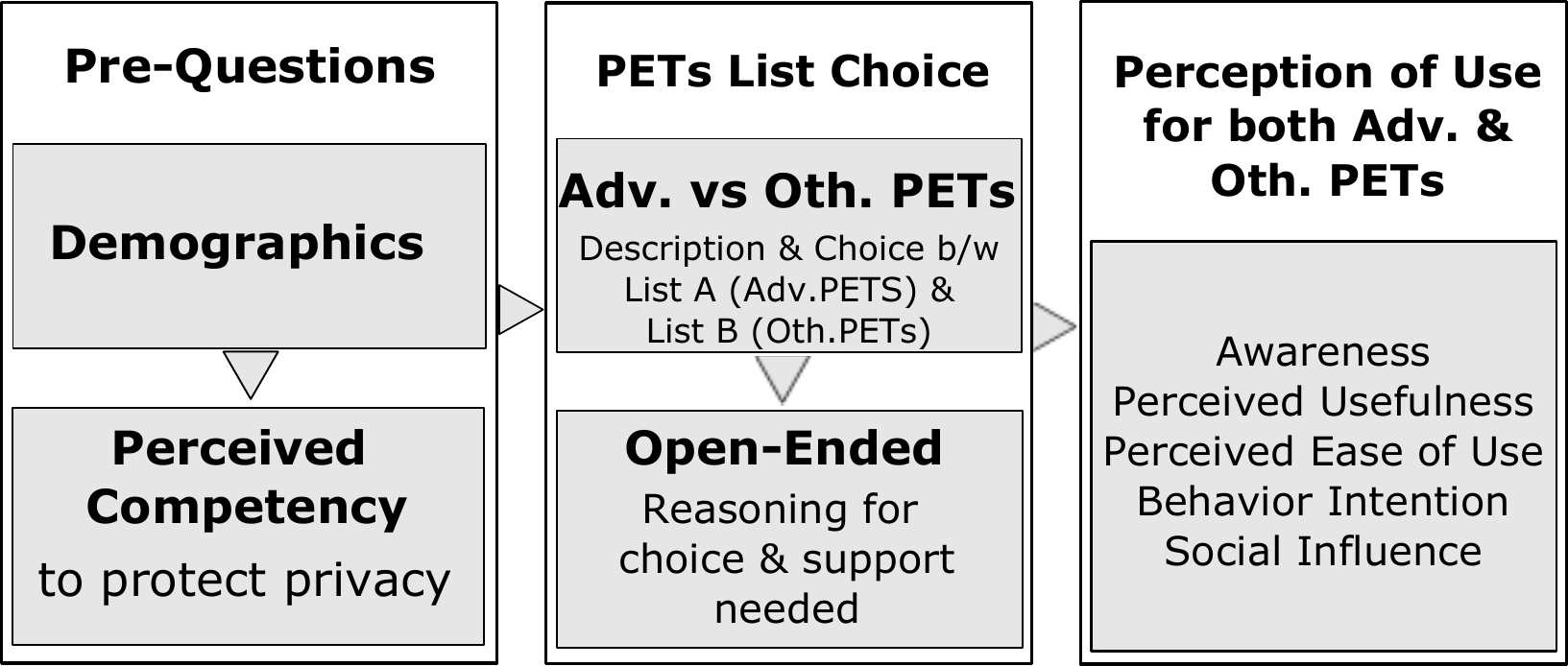}
%[scale=0.7{\textwidth}]
\caption{Study 3 design.}
\label{fig:study3_design}
\vspace{-.3cm}
\end{figure}

Study 3 consisted of the following questionnaires: %\begin{inparaenum}[(a)]
(a) demographics, 
(b) perceived competency in protecting privacy online,
(c) description of two lists of PETs (List A, the Adv.PETs and List B, the Oth.PETs) and questions on which set participants mostly use, as well as support they believe they would need to use the other set, and
(d) awareness, perceived usefulness, ease of use, social influence, behavior intention to compare between the two sets of PETs. %, and
%(e) innovativeness scale.
We depict the procedure in Figure~\ref{fig:study3_design}.

\subsubsection{Measurement Apparatus}
\label{sec:measurement_study3}
This section describes the questions and scales employed in Study 3, where we employed good practice guidelines of security and privacy user studies~\cite{coopamootoo2017niftynine,coopamootoo2017cyber}, in adapting existing scales, as much as possible.

\emph{Choice of PETs, Reasoning \& Support:}
We provided participants with two lists of PETs and asked ``Which of the two lists contains the privacy methods that you most often use to protect your privacy online?''
We then asked them to explain why they `most often' use PETs from one list and why they `very rarely or not at all' use methods from the other list.
We further queried about what would support them to use methods from the list they did not select, in particular, what would help or encourage them.
We provide the full questions and wording in the Appendix~\ref{sec:app_study3_choice}.

\emph{Perceived Competency:}
We adapt Williams \& Deci's~\cite{williams1996internalization,williams1998supporting} perceived competence questionnaire to privacy protection online.
We provide the $4$ items of the questionnaire in Table~\ref{tab:c_alpha_table} of the Appendix~\ref{sec:study3_scale}. The questionnaire is provided with a 7-point Likert scale ranging from $1$ - ``Not true at all" to $7$ - ``Very True" and a midpoint at $4$ with ``Somewhat True".

%\subsubsection{Perception of Use of Technology \& other factors}
\emph{Technology acceptance model components:}
We adapt Davis~\cite{davis1989user, davis1989perceived} scale for measuring the components of TAM, namely
for perceived usefulness, perceived ease of use and intention to use privacy technology, as provided in Table~\ref{tab:c_alpha_table} of the Appendix~\ref{sec:study3_scale}.
The scales were provided with 5-point Likert ranging from $1$ - ``Strongly Disagree" to $5$ - ``Strongly Agree".
%Perceived usefulness is defined as a person’s expectation that using
%the computer will result in improved job performance 

\emph{Awareness \& Social Influence:} % \& innovativeness}
We created a 4-item questionnaire to gauge awareness of PETs,
%developed an awareness of PETs 4-item questionnaire, 
as shown in Table~\ref{tab:c_alpha_table} of the Appendix~\ref{sec:study3_scale}.
We further adapt the social influence scale from Venkatesh et al.~\cite{venkatesh2003user}, where social influence is defined as the degree to which an individual perceives that important others believe he should use the new system.
%We also include a personal innovativeness scale, where personal innovativeness is seen as the inclination of an individual to try out any new information systems~\cite{blake2003innovativeness, chang2005literature}.
These scales are provided on a 5-point Likert ranging from $1$ - ``Strongly Disagree" to $5$ - ``Strongly Agree".

\subsubsection{Limitations}
\label{sec:study3_limits}
\emph{Self-Report Questionnaires:}
Similar to Studies 1 and 2, Study 3's data collection is based on self-reports.
However, in Study 3, we employ standard questionnaires with scales composed of multiple items. 
These scales have been widely used and validated in past research, as referenced in Section~\ref{sec:measurement_study3}.
We provide the assessment of internal consistency obtained in Study 3 for each scale in Table~\ref{tab:c_alpha_table} of the Appendix.

\emph{Additional measures and ordering:}
While we carefully selected questionnaires for the purpose of Study 3, %Our study is limited by the questionnaires used.
additional ones such as to elicit participants' tech-savviness could provide means of verifying the self-reported Perceived Privacy Competency questionnaire, for example.
In addition, 
placing the demographics questionnaire at the fore may consume participants' attention span.
We included attention checks throughout the study.
However, in the future, we would comsider to position the demographics questionnaire at the end of the study. % and also 
%investigate the potential influence of the order of the questionnaires, such as how asking for demographics data and perceived competency first may impact responses to further questions.

\subsection{Quantitative Results}
\label{sec:quantitative_results}
%\subsubsection{Choice of PETs Pattern}
%Henceforth we refer to PETs in the left end of the cluster map as Adv.PETs and those at the right end, Oth.PETs.
In Study $3$, we asked participants to select the list containing the PETs they most often use. 
We detailed the contents of the lists in Section~\ref{sec:list contents}.
Only $17$ of $183$ participants mostly use Adv.PETs, that is the list pertaining to the least used cluster.
All the remaining participants chose the Oth.PETs list.
We distinguish these $2$ groups of users as Adv.Users and Oth.Users.
%(Adv.U) and Oth.Users (Oth.U).

%\subsubsection{User Perceptions}
We compute the internal consistency of the items for each of the perception scales. We provide the Cronbach $\alpha$ values in Table~\ref{tab:c_alpha_table} of the Appendix~\ref{sec:study3_scale}.

We investigate \textbf{RQ3},
\emph{How does Adv.Users' and Oth.Users' perceptions of use differ between Adv.PETs and Oth.PETs?}, and \textbf{RQ5}, \emph{do Adv.Users' and Oth.Users' [awareneness of/perceived social influence to use] differ between Adv.PETs and Oth.PETs?}

We compute pairwise $t$-tests for Adv.PET and Oth.PETs across participant responses for awareness of PETs, perceived usefulness, perceived ease of use, behavior intention and social influence.
We summarise the results in Table~\ref{tab:pairwise}, where Oth.Users showed significantly higher perceptions of Oth.PETs than of Adv.PETs, across all scales with $p<.001$. 
We reject the null hypotheses, 
\const{H_{3o},_0} and \const{H_{5o},_0} that \emph{Oth.Users show no difference in perceptions of use between Adv.PETs and Oth.PETs}, for the five use perceptions shown in Table~\ref{tab:pairwise}.
Adv.Users only reported higher perceived usefulness of Adv.PETs than of Oth.PETs. We accept the null hypotheses \const{H_{3a},_0} and \const{H_{5a},_0} for Adv.User for all user perceptions, except for perceived usefulness.

We investigate \textbf{RQ4}, \emph{Does perceived competency at managing one's privacy online differ between Adv.Users and Oth.Users?}
We compute differences in perceived competence between Adv.Users and Oth.Users, with an independent samples Mann-Whitney U test. %as summarised in Table~\ref{tab:independent-t}. 
We do not observe a statistical significant difference in perceived competency in ensuring privacy protection online between Adv.Users and Oth.Users. 
We therefore accept the null hypothesis \const{H_4,_0} that \emph{there is no difference in perceived privacy competency between Adv.Users and Oth.Users}.

%There is also no difference between the two groups in terms of personal innovativeness.

\begin{table} %[ht] %[p]
\centering
\caption{Within-group pairwise comparison of technology use perception of Adv.PETs vs. Oth.PETs, where Oth.Users have higher perceptions of Oth.PETs than Adv.PETs} %, restricted to significance level $\alpha = .01$}
\label{tab:pairwise}
\footnotesize
\resizebox{\columnwidth}{!}{
\begin{tabular}{lllllllll}
\toprule
\textbf{Use Perception} & \multicolumn{3}{c}{\textbf{Adv.Users}} &&  \multicolumn{3}{c}{\textbf{Oth.Users}}\\
\cline{2-4}
\cline{6-8}
& &$t(16)$ &$p$ &&& $t(165)$ &$p$\\
\midrule
Awareness &&&&& Oth.PETs&-16.290&.000 \\
Perceived Usefulness & Adv.PETs&2.787&.013 && Oth.PETs& -6.999&.000   \\
Perceived Ease of Use & &&&& Oth.PETs&-18.494& .000 \\
Behavior Intention & &&&& Oth.PETs&-15.189 &.000  \\
Social Influence& &&&& Oth.PETs& -9.231 &.000 \\
\bottomrule
\end{tabular}
}
\vspace{-.3cm}
\end{table}

\subsection{Qualitative Results}
\label{sec:qualitative_reports}
%- what did they say?
%We ask participants about the reasons for mostly using Oth.PETs or Adv.PETs and 
%we also query participants' take on the support and encouragement they would need to use Adv.PETs. %motivation
We investigate \textbf{RQ6}, that is \emph{``why do individuals choose Adv.PETs or Oth.PETs? what support would they need to use the other type of PETs?"}
This section describes the coding process and reports responses for (1) users of both types of PETs, (2) users of Adv.PETs and (3) users of Oth.PETs.
Note that List A in the survey refers to Adv.PETs and List B refers to Oth.PETs.

\subsubsection{Codebook Creation}
The sample of $N=183$ participants in Study 3 were asked two open free-form questions: Q1, why they mostly use methods from one list, and Q2, what would support them to use methods from the other list.
Responses were required to be at least $30$ words long, with no maximum.

We facilitated a conventional line by line coding, where $n=50$ responses were coded by one coder to extract concepts from the free-form text. This process has been used in usable privacy research before and is well accepted~\cite{coopamootoo2017whyprivacy}.
The concepts were grouped into $5$ categories for each question, and the coding was validated and refined with discussions with another coder. %as shown in Table~\ref{tab:codebook} of the Appendix
We used the set of categories and concepts to create a codebook. We trained $2$ researchers as coders and further refined the codebook with a final set of $47$ codes within $5$ categories as provided in Table~\ref{tab:codebook} of the Appendix~\ref{sec:study3_codebook}. The `-other' codes, such as \emph{EFF03-other} were added to include concepts not initially catered for in the codebook.

%\subsection{Inter-Rater Reliability}
We evaluate inter-rater reliability via $\%$-agreement and Cohen \textsf{k}~\cite{hallgren2012computing,mcdonald2019reliability} on $100$ responses across the $47$ codes.
We find that the coders were on agreement $96\%$ of the time and there was a substantial agreement with Cohen \textsf{k} of $.837$, $p<.001$.

\subsubsection{Use of both types of PETs}
%PRI05
We note that $35$ of the $183$ participants reported using PETs from both lists, where most of them reported using Oth.PETs primarily and a few Adv.PETs. %how about the few who use mainly Adv,PETs - what are their reasons??
%..............XXX..TODO..XXX...........what `couple' of PETs from both lists were used + what were their reasons for using both lists -- LAYERING, EVERYDAY + CRITICAL SITUATIONS.
The few Adv.PETs that participants reported using are VPN (mentioned by $n=14$), NoScript, DuckDuckGo, Proxy, anti-tracking, Kaspersky and Ghostery.
Note that as described in Section~\ref{sec:procedure_study3}, List A refers to Adv.PET whereas List B refers to Oth.PETs. 

%they use Oth.PETs for everyday use and Adv.PETs for critical conditions
Participants explained that they use Oth.PETs as baseline and Adv.PETs for critical conditions or as a form of layered protection, such as expressed by \textsf{P121} in \emph{`I think methods on List B are things that should use [sic] everybody, \textbf{the bare minimum you should use} while surfing on the world wide web. I also use DuckDuckGo, VPN, NoScript and Ghostery, but the things on the other List are more often'}, 
\textsf{P9} in \emph{`i [sic] do use some from List A, however list B come pretty standard with my tech as well as is free and easy to use.'}
and 
\textsf{P127} in \emph{`I mostly use adblock on very restrictive settings, noscript, no location tracking and private browsing, \textbf{because these methods are the most convenient ones}. I sometimes use VPN and proxies as well, but since this requires more action, \textbf{I only use this in critical cases.}'}

In addition, $2$ participants reported using VPN for other reasons than for privacy protection, with for example \textsf{P112} stating  \emph{`clear cookies and clear the history is a thing i [sic] do every day [\dots] ok sometimes i [sic] use VPN, \textbf{but not to protect myself only to watch movies from the US, which are geo blocked}'} and \textsf{P25} \emph{`I will occasionally use a VPN if I need to access websites that are not available without.'}

\subsubsection{Users of Advanced PETs}
Users of Adv.PETs reasoned that Adv.PETs are more effective in ensuring their privacy online, such as reported by \textsf{P117} in \emph{`List A contains tried and tested methods of protecting privacy, list B is superficial'} or \textsf{P105} in \emph{`list B not effective'}.
They also pointed to the trustworthiness of Adv.PETs, as expressed by \textsf{P110} \emph{`Tor is more trustworthy'} or \textsf{P40} \emph{`List A is [sic] better to rely on'}.

\subsubsection{Users of Other PETs} % - The Six Themes
\label{sec:other_pets}
Overall we find that participants' reasons for mostly using Oth.PETs corroborated with the support and encouragement they believe they would need to use Adv.PETs, as similar themes emerged from the two open-ended questions. We consequently provide the responses for the two open-ended questions based on the themes.
From the $5$ categories of Table~\ref{tab:codebook}, we observe recurrent themes in the responses, namely (1) `\emph{information about}' the PETs, 
(2) `\emph{usability}' type responses, 
(3) level of `\emph{privacy or protection needed}', 
(4) `\emph{cost}' of PET, 
(5) `\emph{trust and reliability}' of PET and 
(6) `\emph{social support}' to use PET. 
%In addition, we find recurrent responses referring to `\emph{trust}' in the technology, and therefore report them as the sixth theme.
We report participants' reasoning for choosing Oth.PETs and support needed to use Adv.PETs across these themes in Table~\ref{tab:qual_responses}.

%The $6$ themes mentioned by participants in their reasoning for choosing one type of PETs over another, and the support they would require to use the other type of PETs were 
%(1) information about the PET, 
%(2) usability,
%(3) privacy and protection needed, 
%(4) cost, 
%(5) social support, and 
%(6) trust and reliability.
%We report on the concepts mentioned for each theme, with the number of participants naming them below.

\begin{table*}[h]
\centering
\footnotesize
\caption{Qualitative Responses from n=166 participants who chose Oth.PETs: \\
reason for choice \& support/encouragement needed to use Adv.PETs}
\label{tab:qual_responses}
\resizebox{.79\textwidth}{!}{
\begin{tabular}{llrcllr} %p{3cm}
\toprule
&\multicolumn{2}{c}{\textbf{Theme 1: Information About}} && \multicolumn{3}{c}{\textbf{Theme 2: Usability}}\\
\cline{1-3} \cline{5-7}
&&\% & &&&\%\\
\textbf{Reasons} & Not know Adv.PETs & 28.92 &&\textbf{Reasons} & Easy to use/install Oth.PETs & 24.10 \\
& Familiar with Oth.PETs &27.11 &&& Adv.PETs complicated to use/install & 7.83 \\
& Technical skills required & 6.02&&& Oth.PETs are readily available & 6.63 \\
& Belief Oth.PETs are for casual/non-tech savvy users & <4 &&& Oth.PETs easy to access & 5.42 \\
& Need to be advised/recommended &<4  &&& Oth.PETs convenient to use & 4.82\\
&&&&& Amount of work vs benefit gained from Adv.PETs & 4.82 \\
&&&&& Oth.PETs integrated in service & <4\\
&&&&& Annoyance or discomfort when using Adv,PETs &<4 \\

&\\
\textbf{Support} & What they are/do & 36.14 &&\textbf{Support}& Ease of use &14.46 \\
& How to use, training & 16.27 &&& Ease of installation and configuration&12.05\\
& Via info channels e.g. SNS, tutorials, trusting list  & 9.04 \\

&\\
%\midrule
&\multicolumn{2}{c}{\textbf{Theme 3: Privacy \& Protection Needed}} && \multicolumn{2}{c}{\textbf{Theme 4: Cost}}\\
\cline{2-3} \cline{5-7}
&&\% && &&\%\\
\textbf{Reasons} & For protection provided (`keeps me private') &15.06 &&\textbf{Reasons}& Having to pay for/cost of Adv.PETS& 14.46\\
& Oth.PETs good enough for privacy needed & 10.84 &&& Oth.PETs are free &4.22 \\
& Successful experience in using Oth.PETs & 6.63 \\
& No need for privacy, nothing to hide & 5.42\\
& Adv.PETs would be extreme/Oth.PETs for everyday use &4.22  \\
& Privacy vs online experience  &<4 \\ %experience hampered with Adv.PETs

&\\
\textbf{Support} & If need specific PET protection functionality (e.g. anonymity) & 9.04  &&\textbf{Support}& If Adv.PETs were free/affordable & 16.87\\
& If Adv.PETs provided more privacy than what they use & 8.43 \\
& If they had a bigger privacy need & 6.02 \\

&\\
%\midrule
&\multicolumn{2}{c}{\textbf{Theme 5: Trust \& Reliability}} && \multicolumn{2}{c}{\textbf{Theme 6: Social Support}}\\
\cline{2-3} \cline{5-7}
&&\% && &&\%\\
\textbf{Reason} & Trust in PETs & 9.04 &&\textbf{Reason}& n/a \\
&\\
\textbf{Support} & If Adv.PETs did not look fake/phoney & 5.42 &&\textbf{Support}&Recommended by reputable company & 5.42  \\
&&&&& If someone they know/trust used it & <4 \\
&&&&& If someone taught them how to use or install & <4 \\

\bottomrule
\end{tabular}
}
\end{table*}

\begin{RedundantContent}
\underline{\textbf{Information About}}
\emph{Motivation to use:} 
Participants reported needing more information about Adv.PETs, 
(a) including what they are and what they do ($60$), %inf1
(b) how to use Adv.PETs, training or further education ($27$), %inf2
(c) via information channels such as social media, tutorial or trustworthy list ($15$). %inf3

\emph{Reason for choosing PETs:}
%\subsubsection{Awareness, Skills Required}
%AWA1, 2
Familiarity with or knowing of the PET was a main reason for participants use of Oth.PETs over Adv.PETs.
In particular, there was $45$ mentions from Oth.U of familiarity and awareness of Oth.PETs and $48$ mentions of not knowing what the PETs in the Adv.PETs list are. %There was only $1$ mention each for Adv.U.

%SKI1 & SKI2
$6$ Oth.U thought Oth.PETs were for more casual, non-tech savvy users whereas $10$ thought using Adv.PETs required a lot of knowledge and technical skills.
%AWA4-other - advised recommended, employment
$3$ Oth.U reported they would need to be advised/recommended to use Adv.PETs, whereas
$2$ Oth.U reported they use Adv.PETs as requirement for their employment.

\underline{\textbf{Usability}}
\emph{Motivation to use:} 
Ease of use, simple or clear solutions ($24$) and ease of install and configuring ($20$) Adv.PETs were the main usability criteria mentioned for support.

\emph{Reason for choosing PETs:}
The main usability related reasoning for using Oth.PETs was 
(a) that they were easy of use and easy to install ($40$), %usr3
(b) that Adv.PETs were complicated to use, setup or users had a tough time using ($13$),  %usr4
(c) that Oth.PETs were readily available ($11$), integrated in service ($6$) and %usr1 +usr8-other
(d) were convenient to use ($8$), and %usr5
(e) ease to access ($9$).  %usr8-other

%usr8-other
In addition, $8$ Oth.U reported that the amount of work required versus the benefit gained and having to download and install software was a reason for not using Adv.PETs and $5$ reported annoyance or discomfort in using Adv.PETs. %because things break

\underline{\textbf{Cost}} %inc2
\emph{Motivation to use:} 
$28$ participants reported that they would use Adv.PETs if they were free of cost or were affordable.

\emph{Reason for choosing PETs:}
$24$ Oth.U reported that having to pay for or the cost of Adv.PETs was a main barrier to use, and $7$ claiming to choose Oth.PETs because they were free. %USR 6+7

\underline{\textbf{Privacy \& Protection Needed}}
\emph{Motivation to use:}
%PNE
Participants reported they would use Adv.PETs more if they had a bigger privacy need ($10$), %PNE1
needed specific PETs protection functionality such as anonymity ($15$), %PNE2 name the specific functionalities
or if Adv.PETs provide more privacy protection than the one they are currently using ($14$).%an incentive, more helpful %%PNE3

\emph{Reason for choosing PETs:}
%\subsection{Privacy Protection Provided, `Enoughness' of Privacy \& Trust in Technology}
%PRI03
$30$ participants (5 Adv.U and 25 Oth.U) provided reasoning pertaining to the privacy protection provided by the PETs in their preferred list, such as \emph{``keep me private"} or \emph{``increase my privacy"}.

%problematic 
%PRI06 other
Of Oth.U, $4$ explained that using Adv.PETs would be ``extreme" or ``over the top", 
%PRI01
$9$ reported ``not needing privacy" or that they ``have nothing to hide", 
%PRI02
$5$ reported that using Adv.PETs hampers their online experience, 
%EFF03-other
$3$ said they would rather use Oth.PETs for everyday use.

%EFF01 & EFF02
$18$ Oth.U %versus $1$ Adv.U 
reported that the list they use is ``good enough for the privacy they need" while $11$ %versus $1$ Adv.U 
explained that they have successful experience with Oth.PETs (``use it and it works, so continue using"). 

\underline{\textbf{Social Support}} %soc
\emph{Motivation:}
Participants reported they would use Adv.PETs if 
(a) if someone they knew used it, or someone they trusted recommended it ($6$),
(b) if someone teaches them or installs the PET for them ($6$), and 
(c) if it was recommended by a reputable company or accreditated ($9$).
There were no reasoning fitting this theme for choice of PETs.

\underline{\textbf{Trust \& Reliability}}
\emph{Motivation:} 
$9$ participants reported that they would use Adv.PETs if they knew it did not pose a security threat, or that it was not fake or ``phoney". %inc1

\emph{Reason for choosing PETs:}
%EFF03 - other trust, reliable, best method
$15$ participants overall referred to `trust' in the PET, its reliability or accuracy as reasons for using them.

%$4$ participants said PETs were not effective (which PETs???).
\end{RedundantContent}

%**************************************************************
\section{Discussion}
\label{sec:discussion}
We organise this section into four main sections, with each providing a brief summary of findings, followed with a discussion of the implications of our findings as well as their relation to previous research.
We also highlight lines of investigation for future research.

\subsection{Usage Pattern of Privacy Methods}
\emph{Summary of findings:} 
We visualised privacy methods use via a cluster map, and relations between methods and country of use via a spatial map.
We identified three clusters pertaining to three distinct privacy method use patterns, %, that depict a gradation in skills and in concerns.
%\subsubsection{What do the clusters show?}
%#88The clusters contribute differently to the first dimension, which we interpreted as the `Popularity of Privacy Method Dimension', 
with the left and right clusters respectively showing the least and most used methods.
%The middle cluster 
%*Privacy methods in the same cluster are as similar as possible in terms of user preference whereas privacy methods in different clusters are as dissimilar as possible. %the basic idea behind k-means clustering consists of defining clusters so that the total intra-cluster variation (known as within-cluster variation) is minimised.
%what else does the clustering show?

%\subsubsection{Cluster Transitions and Use}
\emph{Interpretation of Clusters.}
%We perceived that the right cluster may require user compromise of their sharing for privacy where users lose sharing functionality, whereas the left cluster mostly contains standalone PET examples that add an additional layer of technology for privacy. 
A visual inspection of the clusters reveals insights into participants' privacy methods preference, such as, 
%For instance, 
that the right cluster contains simple, easy to use, and convenient methods of protection, that can be said to be more inclusive of skill levels and are more mainstream, while the leftmost cluster contains more advanced PETs. %than those of the left cluster.

\begin{RedundantContent}
A visual inspection of the clusters reveals other information emerging from the natural clustering, including
insights into participants' privacy methods use habits and preference.
For instance, the rightmost cluster contains simple, easy to use, and convenient methods of protection, 
%that can be said to be more inclusive of skill levels and are more mainstream, 
while the leftmost cluster contains more advanced PETs.
%the right cluster may require lower user competency in understanding the technology and in interaction, than the left one.
\end{RedundantContent}

%**(2) the right cluster may require lower user competency in understanding the technology and in interaction, than the left one.
%***(3) the right cluster may require compromise sharing functionality, whereas the left cluster contain mostly standalone PET examples that add an additional layer of technology for privacy.
%of their sharing for privacy where users lose sharing functionality, whereas the left cluster contain mostly standalone PET examples that add an additional layer of technology for privacy, 
%***(4) the methods in the right cluster pertains to using privacy methods as a secondary goal (during sharing interaction), whereas the left cluster pertains more to using the technology as primary goal for privacy.

%However, participants in Study 3 reported that Adv.PETs can both interfere with online experience as well as provided additional protection such as the VPN.
%In addition, the methods in the right cluster pertains to using privacy methods as a secondary goal (during sharing interaction), whereas the left cluster pertains more to using the technology as primary goal for privacy.

\emph{Use of the classification.}
The results of the clustering may be used to support users with recommendations, such as suggesting other privacy methods in the same cluster, with the knowledge that users with similar methods preference used these other methods, and
the underlying assumption that methods in the same cluster share similarity in ease of use and perceived required skill.
%of the similarity in ease and competency required to use methods in the same cluster.
%**We could also use the clustering to recommend privacy methods with similar functionality in the same or different cluster, where shifting to a method in a different cluster may require putting in place support systems for the user, such as skill building strategies. %, such as guidance followed with feedback on skills upgrade.
The classification is also in itself a methodological tool, that can facilitate further user-centric investigation of PETs.  %***including those into collections of PETs. % rather than individual ones. 
%***Examples include perception of use of PETs types (as in Study 3), or use interaction of PET types.

\emph{Cluster Transitions.}
We consider questions for future research into 
%the conditions that 
facilitating a transition from using a right-cluster method to a left-cluster method, %where
%important questions can be raised for future research, 
such as %`how to facilitate the transition from right to left?', 
`under what conditions do users transition right to left? what is the influence of more privacy concern, more skill, a realisation that simple PETs are not enough or the influence by social contacts?'
`how to support users to shift from right to left or consolidate his position in the current cluster?'
%`how to know that an individual is at the edge of a cluster and requires m, 
`how can the methods in the different clusters be used in a layered approach for more complete protection?'

\begin{RedundantContent}
%For an individual with methods preference falling in the right cluster to shift to a method in the left cluster, she 
It might need an upgrade in skills, or a change of habit, where an individual's preference may transition first to the middle cluster with methods such as `not store information', `not save password', `give fake info', 
%`not use Facebook', 
having an anonymous profile, being more aware via the `terms and conditions' and shift towards `encryption' and 'VPN' on the left edge of the middle cluster. Note that the middle cluster has a higher $\%$ of non-technology methods. 
This transition from right to left may also signal higher privacy concerns, or a realisation that the methods in the right cluster are not enough in providing the protection desired.
It is also possible that individuals' use of PETs follows a natural progression from the right to the left cluster.
\end{RedundantContent}

\begin{RedundantContent}
For an individual with methods preference falling in the right cluster to shift to a method in the left cluster, she might need an upgrade in perceived competency and actual skills, or a change of habit, where the individual's method may transition first to the middle cluster with methods such as `not store information', `not save password', `give fake info', 
%`not use Facebook', 
having an anonymous profile, being more aware via the `terms and conditions' and shift towards `encryption' and 'vpn' on the left edge of the middle cluster. Note that the middle cluster has a higher $\%$ of user-defined methods. 

A transition from using a method in the right to the left cluster
may signal higher privacy concerns, or a realisation that the methods in the right cluster are not enough or are not complete in providing the protection desired.

It is also possible that individuals' use of privacy technology follows a natural progression from the right to the left cluster, where important questions can be raised for future research `how to facilitate the transition from right to left?', `how to know that an individual is at the edge of a cluster and requires more support to shift from right to left or consolidate his position in the current cluster?', `how can the methods in the different clusters be used in a layered approach for more complete protection?'

Disruptive (lose functionality), sharing-compromise PETs vs PETs to add a layer of privacy\\
Aim of cluster analysis:  Identifying the set of objects with similar characteristics (what characteristics are we talking about? similar use preference?)
The objects in the same group are more similar to each other (in terms of privacy goals primary versus secondary, design, user skills needed to use) than to those in other groups
\end{RedundantContent}

\emph{Cross-National Visualisation.}
The spatial plot of privacy methods in relation to country (of Figure~\ref{fig:symmetric_plot}) clearly shows a gravitation of German participants towards more advanced and active privacy methods, compared to US and UK participants. 
This may be a reflection of Germans' privacy perceptions, where % and actions. %, as well as data protection advertising. % and culture.
%In particular, 
previous research reported that Germans, compared to participants of other countries, 
found controlling access to personal data, private realms and data protection more important~\cite{soffer2014privacy},
% than disclosing data~\cite{soffer2014privacy},
were among those most sensitive to the duration and quantity of data collection (location data)~\cite{cvrcek2006study}, 
or %expected more damage and 
attributed a higher probability to privacy violations on SNS~\cite{krasnova2010privacy}.
%With the country-specific lens, 
Figure~\ref{fig:symmetric_plot} may also support discussions and further investigations of the effectiveness of privacy campaigns, of the role of the media or social connections in a particular country. %, in particular `why do the US and UK have different patterns? why are their patterns different from Germany?'

%XXX--- review of previous research reference to Germany ---XXX
%cvrcek2006 - germans among those most sensitive to extension in duration and quantity of data collected
%krasnova2010 - Germany vs US - German users expect more damage and attribute higher probability to privacy violations - on SNS
%soffer2014 - controlling access to personal data, private realms and data protection seems to be most important to German

\subsection{Non-Technology Methods in Top 10}
%\subsubsection{Most Preferred Methods}
%DE and US participants reported using $3$ and $2$ more privacy methods on average than UK participants respectively, where although there are similarities in that $4$ items are among the top $10$ most used privacy methods in countries, they differ across $23$ methods with the US reporting higher usage in $17$ of them.
\emph{Summary of findings:} 
Among the similarities in privacy methods usage across the three countries (from Study 2, Table~\ref{tab:top10}), we find that non-technology methods %\begin{inparaenum}[(1)]
(1) of being careful of websites, 
(2) to limit sharing, 
(3) research before engaging (2 out of 3 countries), 
(4) not subscribe to newsletters, and 
(5) not access accounts in public places, appear in the most used methods across countries. 
%\end{inparaenum} 

For the three countries, these non-technology methods 
made up $4$ or $5$ of the top $10$ most preferred privacy methods
%$5$ user-defined methods are ranked in the top $10$ most preferred, 
thereby demonstrating that users rely more on their own means to protect themselves than privacy technologies.
%%%why do people prefer to use their own methods rather than PETS?
%% are they concerned about privacy?
%% do they know of /aware of PETS?
%% is it that PETS do not give people feedback, hence don't know what happened after use?
%% if they failed to use it once, they won't use anymoe? 'initial contact' effect
This raises questions about the reasons for reliance on non-technology methods rather than PETs, such as 
`are users concerned enough and aware of PETs to use them?
%have they been turned off by PETS before rather than have a positive contact?
how were their previous experience with PETs?' %We answer these questions via Study 3.

\subsection{Perception of Use}
\emph{Summary of findings:}  9\% of participants in Study 3 reported to use Adv.PETs, 91\% to use Oth.PETs and 19\% to use PETs from both lists.   
%with reasoning provided in Section~\ref{sec:qualitative_reports} and Table~\ref{tab:qual_responses}.
%***** relation to previous research ***
%*********Oomen paper survey - majority use standard security measures, only a third use advanced protection strategies\\
%*********Gerber2019 - almost no one mentionned advanced PETs\\
The low preference for Adv.PETs corresponds to previous research findings of no mention or of low use of Adv.PETs~\cite{oomen2008privacy,gerber2019johnny}.

Study 3 enabled us to evaluate perception of use of PETs within different clusters.
From Table~\ref{tab:pairwise}, it is clear that Oth.Users have higher use perceptions of Oth.PETs than Adv.PETs, which may
%The comparison of perception of use of Oth.PETs versus Adv.PETs (in Table~\ref{tab:pairwise}) 
highlight a perceptive barrier to individuals' use of Adv.PETs.
Similar to previous research of the importance of perceived usefulness for the adoption of anonymous credentials~\cite{benenson2015user,harborth2018examining}, our respondents who reported to use Adv.PETs showed higher perceived usefulness of Adv.PETs than Oth.PETs.

%Whether people use Adv.PETs or not, their perceived competency in protecting their privacy online is not different.
In addition. users of Adv.PETs and Oth.PETs did not report a difference in their perceived competency in protecting their privacy online.
%We find that competency is correlated with perceived ease of use, usefulness, as well as behavior intention with regards to Oth.PETs for Oth.U, with $r=.226$, $r=.178$ and $r=.225$ respectively.
%As discussed in the background section, consistent with self-determination theory~\cite{deci1985conceptualizations}, 
While perceived competence may predict continuance or persistence in behavior~\cite{roca2008understanding, oduor2017commitment},
given that individuals may feel more motivated to perform a task when they feel competent in, 
%This means that 
it may be complex to change individuals' behavior patterns from habitual use of Oth.PETs. % in particular when they find Oth.PETs useful, feel competent in protecting their privacy, and not know why they would need Adv.PETs.

%The quantitative results about perception of use also complement the qualitative responses provided by the participants, discussed below.

%\vspace{-.22cm}
\subsection{PETs Choice and Support Needed}
%\subsubsection{Building on previous research}
%++plus provide a non-contextual collective view, + provide a cross-national investigation
\emph{Summary of findings.}
The reported responses of Oth.Users in 
Study 3 support and extend the findings of previous research on factors influencing the active use of PETs (summarised earlier in Table~\ref{tab:behavior_review} of Section~\ref{sec:factors_on_behavior}). 
%Study 3 elaborates on the reasons for choosing Oth.PETs and 
%adds new reports and support needed across the 
We report Oth.Users' responses in Table~\ref{tab:qual_responses} and discuss them below.

Our findings contribute to previous research with a collective perspective that is neither specific to particular PETs nor contextual to use scenario, such as social network or e-commerce. % or health technology.
Our methodology may therefore help to identify general beliefs and biases of users that act as obstacles to the use of Adv.PETs. In particular, Table~\ref{tab:qual_responses} depicts a picture of `don't know, therefore don't use Adv.PETs', and yet have perceptions that Adv.PETs are not easy to use/install, are costly, or are not trustworthy.
Therefore, further to being aware of Adv.PETs, individuals may also have to overcome the biased perceptions they hold about these PETs.

%****Overall, while individuals' are hardly more concerned about their privacy than a few years a ago, and some claim to take certain actions to protect their data, most use Oth.PETs and are pretty satisfied in doing so: they are more aware of Oth.PETs and consequently feel these are more useful, easy to use, socially influenced to use and intend to continue using Oth.PETs compared to Adv.PETs.
%perceptual blocks
%to end up using Adv.PETs, apart from

%\subsubsection{Methodology helps identify biased perception}
%**Support needed: the advantage of the methodology involving a collection of PETs, that people don't know, don't use - it gives us a view of their beliefs/blocks ...... their biases even though they don't know about the tech -- they don't know about Adv.PETs, yet have a bias that they are not ease to use/install, are costly, not trustworthy etc \\
%- while not being aware/know of Adv.PETs is an important reason not to use them, people have an expectancy/bias of them being not easy to use/configure, not free/costly, not trustworthy etc — therefore, first step (as Renaud/volkamer staircase) — know about PETs, then remove those bias
%---cost is another bias?

%++plus find other reasons for people to not use Adv.PETs, or support they would need, 
%In particular, \begin{enumerate}
\emph{Awareness of PETs (from Theme 1: Information about).}
Our participants reported to not knowing about Adv.PETs and would benefit with an understanding of what they are and what they do. This is similar to previous reports on awareness of protection tools or inability to use protection for tracking or end-to-end encryption~\cite{renaud2014doesn,shirazi2014deters}.
In addition, we also found beliefs of technical skills requirement, and requiring training or information via social media and other channels.

\emph{Usability (Theme 2).}
Our participants also expressed a belief that Adv.PETs are not easy to use or to install, with other usability-related beliefs listed in Table~\ref{tab:qual_responses}.
Ease of use for anonymity technology~\cite{harborth2018examining,benenson2015user} and usability for secure communication~\cite{abu2017obstacles,renaud2014doesn} were also reported in previous research. 
However, although usability was previously found to not be more important than other factors (such as risk of sharing in social network~\cite{garg2014privacy}, contextual aspects of messaging tool~\cite{abu2017obstacles} or perceived usefulness~\cite{benenson2015user}), for our participants, one of the most important reason for 
using Oth.PETs, rather than Adv.PETs, was ease of use and configuration, 
%not using Adv.PETs was ease of use and configuration, 
which may be a biased perception about Adv.PETs' usability. % outside a particular context of use.
%This is their belief that Adv.PETs would not be ease to use or configure as opposed to actual usability evaluation in particular context such as secure communication or E2E.

\emph{Usefulness (Theme 3: Privacy \& Protection Needed).}
Our participants reported to use Oth.PETs for the protection provided, that Oth.PETs are good enough, that they had a successful use experience with Oth.PETs, 
%In addition, our participants reported that they would 
and that if they were to use Adv.PETs, it would be for other specific protection functionalities that are not provided by Oth.PETs.
While this points to their satisfaction in using Oth.PETs, being able to compare the privacy protection provided between Oth.PETs and Adv.PETs, may provide users with a wider choice.

\emph{Need for Privacy (Theme 3: Privacy \& Protection Needed).}
Some of our participants reported to not needing privacy/have nothing to hide, or that using Adv.PETs would be extreme. %, and that they may use Adv.PET if they had a bigger privacy need. 
These loosely match previous reports of no perceived need to act for tracking protection and end-to-end encryption technologies~\cite{renaud2014doesn,shirazi2014deters}.
%A few Oth.Users also mentioned using Adv.PETs in critical cases 
This depicts how ill-informed users may be with regards to their privacy online, in particular not knowing how much more protection Adv.PETs can offer them, as well as not realising the potential extent of privacy loss and online harms in the data-centric web.

\emph{Cost (Theme 4).}
Our participants referenced the perceived monetary cost Adv.PETs to be a deterrent to their use.
In comparison, previous research investigations in relation to cost included %cost in relation to privacy via
the perceived value of information~\cite{cvrcek2006study}, monetary reward for disclosure~\cite{hann2002online}, cost and benefits tradeoffs of disclosure~\cite{dinev2006extended}, cost of privacy breaches~\cite{acquisti2006there} or the time spent reading privacy policies~\cite{mcdonald2008cost}.
%the privacy calculus and the cost of disclosure~\cite{}, 
%.....xxx was cost only about cost of disclosure in previous research? how about effort? .....xxx....

\emph{Trust \& Social Support (Themes 5 \& 6).}
PETs' trustworthiness and recommendation (either socially or accredited via a reputable company), were also deemed important, as well as social support to recommend or teach how to use Adv.PETs. %, were thought to be important for the use of Adv. PETs. 
Trust has also been considered before in investigations involving the active use of privacy controls~\cite{garg2014privacy} %in relation to perceived risks
and adoption of PETs~\cite{harborth2018examining,benenson2015user}, and in relation to accessible and salient privacy information in websites~\cite{tsai2011effect},
%and more widely with regards to disclosure~\cite{joinson2010privacy,fogel2009internet},
%We also noted reports related to trust in PETs, as well as requirements for social support. 
while social influence had links with the use of privacy~\cite{abu2017obstacles,gerber2019johnny} and security~\cite{das2015role} controls.
%has been linked with use of privacy controls with regards to secure messaging~\cite{abu2017obstacles,} or social networks~\cite{gerber2019johnny}, as well as influencing adoption of security features~\cite{das2015role}.
While neutral accredited lists already exist via privacy-promoting organisations such as the Electronic Frontier Foundation, the influence of social connections and support from those individuals know and/or trust may have a `multiplier' effect in increasing use of Adv.PETs.
Social influence and support can facilitate integration of PETs into daily life, expand use to more casual users, as well as help sustain use. %build and sustain a more privacy empowered culture online. 
%were named as support to use Adv.PETs, these

%For those who find it hard --- ---- OxIS - non-users find it hard to see the benefits of going online and we need to pay attention to the value of support for non-users to experience the Internet.

\begin{RedundantContent}
\subsubsection{`Bigger' privacy need/not aware enough of potential for online harms?}
Participants reported they would use Adv.PETs if they had a bigger privacy need, needed them for specific protection functionality, or if the PETs provided them with more privacy.
This depicts how ill-informed users may be with regards to their privacy online, in particular not knowing how much more protection Adv.PETs can offer them, as well as not realising the potential extent of privacy loss and online harms with the data-centric web.

There were also statements that using Adv.PETs would be over the top or extreme, that they would hamper online experience, and Oth.PETs were believed to be good enough for everyday use. This highlights a challenge for Adv.PETs designers, in particular how to build Adv.PETs such that they can be integrated within everyday usage.

In addition,
(1) information about the protection functionality provided by the PETs and how they can be useful, 
(2) their trustworthiness and recommendation (either socially or accredited via a reputable company), were deemed important.
Furthermore, 
(3) reducing the burden of PET selection, installation and use via PETs recommendations, 
(4) less complex designs, as well as 
(5) social support to recommend or teach usage of PETs, were thought to be important for the use of Adv. PETs. 
\end{RedundantContent}

\section{Conclusion}
This paper provides a large scale, cross-national investigation of the use of a collection of privacy methods online, derives patterns of use and follows-up with an investigation of perception of use of PETs between the patterns.
%It also offers a cross-national investigation.
The usage classification, while contributing to our knowledge of PETs use, also provides a methodological contribution to the field. It can further be used in future user-centric investigations of PETs.

The clusters raise a number of questions about both their components and the conditions for transitioning to the least popular cluster. 
In addition, by investigating individuals' own perception of use of PETs as well as the support they believe they would need to use more advanced PETs, we highlighted a few themes that may help to expand the use of advanced PETs.
We recommend investigations of the questions raised in the discussion section, as future work, as well as for designers to consider the reasoning and support needed by individuals to use PETs (in particular, more advanced PETs), as provided in this paper. 

\begin{acks}
This research was supported by a Newcastle University Research Fellowship (Academic Track). 
I am thankful for the feedback provided by colleagues, all from Newcastle University, namely, 
Prof. Aad van Moorsel, Dr. Jaume Bacardit, Dr. Changyu Dong and Dr. Magdalene Ng.
I am also grateful for the feedback and comments provided by the anonymous reviewers of CCS'20 .
\end{acks}

\balance
\bibliographystyle{ACM-Reference-Format}
\bibliography{repository,cfs,laser,techuse,psychology_affect,passwords,methods_resources,fearappeals,emotion,anger,mturk,empowerment,se,privacy,surveys,interventions,psychometry}

%%% -*-BibTeX-*-
%%% Do NOT edit. File created by BibTeX with style
%%% ACM-Reference-Format-Journals [18-Jan-2012].

\begin{thebibliography}{00}

%%% ====================================================================
%%% NOTE TO THE USER: you can override these defaults by providing
%%% customized versions of any of these macros before the \bibliography
%%% command.  Each of them MUST provide its own final punctuation,
%%% except for \shownote{}, \showDOI{}, and \showURL{}.  The latter two
%%% do not use final punctuation, in order to avoid confusing it with
%%% the Web address.
%%%
%%% To suppress output of a particular field, define its macro to expand
%%% to an empty string, or better, \unskip, like this:
%%%
%%% \newcommand{\showDOI}[1]{\unskip}   % LaTeX syntax
%%%
%%% \def \showDOI #1{\unskip}           % plain TeX syntax
%%%
%%% ====================================================================

\ifx \showCODEN    \undefined \def \showCODEN     #1{\unskip}     \fi
\ifx \showDOI      \undefined \def \showDOI       #1{#1}\fi
\ifx \showISBNx    \undefined \def \showISBNx     #1{\unskip}     \fi
\ifx \showISBNxiii \undefined \def \showISBNxiii  #1{\unskip}     \fi
\ifx \showISSN     \undefined \def \showISSN      #1{\unskip}     \fi
\ifx \showLCCN     \undefined \def \showLCCN      #1{\unskip}     \fi
\ifx \shownote     \undefined \def \shownote      #1{#1}          \fi
\ifx \showarticletitle \undefined \def \showarticletitle #1{#1}   \fi
\ifx \showURL      \undefined \def \showURL       {\relax}        \fi
% The following commands are used for tagged output and should be
% invisible to TeX
\providecommand\bibfield[2]{#2}
\providecommand\bibinfo[2]{#2}
\providecommand\natexlab[1]{#1}
\providecommand\showeprint[2][]{arXiv:#2}

\bibitem[\protect\citeauthoryear{Abu-Salma, Sasse, Bonneau, Danilova,
  Naiakshina, and Smith}{Abu-Salma et~al\mbox{.}}{2017}]%
        {abu2017obstacles}
\bibfield{author}{\bibinfo{person}{Ruba Abu-Salma}, \bibinfo{person}{M~Angela
  Sasse}, \bibinfo{person}{Joseph Bonneau}, \bibinfo{person}{Anastasia
  Danilova}, \bibinfo{person}{Alena Naiakshina}, {and} \bibinfo{person}{Matthew
  Smith}.} \bibinfo{year}{2017}\natexlab{}.
\newblock \showarticletitle{Obstacles to the adoption of secure communication
  tools}. In \bibinfo{booktitle}{{\em 2017 IEEE Symposium on Security and
  Privacy (SP)}}. IEEE, \bibinfo{pages}{137--153}.
\newblock


\bibitem[\protect\citeauthoryear{Acquisti, Friedman, and Telang}{Acquisti
  et~al\mbox{.}}{2006}]%
        {acquisti2006there}
\bibfield{author}{\bibinfo{person}{Alessandro Acquisti}, \bibinfo{person}{Allan
  Friedman}, {and} \bibinfo{person}{Rahul Telang}.}
  \bibinfo{year}{2006}\natexlab{}.
\newblock \showarticletitle{Is there a cost to privacy breaches? An event
  study}.
\newblock \bibinfo{journal}{{\em ICIS 2006 Proceedings\/}}
  (\bibinfo{year}{2006}), \bibinfo{pages}{94}.
\newblock


\bibitem[\protect\citeauthoryear{Acquisti and Gross}{Acquisti and
  Gross}{2006}]%
        {AcqGro2006}
\bibfield{author}{\bibinfo{person}{Alessandro Acquisti} {and}
  \bibinfo{person}{Ralph Gross}.} \bibinfo{year}{2006}\natexlab{}.
\newblock \showarticletitle{Imagined communities: Awareness, information
  sharing, and privacy on the Facebook}. In \bibinfo{booktitle}{{\em Privacy
  enhancing technologies}}. Springer, \bibinfo{pages}{36--58}.
\newblock


\bibitem[\protect\citeauthoryear{Acquisti and Grossklags}{Acquisti and
  Grossklags}{2005}]%
        {AcqGro2005}
\bibfield{author}{\bibinfo{person}{Alessandro Acquisti} {and}
  \bibinfo{person}{Jens Grossklags}.} \bibinfo{year}{2005}\natexlab{}.
\newblock \showarticletitle{Privacy and rationality in individual decision
  making}.
\newblock \bibinfo{journal}{{\em IEEE Security \& Privacy\/}}
  \bibinfo{volume}{2} (\bibinfo{year}{2005}), \bibinfo{pages}{24--30}.
\newblock


\bibitem[\protect\citeauthoryear{Auxier, Rainie, Anderson, Perrin, Kumar, and
  Turner}{Auxier et~al\mbox{.}}{2019}]%
        {auxier2019americans}
\bibfield{author}{\bibinfo{person}{Brooke Auxier}, \bibinfo{person}{Lee
  Rainie}, \bibinfo{person}{Monica Anderson}, \bibinfo{person}{Andrew Perrin},
  \bibinfo{person}{Madhu Kumar}, {and} \bibinfo{person}{Erica Turner}.}
  \bibinfo{year}{2019}\natexlab{}.
\newblock \showarticletitle{Americans and privacy: Concerned, confused and
  feeling lack of control over their personal information}.
\newblock \bibinfo{journal}{{\em Pew Research Center: Internet, Science \& Tech
  (blog). November\/}}  \bibinfo{volume}{15} (\bibinfo{year}{2019}),
  \bibinfo{pages}{2019}.
\newblock


\bibitem[\protect\citeauthoryear{Bagchi, Cerveny, Hart, and Peterson}{Bagchi
  et~al\mbox{.}}{2003}]%
        {bagchi2003influence}
\bibfield{author}{\bibinfo{person}{Kallol Bagchi}, \bibinfo{person}{Robert
  Cerveny}, \bibinfo{person}{Paul Hart}, {and} \bibinfo{person}{Mark
  Peterson}.} \bibinfo{year}{2003}\natexlab{}.
\newblock \showarticletitle{The influence of national culture in information
  technology product adoption}.
\newblock \bibinfo{journal}{{\em AMCIS 2003 Proceedings\/}}
  (\bibinfo{year}{2003}), \bibinfo{pages}{119}.
\newblock


\bibitem[\protect\citeauthoryear{Balijepally, Mangalaraj, and
  Iyengar}{Balijepally et~al\mbox{.}}{2011}]%
        {balijepally2011we}
\bibfield{author}{\bibinfo{person}{VenuGopal Balijepally},
  \bibinfo{person}{George Mangalaraj}, {and} \bibinfo{person}{Kishen Iyengar}.}
  \bibinfo{year}{2011}\natexlab{}.
\newblock \showarticletitle{Are we wielding this hammer correctly? A reflective
  review of the application of cluster analysis in information systems
  research}.
\newblock \bibinfo{journal}{{\em Journal of the Association for Information
  Systems\/}} \bibinfo{volume}{12}, \bibinfo{number}{5} (\bibinfo{year}{2011}),
  \bibinfo{pages}{1}.
\newblock


\bibitem[\protect\citeauthoryear{Barnes}{Barnes}{2006}]%
        {barnes2006privacy}
\bibfield{author}{\bibinfo{person}{Susan~B Barnes}.}
  \bibinfo{year}{2006}\natexlab{}.
\newblock \showarticletitle{A privacy paradox: Social networking in the United
  States}.
\newblock \bibinfo{journal}{{\em First Monday\/}} \bibinfo{volume}{11},
  \bibinfo{number}{9} (\bibinfo{year}{2006}).
\newblock


\bibitem[\protect\citeauthoryear{Barth and De~Jong}{Barth and De~Jong}{2017}]%
        {barth2017privacy}
\bibfield{author}{\bibinfo{person}{Susanne Barth} {and}
  \bibinfo{person}{Menno~DT De~Jong}.} \bibinfo{year}{2017}\natexlab{}.
\newblock \showarticletitle{The privacy paradox--Investigating discrepancies
  between expressed privacy concerns and actual online behavior--A systematic
  literature review}.
\newblock \bibinfo{journal}{{\em Telematics and informatics\/}}
  \bibinfo{volume}{34}, \bibinfo{number}{7} (\bibinfo{year}{2017}),
  \bibinfo{pages}{1038--1058}.
\newblock


\bibitem[\protect\citeauthoryear{Bellman, Johnson, Kobrin, and Lohse}{Bellman
  et~al\mbox{.}}{2004}]%
        {bellman2004international}
\bibfield{author}{\bibinfo{person}{Steven Bellman}, \bibinfo{person}{Eric~J
  Johnson}, \bibinfo{person}{Stephen~J Kobrin}, {and} \bibinfo{person}{Gerald~L
  Lohse}.} \bibinfo{year}{2004}\natexlab{}.
\newblock \showarticletitle{International differences in information privacy
  concerns: A global survey of consumers}.
\newblock \bibinfo{journal}{{\em The Information Society\/}}
  \bibinfo{volume}{20}, \bibinfo{number}{5} (\bibinfo{year}{2004}),
  \bibinfo{pages}{313--324}.
\newblock


\bibitem[\protect\citeauthoryear{Bendixen}{Bendixen}{1996}]%
        {bendixen1996practical}
\bibfield{author}{\bibinfo{person}{Mike Bendixen}.}
  \bibinfo{year}{1996}\natexlab{}.
\newblock \showarticletitle{A practical guide to the use of correspondence
  analysis in marketing research}.
\newblock \bibinfo{journal}{{\em Marketing Research On-Line\/}}
  \bibinfo{volume}{1}, \bibinfo{number}{1} (\bibinfo{year}{1996}),
  \bibinfo{pages}{16--36}.
\newblock


\bibitem[\protect\citeauthoryear{Benenson, Girard, and Krontiris}{Benenson
  et~al\mbox{.}}{2015}]%
        {benenson2015user}
\bibfield{author}{\bibinfo{person}{Zinaida Benenson}, \bibinfo{person}{Anna
  Girard}, {and} \bibinfo{person}{Ioannis Krontiris}.}
  \bibinfo{year}{2015}\natexlab{}.
\newblock \showarticletitle{User Acceptance Factors for Anonymous Credentials:
  An Empirical Investigation.}. In \bibinfo{booktitle}{{\em WEIS}}.
\newblock


\bibitem[\protect\citeauthoryear{Blank, Dutton, and Lefkowitz}{Blank
  et~al\mbox{.}}{2019}]%
        {blank2019perceived}
\bibfield{author}{\bibinfo{person}{Grant Blank}, \bibinfo{person}{William~H
  Dutton}, {and} \bibinfo{person}{Julia Lefkowitz}.}
  \bibinfo{year}{2019}\natexlab{}.
\newblock \showarticletitle{Perceived Threats to Privacy Online: The Internet
  in Britain, the Oxford Internet Survey, 2019}.
\newblock  (\bibinfo{year}{2019}).
\newblock


\bibitem[\protect\citeauthoryear{Chandramouli, Goldstein, Jin, Raman, and
  Duan}{Chandramouli et~al\mbox{.}}{2013}]%
        {chandramouli2013real}
\bibfield{author}{\bibinfo{person}{Badrish Chandramouli},
  \bibinfo{person}{Jonathan Goldstein}, \bibinfo{person}{Xin Jin},
  \bibinfo{person}{Balan~Sethu Raman}, {and} \bibinfo{person}{Songyun Duan}.}
  \bibinfo{year}{2013}\natexlab{}.
\newblock \bibinfo{title}{Real-time-ready behavioral targeting in a large-scale
  advertisement system}.
\newblock   (\bibinfo{date}{May~14} \bibinfo{year}{2013}).
\newblock
\newblock
\shownote{US Patent 8,442,863.}


\bibitem[\protect\citeauthoryear{Chellappa and Sin}{Chellappa and Sin}{2005}]%
        {chellappa2005personalization}
\bibfield{author}{\bibinfo{person}{Ramnath~K Chellappa} {and}
  \bibinfo{person}{Raymond~G Sin}.} \bibinfo{year}{2005}\natexlab{}.
\newblock \showarticletitle{Personalization versus privacy: An empirical
  examination of the online consumer's dilemma}.
\newblock \bibinfo{journal}{{\em Information Technology and Management\/}}
  \bibinfo{volume}{6}, \bibinfo{number}{2-3} (\bibinfo{year}{2005}),
  \bibinfo{pages}{181--202}.
\newblock


\bibitem[\protect\citeauthoryear{Cho, Rivera-S{\'a}nchez, and Lim}{Cho
  et~al\mbox{.}}{2009}]%
        {cho2009multinational}
\bibfield{author}{\bibinfo{person}{Hichang Cho}, \bibinfo{person}{Milagros
  Rivera-S{\'a}nchez}, {and} \bibinfo{person}{Sun~Sun Lim}.}
  \bibinfo{year}{2009}\natexlab{}.
\newblock \showarticletitle{A multinational study on online privacy: global
  concerns and local responses}.
\newblock \bibinfo{journal}{{\em New media \& society\/}} \bibinfo{volume}{11},
  \bibinfo{number}{3} (\bibinfo{year}{2009}), \bibinfo{pages}{395--416}.
\newblock


\bibitem[\protect\citeauthoryear{Coles-Kemp and Kani-Zabihi}{Coles-Kemp and
  Kani-Zabihi}{2011}]%
        {coles2011practice}
\bibfield{author}{\bibinfo{person}{Lizzie Coles-Kemp} {and}
  \bibinfo{person}{Elahe Kani-Zabihi}.} \bibinfo{year}{2011}\natexlab{}.
\newblock \showarticletitle{Practice makes perfect: motivating confident
  privacy protection practices}. In \bibinfo{booktitle}{{\em 2011 IEEE Third
  International Conference on Privacy, Security, Risk and Trust and 2011 IEEE
  Third International Conference on Social Computing}}. IEEE,
  \bibinfo{pages}{866--871}.
\newblock


\bibitem[\protect\citeauthoryear{Coopamootoo and Gro{\ss}}{Coopamootoo and
  Gro{\ss}}{2017a}]%
        {coopamootoo2017niftynine}
\bibfield{author}{\bibinfo{person}{Kovila~P.L. Coopamootoo} {and}
  \bibinfo{person}{Thomas Gro{\ss}}.} \bibinfo{year}{2017}\natexlab{a}.
\newblock \bibinfo{booktitle}{{\em A Codebook for Evidence-Based Research: The
  Nifty Nine Completeness Indicators v1.1}}.
\newblock \bibinfo{type}{Technical Report} 1514.
  \bibinfo{institution}{Newcastle University}.
\newblock


\bibitem[\protect\citeauthoryear{Coopamootoo and Gro{\ss}}{Coopamootoo and
  Gro{\ss}}{2017b}]%
        {coopamootoo2017cyber}
\bibfield{author}{\bibinfo{person}{Kovila~PL Coopamootoo} {and}
  \bibinfo{person}{Thomas Gro{\ss}}.} \bibinfo{year}{2017}\natexlab{b}.
\newblock \showarticletitle{Cyber Security and Privacy Experiments: A Design
  and Reporting Toolkit}. In \bibinfo{booktitle}{{\em IFIP International Summer
  School on Privacy and Identity Management}}. Springer,
  \bibinfo{pages}{243--262}.
\newblock


\bibitem[\protect\citeauthoryear{Coopamootoo and Gro{\ss}}{Coopamootoo and
  Gro{\ss}}{2017c}]%
        {coopamootoo2017whyprivacy}
\bibfield{author}{\bibinfo{person}{Kovila~PL Coopamootoo} {and}
  \bibinfo{person}{Thomas Gro{\ss}}.} \bibinfo{year}{2017}\natexlab{c}.
\newblock \showarticletitle{Why Privacy is All But Forgotten - An Empirical
  Study of Privacy and Sharing Attitude}.
\newblock \bibinfo{journal}{{\em Proceedings on Privacy Enhancing
  Technologies\/}}  \bibinfo{volume}{4} (\bibinfo{year}{2017}),
  \bibinfo{pages}{39--60}.
\newblock


\bibitem[\protect\citeauthoryear{Coventry, Jeske, and Briggs}{Coventry
  et~al\mbox{.}}{2014}]%
        {coventry2014perceptions}
\bibfield{author}{\bibinfo{person}{Lynne Coventry}, \bibinfo{person}{Debora
  Jeske}, {and} \bibinfo{person}{Pamela Briggs}.}
  \bibinfo{year}{2014}\natexlab{}.
\newblock \showarticletitle{Perceptions and actions: Combining privacy and risk
  perceptions to better understand user behaviour}.
\newblock  (\bibinfo{year}{2014}).
\newblock


\bibitem[\protect\citeauthoryear{Cvrcek, Kumpost, Matyas, and Danezis}{Cvrcek
  et~al\mbox{.}}{2006}]%
        {cvrcek2006study}
\bibfield{author}{\bibinfo{person}{Dan Cvrcek}, \bibinfo{person}{Marek
  Kumpost}, \bibinfo{person}{Vashek Matyas}, {and} \bibinfo{person}{George
  Danezis}.} \bibinfo{year}{2006}\natexlab{}.
\newblock \showarticletitle{A study on the value of location privacy}. In
  \bibinfo{booktitle}{{\em Proceedings of the 5th ACM workshop on Privacy in
  electronic society}}. \bibinfo{pages}{109--118}.
\newblock


\bibitem[\protect\citeauthoryear{Das, Kramer, Dabbish, and Hong}{Das
  et~al\mbox{.}}{2015}]%
        {das2015role}
\bibfield{author}{\bibinfo{person}{Sauvik Das}, \bibinfo{person}{Adam~DI
  Kramer}, \bibinfo{person}{Laura~A Dabbish}, {and} \bibinfo{person}{Jason~I
  Hong}.} \bibinfo{year}{2015}\natexlab{}.
\newblock \showarticletitle{The role of social influence in security feature
  adoption}. In \bibinfo{booktitle}{{\em Proceedings of the 18th ACM conference
  on computer supported cooperative work \& social computing}}.
  \bibinfo{pages}{1416--1426}.
\newblock


\bibitem[\protect\citeauthoryear{Davis}{Davis}{1989}]%
        {davis1989perceived}
\bibfield{author}{\bibinfo{person}{Fred~D Davis}.}
  \bibinfo{year}{1989}\natexlab{}.
\newblock \showarticletitle{Perceived usefulness, perceived ease of use, and
  user acceptance of information technology}.
\newblock \bibinfo{journal}{{\em MIS quarterly\/}} (\bibinfo{year}{1989}),
  \bibinfo{pages}{319--340}.
\newblock


\bibitem[\protect\citeauthoryear{Davis, Bagozzi, and Warshaw}{Davis
  et~al\mbox{.}}{1989}]%
        {davis1989user}
\bibfield{author}{\bibinfo{person}{Fred~D Davis}, \bibinfo{person}{Richard~P
  Bagozzi}, {and} \bibinfo{person}{Paul~R Warshaw}.}
  \bibinfo{year}{1989}\natexlab{}.
\newblock \showarticletitle{User acceptance of computer technology: a
  comparison of two theoretical models}.
\newblock \bibinfo{journal}{{\em Management science\/}} \bibinfo{volume}{35},
  \bibinfo{number}{8} (\bibinfo{year}{1989}), \bibinfo{pages}{982--1003}.
\newblock


\bibitem[\protect\citeauthoryear{Dienlin and Trepte}{Dienlin and
  Trepte}{2015}]%
        {dienlin2015privacy}
\bibfield{author}{\bibinfo{person}{Tobias Dienlin} {and}
  \bibinfo{person}{Sabine Trepte}.} \bibinfo{year}{2015}\natexlab{}.
\newblock \showarticletitle{Is the privacy paradox a relic of the past? An
  in-depth analysis of privacy attitudes and privacy behaviors}.
\newblock \bibinfo{journal}{{\em European Journal of Social Psychology\/}}
  \bibinfo{volume}{45}, \bibinfo{number}{3} (\bibinfo{year}{2015}),
  \bibinfo{pages}{285--297}.
\newblock


\bibitem[\protect\citeauthoryear{Dinev, Albano, Xu, D'Atri, and Hart}{Dinev
  et~al\mbox{.}}{2016}]%
        {dinev2016individuals}
\bibfield{author}{\bibinfo{person}{Tamara Dinev}, \bibinfo{person}{Valentina
  Albano}, \bibinfo{person}{Heng Xu}, \bibinfo{person}{Alessandro D'Atri},
  {and} \bibinfo{person}{Paul Hart}.} \bibinfo{year}{2016}\natexlab{}.
\newblock \showarticletitle{Individuals' attitudes towards electronic health
  records: A privacy calculus perspective}.
\newblock In \bibinfo{booktitle}{{\em Advances in healthcare informatics and
  analytics}}. \bibinfo{publisher}{Springer}, \bibinfo{pages}{19--50}.
\newblock


\bibitem[\protect\citeauthoryear{Dinev, Bellotto, Hart, Russo, Serra, and
  Colautti}{Dinev et~al\mbox{.}}{2006}]%
        {dinev2006privacy}
\bibfield{author}{\bibinfo{person}{Tamara Dinev}, \bibinfo{person}{Massimo
  Bellotto}, \bibinfo{person}{Paul Hart}, \bibinfo{person}{Vincenzo Russo},
  \bibinfo{person}{Ilaria Serra}, {and} \bibinfo{person}{Christian Colautti}.}
  \bibinfo{year}{2006}\natexlab{}.
\newblock \showarticletitle{Privacy calculus model in e-commerce--a study of
  Italy and the United States}.
\newblock \bibinfo{journal}{{\em European Journal of Information Systems\/}}
  \bibinfo{volume}{15}, \bibinfo{number}{4} (\bibinfo{year}{2006}),
  \bibinfo{pages}{389--402}.
\newblock


\bibitem[\protect\citeauthoryear{Dinev and Hart}{Dinev and Hart}{2006}]%
        {dinev2006extended}
\bibfield{author}{\bibinfo{person}{Tamara Dinev} {and} \bibinfo{person}{Paul
  Hart}.} \bibinfo{year}{2006}\natexlab{}.
\newblock \showarticletitle{An extended privacy calculus model for e-commerce
  transactions}.
\newblock \bibinfo{journal}{{\em Information systems research\/}}
  \bibinfo{volume}{17}, \bibinfo{number}{1} (\bibinfo{year}{2006}),
  \bibinfo{pages}{61--80}.
\newblock


\bibitem[\protect\citeauthoryear{Ebbert and Dutke}{Ebbert and Dutke}{2020}]%
        {ebbert2020patterns}
\bibfield{author}{\bibinfo{person}{Daniel Ebbert} {and}
  \bibinfo{person}{Stephan Dutke}.} \bibinfo{year}{2020}\natexlab{}.
\newblock \showarticletitle{Patterns in students' usage of lecture recordings:
  a cluster analysis of self-report data}.
\newblock \bibinfo{journal}{{\em Research in Learning Technology\/}}
  \bibinfo{volume}{28} (\bibinfo{year}{2020}).
\newblock


\bibitem[\protect\citeauthoryear{Edelman and Luca}{Edelman and Luca}{2014}]%
        {edelman2014digital}
\bibfield{author}{\bibinfo{person}{Benjamin~G Edelman} {and}
  \bibinfo{person}{Michael Luca}.} \bibinfo{year}{2014}\natexlab{}.
\newblock \showarticletitle{Digital discrimination: The case of Airbnb. com}.
\newblock \bibinfo{journal}{{\em Harvard Business School NOM Unit Working
  Paper\/}} \bibinfo{number}{14-054} (\bibinfo{year}{2014}).
\newblock


\bibitem[\protect\citeauthoryear{ENISA}{ENISA}{2020}]%
        {ENISA2020}
\bibfield{author}{\bibinfo{person}{ENISA}.} \bibinfo{year}{2020}\natexlab{}.
\newblock \bibinfo{title}{Privacy Enhancing Technologies}.
\newblock   (\bibinfo{year}{2020}).
\newblock
\showURL{%
\url{https://www.enisa.europa.eu/topics/data-protection/privacy-enhancing-technologies}}


\bibitem[\protect\citeauthoryear{Erumban and De~Jong}{Erumban and
  De~Jong}{2006}]%
        {erumban2006cross}
\bibfield{author}{\bibinfo{person}{Abdul~Azeez Erumban} {and}
  \bibinfo{person}{Simon~B De~Jong}.} \bibinfo{year}{2006}\natexlab{}.
\newblock \showarticletitle{Cross-country differences in ICT adoption: A
  consequence of Culture?}
\newblock \bibinfo{journal}{{\em Journal of world business\/}}
  \bibinfo{volume}{41}, \bibinfo{number}{4} (\bibinfo{year}{2006}),
  \bibinfo{pages}{302--314}.
\newblock


\bibitem[\protect\citeauthoryear{Eurobarometer}{Eurobarometer}{2019}]%
        {eurobarometer2019european}
\bibfield{author}{\bibinfo{person}{Special Eurobarometer}.}
  \bibinfo{year}{2019}\natexlab{}.
\newblock \showarticletitle{The General Data Protection Regulation - Special
  Eurobarometer 487a}.
\newblock \bibinfo{journal}{{\em Special Eurobarometer\/}}
  (\bibinfo{year}{2019}).
\newblock


\bibitem[\protect\citeauthoryear{Forbes-Insights}{Forbes-Insights}{2019}]%
        {forbes2019}
\bibfield{author}{\bibinfo{person}{Forbes-Insights}.}
  \bibinfo{year}{2019}\natexlab{}.
\newblock \bibinfo{title}{Rethinking Privacy in the AI Era}.
\newblock   (\bibinfo{year}{2019}).
\newblock
\showURL{%
\url{https://www.forbes.com/sites/insights-intelai/2019/03/27/rethinking-privacy-for-the-ai-era/}}


\bibitem[\protect\citeauthoryear{Garg, Benton, and Camp}{Garg
  et~al\mbox{.}}{2014}]%
        {garg2014privacy}
\bibfield{author}{\bibinfo{person}{Vaibhav Garg}, \bibinfo{person}{Kevin
  Benton}, {and} \bibinfo{person}{L~Jean Camp}.}
  \bibinfo{year}{2014}\natexlab{}.
\newblock \showarticletitle{The privacy paradox: a Facebook case study}. In
  \bibinfo{booktitle}{{\em 2014 TPRC conference paper}}.
\newblock


\bibitem[\protect\citeauthoryear{Garrett}{Garrett}{2019}]%
        {garrett2019social}
\bibfield{author}{\bibinfo{person}{R~Kelly Garrett}.}
  \bibinfo{year}{2019}\natexlab{}.
\newblock \showarticletitle{Social media's contribution to political
  misperceptions in US Presidential elections}.
\newblock \bibinfo{journal}{{\em PloS one\/}} \bibinfo{volume}{14},
  \bibinfo{number}{3} (\bibinfo{year}{2019}), \bibinfo{pages}{e0213500}.
\newblock


\bibitem[\protect\citeauthoryear{Gerber, Gerber, and Volkamer}{Gerber
  et~al\mbox{.}}{2018}]%
        {gerber2018explaining}
\bibfield{author}{\bibinfo{person}{Nina Gerber}, \bibinfo{person}{Paul Gerber},
  {and} \bibinfo{person}{Melanie Volkamer}.} \bibinfo{year}{2018}\natexlab{}.
\newblock \showarticletitle{Explaining the privacy paradox: A systematic review
  of literature investigating privacy attitude and behavior}.
\newblock \bibinfo{journal}{{\em Computers \& Security\/}}
  \bibinfo{volume}{77} (\bibinfo{year}{2018}), \bibinfo{pages}{226--261}.
\newblock


\bibitem[\protect\citeauthoryear{Gerber, Zimmermann, and Volkamer}{Gerber
  et~al\mbox{.}}{2019}]%
        {gerber2019johnny}
\bibfield{author}{\bibinfo{person}{Nina Gerber}, \bibinfo{person}{Verena
  Zimmermann}, {and} \bibinfo{person}{Melanie Volkamer}.}
  \bibinfo{year}{2019}\natexlab{}.
\newblock \showarticletitle{Why Johnny Fails to Protect his Privacy}. In
  \bibinfo{booktitle}{{\em 2019 IEEE European Symposium on Security and Privacy
  Workshops (EuroS\&PW)}}. IEEE, \bibinfo{pages}{109--118}.
\newblock


\bibitem[\protect\citeauthoryear{Greenacre}{Greenacre}{2017}]%
        {greenacre2017correspondence}
\bibfield{author}{\bibinfo{person}{Michael Greenacre}.}
  \bibinfo{year}{2017}\natexlab{}.
\newblock \bibinfo{booktitle}{{\em Correspondence analysis in practice}}.
\newblock \bibinfo{publisher}{Chapman and Hall/CRC}.
\newblock


\bibitem[\protect\citeauthoryear{Greenacre}{Greenacre}{1988}]%
        {greenacre1988clustering}
\bibfield{author}{\bibinfo{person}{Michael~J Greenacre}.}
  \bibinfo{year}{1988}\natexlab{}.
\newblock \showarticletitle{Clustering the rows and columns of a contingency
  table}.
\newblock \bibinfo{journal}{{\em Journal of Classification\/}}
  \bibinfo{volume}{5}, \bibinfo{number}{1} (\bibinfo{year}{1988}),
  \bibinfo{pages}{39--51}.
\newblock


\bibitem[\protect\citeauthoryear{Hair, Black, Babin, Anderson, Tatham,
  et~al\mbox{.}}{Hair et~al\mbox{.}}{1998}]%
        {hair1998multivariate}
\bibfield{author}{\bibinfo{person}{Joseph~F Hair}, \bibinfo{person}{William~C
  Black}, \bibinfo{person}{Barry~J Babin}, \bibinfo{person}{Rolph~E Anderson},
  \bibinfo{person}{Ronald~L Tatham}, {et~al\mbox{.}}}
  \bibinfo{year}{1998}\natexlab{}.
\newblock \bibinfo{booktitle}{{\em Multivariate data analysis}}.
  Vol.~\bibinfo{volume}{5}.
\newblock \bibinfo{publisher}{Prentice hall Upper Saddle River, NJ}.
\newblock


\bibitem[\protect\citeauthoryear{Hallgren}{Hallgren}{2012}]%
        {hallgren2012computing}
\bibfield{author}{\bibinfo{person}{Kevin~A Hallgren}.}
  \bibinfo{year}{2012}\natexlab{}.
\newblock \showarticletitle{Computing inter-rater reliability for observational
  data: an overview and tutorial}.
\newblock \bibinfo{journal}{{\em Tutorials in quantitative methods for
  psychology\/}} \bibinfo{volume}{8}, \bibinfo{number}{1}
  (\bibinfo{year}{2012}), \bibinfo{pages}{23}.
\newblock


\bibitem[\protect\citeauthoryear{Hallinan, Friedewald, and McCarthy}{Hallinan
  et~al\mbox{.}}{2012}]%
        {hallinan2012citizens}
\bibfield{author}{\bibinfo{person}{Dara Hallinan}, \bibinfo{person}{Michael
  Friedewald}, {and} \bibinfo{person}{Paul McCarthy}.}
  \bibinfo{year}{2012}\natexlab{}.
\newblock \showarticletitle{Citizens' perceptions of data protection and
  privacy in Europe}.
\newblock \bibinfo{journal}{{\em Computer law \& security review\/}}
  \bibinfo{volume}{28}, \bibinfo{number}{3} (\bibinfo{year}{2012}),
  \bibinfo{pages}{263--272}.
\newblock


\bibitem[\protect\citeauthoryear{Hann, Hui, Lee, and Png}{Hann
  et~al\mbox{.}}{2002}]%
        {hann2002online}
\bibfield{author}{\bibinfo{person}{Il-Horn Hann}, \bibinfo{person}{Kai-Lung
  Hui}, \bibinfo{person}{Tom Lee}, {and} \bibinfo{person}{I Png}.}
  \bibinfo{year}{2002}\natexlab{}.
\newblock \showarticletitle{Online information privacy: Measuring the
  cost-benefit trade-off}.
\newblock \bibinfo{journal}{{\em ICIS 2002 proceedings\/}}
  (\bibinfo{year}{2002}), \bibinfo{pages}{1}.
\newblock


\bibitem[\protect\citeauthoryear{Harborth and Pape}{Harborth and Pape}{2018}]%
        {harborth2018examining}
\bibfield{author}{\bibinfo{person}{David Harborth} {and}
  \bibinfo{person}{Sebastian Pape}.} \bibinfo{year}{2018}\natexlab{}.
\newblock \showarticletitle{Examining technology use factors of
  privacy-enhancing technologies: the role of perceived anonymity and trust}.
\newblock  (\bibinfo{year}{2018}).
\newblock


\bibitem[\protect\citeauthoryear{Hargittai et~al\mbox{.}}{Hargittai
  et~al\mbox{.}}{2010}]%
        {hargittai2010facebook}
\bibfield{author}{\bibinfo{person}{Eszter Hargittai} {et~al\mbox{.}}}
  \bibinfo{year}{2010}\natexlab{}.
\newblock \showarticletitle{Facebook privacy settings: Who cares?}
\newblock \bibinfo{journal}{{\em First Monday\/}} (\bibinfo{year}{2010}).
\newblock


\bibitem[\protect\citeauthoryear{Hofstede}{Hofstede}{1984}]%
        {hofstede1984culture}
\bibfield{author}{\bibinfo{person}{Geert Hofstede}.}
  \bibinfo{year}{1984}\natexlab{}.
\newblock \bibinfo{booktitle}{{\em Culture's consequences: International
  differences in work-related values}}. Vol.~\bibinfo{volume}{5}.
\newblock \bibinfo{publisher}{sage}.
\newblock


\bibitem[\protect\citeauthoryear{Hofstede}{Hofstede}{2001}]%
        {hofstede2001culture}
\bibfield{author}{\bibinfo{person}{Geert Hofstede}.}
  \bibinfo{year}{2001}\natexlab{}.
\newblock \bibinfo{booktitle}{{\em Culture's consequences: Comparing values,
  behaviors, institutions and organizations across nations}}.
\newblock \bibinfo{publisher}{Sage publications}.
\newblock


\bibitem[\protect\citeauthoryear{Huang and Bashir}{Huang and Bashir}{2016}]%
        {huang2016privacy}
\bibfield{author}{\bibinfo{person}{Hsiao-Ying Huang} {and}
  \bibinfo{person}{Masooda Bashir}.} \bibinfo{year}{2016}\natexlab{}.
\newblock \showarticletitle{Privacy by region: Evaluation online users' privacy
  perceptions by geographical region}. In \bibinfo{booktitle}{{\em 2016 Future
  Technologies Conference (FTC)}}. IEEE, \bibinfo{pages}{968--977}.
\newblock


\bibitem[\protect\citeauthoryear{Index-Mundi}{Index-Mundi}{2019}]%
        {indexmundi}
\bibfield{author}{\bibinfo{person}{Index-Mundi}.}
  \bibinfo{year}{2019}\natexlab{}.
\newblock \bibinfo{title}{Germany Demographics Profile 2019}.
\newblock   (\bibinfo{year}{2019}).
\newblock
\showURL{%
\url{https://www.indexmundi.com/germany/demographics_profile.html}}


\bibitem[\protect\citeauthoryear{Kassambara and Mundt}{Kassambara and
  Mundt}{2017}]%
        {kassambara2017package}
\bibfield{author}{\bibinfo{person}{Alboukadel Kassambara} {and}
  \bibinfo{person}{Fabian Mundt}.} \bibinfo{year}{2017}\natexlab{}.
\newblock \showarticletitle{Package `factoextra'}.
\newblock \bibinfo{journal}{{\em Extract and visualize the results of
  multivariate data analyses\/}}  \bibinfo{volume}{76} (\bibinfo{year}{2017}).
\newblock


\bibitem[\protect\citeauthoryear{Kaufman and Rousseeuw}{Kaufman and
  Rousseeuw}{2009}]%
        {kaufman2009finding}
\bibfield{author}{\bibinfo{person}{Leonard Kaufman} {and}
  \bibinfo{person}{Peter~J Rousseeuw}.} \bibinfo{year}{2009}\natexlab{}.
\newblock \bibinfo{booktitle}{{\em Finding groups in data: an introduction to
  cluster analysis}}. Vol.~\bibinfo{volume}{344}.
\newblock \bibinfo{publisher}{John Wiley \& Sons}.
\newblock


\bibitem[\protect\citeauthoryear{Kehr, Kowatsch, Wentzel, and Fleisch}{Kehr
  et~al\mbox{.}}{2015}]%
        {kehr2015blissfully}
\bibfield{author}{\bibinfo{person}{Flavius Kehr}, \bibinfo{person}{Tobias
  Kowatsch}, \bibinfo{person}{Daniel Wentzel}, {and} \bibinfo{person}{Elgar
  Fleisch}.} \bibinfo{year}{2015}\natexlab{}.
\newblock \showarticletitle{Blissfully ignorant: the effects of general privacy
  concerns, general institutional trust, and affect in the privacy calculus}.
\newblock \bibinfo{journal}{{\em Information Systems Journal\/}}
  \bibinfo{volume}{25}, \bibinfo{number}{6} (\bibinfo{year}{2015}),
  \bibinfo{pages}{607--635}.
\newblock


\bibitem[\protect\citeauthoryear{Kokolakis}{Kokolakis}{2017}]%
        {kokolakis2017privacy}
\bibfield{author}{\bibinfo{person}{Spyros Kokolakis}.}
  \bibinfo{year}{2017}\natexlab{}.
\newblock \showarticletitle{Privacy attitudes and privacy behaviour: A review
  of current research on the privacy paradox phenomenon}.
\newblock \bibinfo{journal}{{\em Computers \& Security\/}}
  \bibinfo{volume}{64} (\bibinfo{year}{2017}), \bibinfo{pages}{122--134}.
\newblock


\bibitem[\protect\citeauthoryear{Krasnova and Veltri}{Krasnova and
  Veltri}{2010}]%
        {krasnova2010privacy}
\bibfield{author}{\bibinfo{person}{Hanna Krasnova} {and}
  \bibinfo{person}{Natasha~F Veltri}.} \bibinfo{year}{2010}\natexlab{}.
\newblock \showarticletitle{Privacy calculus on social networking sites:
  Explorative evidence from Germany and USA}. In \bibinfo{booktitle}{{\em 2010
  43rd Hawaii international conference on system sciences}}. IEEE,
  \bibinfo{pages}{1--10}.
\newblock


\bibitem[\protect\citeauthoryear{Lallmahamood}{Lallmahamood}{1970}]%
        {lallmahamood1970examination}
\bibfield{author}{\bibinfo{person}{Muniruddeen Lallmahamood}.}
  \bibinfo{year}{1970}\natexlab{}.
\newblock \showarticletitle{An Examination of Individual{\~A}{{\textcent}}
  {\^A} {\^A} s Perceived Security and Privacy of the Internet in Malaysia and
  the Influence of This on Their Intention to Use E-Commerce: Using An
  Extension of the Technology Acceptance Model}.
\newblock \bibinfo{journal}{{\em The Journal of Internet Banking and
  Commerce\/}} \bibinfo{volume}{12}, \bibinfo{number}{3}
  (\bibinfo{year}{1970}), \bibinfo{pages}{1--26}.
\newblock


\bibitem[\protect\citeauthoryear{Lally, Van~Jaarsveld, Potts, and Wardle}{Lally
  et~al\mbox{.}}{2010}]%
        {lally2010habits}
\bibfield{author}{\bibinfo{person}{Phillippa Lally},
  \bibinfo{person}{Cornelia~HM Van~Jaarsveld}, \bibinfo{person}{Henry~WW
  Potts}, {and} \bibinfo{person}{Jane Wardle}.}
  \bibinfo{year}{2010}\natexlab{}.
\newblock \showarticletitle{How are habits formed: Modelling habit formation in
  the real world}.
\newblock \bibinfo{journal}{{\em European journal of social psychology\/}}
  \bibinfo{volume}{40}, \bibinfo{number}{6} (\bibinfo{year}{2010}),
  \bibinfo{pages}{998--1009}.
\newblock


\bibitem[\protect\citeauthoryear{Lankton, McKnight, and Tripp}{Lankton
  et~al\mbox{.}}{2017}]%
        {lankton2017facebook}
\bibfield{author}{\bibinfo{person}{Nancy~K Lankton},
  \bibinfo{person}{D~Harrison McKnight}, {and} \bibinfo{person}{John~F Tripp}.}
  \bibinfo{year}{2017}\natexlab{}.
\newblock \showarticletitle{Facebook privacy management strategies: A cluster
  analysis of user privacy behaviors}.
\newblock \bibinfo{journal}{{\em Computers in Human Behavior\/}}
  \bibinfo{volume}{76} (\bibinfo{year}{2017}), \bibinfo{pages}{149--163}.
\newblock


\bibitem[\protect\citeauthoryear{Legris, Ingham, and Collerette}{Legris
  et~al\mbox{.}}{2003}]%
        {legris2003people}
\bibfield{author}{\bibinfo{person}{Paul Legris}, \bibinfo{person}{John Ingham},
  {and} \bibinfo{person}{Pierre Collerette}.} \bibinfo{year}{2003}\natexlab{}.
\newblock \showarticletitle{Why do people use information technology? A
  critical review of the technology acceptance model}.
\newblock \bibinfo{journal}{{\em Information \& management\/}}
  \bibinfo{volume}{40}, \bibinfo{number}{3} (\bibinfo{year}{2003}),
  \bibinfo{pages}{191--204}.
\newblock


\bibitem[\protect\citeauthoryear{Lemay, Doleck, and Bazelais}{Lemay
  et~al\mbox{.}}{2017}]%
        {lemay2017passion}
\bibfield{author}{\bibinfo{person}{David~John Lemay}, \bibinfo{person}{Tenzin
  Doleck}, {and} \bibinfo{person}{Paul Bazelais}.}
  \bibinfo{year}{2017}\natexlab{}.
\newblock \showarticletitle{``Passion and concern for privacy'' as factors
  affecting snapchat use: A situated perspective on technology acceptance}.
\newblock \bibinfo{journal}{{\em Computers in Human Behavior\/}}
  \bibinfo{volume}{75} (\bibinfo{year}{2017}), \bibinfo{pages}{264--271}.
\newblock


\bibitem[\protect\citeauthoryear{Liu, Li, Xiong, Gao, and Wu}{Liu
  et~al\mbox{.}}{2010}]%
        {liu2010understanding}
\bibfield{author}{\bibinfo{person}{Yanchi Liu}, \bibinfo{person}{Zhongmou Li},
  \bibinfo{person}{Hui Xiong}, \bibinfo{person}{Xuedong Gao}, {and}
  \bibinfo{person}{Junjie Wu}.} \bibinfo{year}{2010}\natexlab{}.
\newblock \showarticletitle{Understanding of internal clustering validation
  measures}. In \bibinfo{booktitle}{{\em 2010 IEEE International Conference on
  Data Mining}}. IEEE, \bibinfo{pages}{911--916}.
\newblock


\bibitem[\protect\citeauthoryear{MacQueen et~al\mbox{.}}{MacQueen
  et~al\mbox{.}}{1967}]%
        {macqueen1967some}
\bibfield{author}{\bibinfo{person}{James MacQueen} {et~al\mbox{.}}}
  \bibinfo{year}{1967}\natexlab{}.
\newblock \showarticletitle{Some methods for classification and analysis of
  multivariate observations}. In \bibinfo{booktitle}{{\em Proceedings of the
  fifth Berkeley symposium on mathematical statistics and probability}},
  Vol.~\bibinfo{volume}{1}. Oakland, CA, USA, \bibinfo{pages}{281--297}.
\newblock


\bibitem[\protect\citeauthoryear{Madden, Rainie, Zickuhr, Duggan, and
  Smith}{Madden et~al\mbox{.}}{2014}]%
        {madden2014public}
\bibfield{author}{\bibinfo{person}{Mary Madden}, \bibinfo{person}{Lee Rainie},
  \bibinfo{person}{Kathryn Zickuhr}, \bibinfo{person}{Maeve Duggan}, {and}
  \bibinfo{person}{Aaron Smith}.} \bibinfo{year}{2014}\natexlab{}.
\newblock \showarticletitle{Public perceptions of privacy and security in the
  post-Snowden era}.
\newblock \bibinfo{journal}{{\em Pew Research Center\/}}  \bibinfo{volume}{12}
  (\bibinfo{year}{2014}).
\newblock


\bibitem[\protect\citeauthoryear{Marshall, Cardon, Norris, Goreva, and
  D'Souza}{Marshall et~al\mbox{.}}{2008}]%
        {marshall2008social}
\bibfield{author}{\bibinfo{person}{Bryan~A Marshall}, \bibinfo{person}{Peter~W
  Cardon}, \bibinfo{person}{Daniel~T Norris}, \bibinfo{person}{Natalya Goreva},
  {and} \bibinfo{person}{Ryan D'Souza}.} \bibinfo{year}{2008}\natexlab{}.
\newblock \showarticletitle{Social networking websites in India and the United
  States: A cross-national comparison of online privacy and communication}.
\newblock \bibinfo{journal}{{\em Issues in Information Systems\/}}
  \bibinfo{volume}{9}, \bibinfo{number}{2} (\bibinfo{year}{2008}),
  \bibinfo{pages}{87--94}.
\newblock


\bibitem[\protect\citeauthoryear{McDonald and Cranor}{McDonald and
  Cranor}{2008}]%
        {mcdonald2008cost}
\bibfield{author}{\bibinfo{person}{Aleecia~M McDonald} {and}
  \bibinfo{person}{Lorrie~Faith Cranor}.} \bibinfo{year}{2008}\natexlab{}.
\newblock \showarticletitle{The cost of reading privacy policies}.
\newblock \bibinfo{journal}{{\em Isjlp\/}}  \bibinfo{volume}{4}
  (\bibinfo{year}{2008}), \bibinfo{pages}{543}.
\newblock


\bibitem[\protect\citeauthoryear{McDonald, Schoenebeck, and Forte}{McDonald
  et~al\mbox{.}}{2019}]%
        {mcdonald2019reliability}
\bibfield{author}{\bibinfo{person}{Nora McDonald}, \bibinfo{person}{Sarita
  Schoenebeck}, {and} \bibinfo{person}{Andrea Forte}.}
  \bibinfo{year}{2019}\natexlab{}.
\newblock \showarticletitle{Reliability and inter-rater reliability in
  qualitative research: Norms and guidelines for CSCW and HCI practice}.
\newblock \bibinfo{journal}{{\em Proceedings of the ACM on Human-Computer
  Interaction\/}} \bibinfo{volume}{3}, \bibinfo{number}{CSCW}
  (\bibinfo{year}{2019}), \bibinfo{pages}{1--23}.
\newblock


\bibitem[\protect\citeauthoryear{Miltgen, Popovi{\v{c}}, and Oliveira}{Miltgen
  et~al\mbox{.}}{2013}]%
        {miltgen2013determinants}
\bibfield{author}{\bibinfo{person}{Caroline~Lancelot Miltgen},
  \bibinfo{person}{Ale{\v{s}} Popovi{\v{c}}}, {and} \bibinfo{person}{Tiago
  Oliveira}.} \bibinfo{year}{2013}\natexlab{}.
\newblock \showarticletitle{Determinants of end-user acceptance of biometrics:
  Integrating the ``Big 3'' of technology acceptance with privacy context}.
\newblock \bibinfo{journal}{{\em Decision Support Systems\/}}
  \bibinfo{volume}{56} (\bibinfo{year}{2013}), \bibinfo{pages}{103--114}.
\newblock


\bibitem[\protect\citeauthoryear{Miltgen and Smith}{Miltgen and Smith}{2015}]%
        {miltgen2015exploring}
\bibfield{author}{\bibinfo{person}{Caroline~Lancelot Miltgen} {and}
  \bibinfo{person}{H~Jeff Smith}.} \bibinfo{year}{2015}\natexlab{}.
\newblock \showarticletitle{Exploring information privacy regulation, risks,
  trust, and behavior}.
\newblock \bibinfo{journal}{{\em Information \& Management\/}}
  \bibinfo{volume}{52}, \bibinfo{number}{6} (\bibinfo{year}{2015}),
  \bibinfo{pages}{741--759}.
\newblock


\bibitem[\protect\citeauthoryear{Murgia and Harlow}{Murgia and Harlow}{2019}]%
        {murgia2019how}
\bibfield{author}{\bibinfo{person}{Madhumita Murgia} {and} \bibinfo{person}{Max
  Harlow}.} \bibinfo{year}{2019}\natexlab{}.
\newblock \bibinfo{title}{How top health websites are sharing sensitive data
  with advertisers}.
\newblock   (\bibinfo{year}{2019}).
\newblock
\showURL{%
\url{https://www.ft.com/content/0fbf4d8e-022b-11ea-be59-e49b2a136b8d}}


\bibitem[\protect\citeauthoryear{Namara, Wilkinson, Caine, and
  Knijnenburg}{Namara et~al\mbox{.}}{2020}]%
        {namara2020emotional}
\bibfield{author}{\bibinfo{person}{Moses Namara}, \bibinfo{person}{Daricia
  Wilkinson}, \bibinfo{person}{Kelly Caine}, {and} \bibinfo{person}{Bart~P
  Knijnenburg}.} \bibinfo{year}{2020}\natexlab{}.
\newblock \showarticletitle{Emotional and Practical Considerations Towards the
  Adoption and Abandonment of VPNs as a Privacy-Enhancing Technology}.
\newblock \bibinfo{journal}{{\em Proceedings on Privacy Enhancing
  Technologies\/}} \bibinfo{volume}{2020}, \bibinfo{number}{1}
  (\bibinfo{year}{2020}), \bibinfo{pages}{83--102}.
\newblock


\bibitem[\protect\citeauthoryear{Norberg, Horne, and Horne}{Norberg
  et~al\mbox{.}}{2007}]%
        {norberg2007privacy}
\bibfield{author}{\bibinfo{person}{Patricia~A Norberg},
  \bibinfo{person}{Daniel~R Horne}, {and} \bibinfo{person}{David~A Horne}.}
  \bibinfo{year}{2007}\natexlab{}.
\newblock \showarticletitle{The privacy paradox: Personal information
  disclosure intentions versus behaviors}.
\newblock \bibinfo{journal}{{\em Journal of Consumer Affairs\/}}
  \bibinfo{volume}{41}, \bibinfo{number}{1} (\bibinfo{year}{2007}),
  \bibinfo{pages}{100--126}.
\newblock


\bibitem[\protect\citeauthoryear{Oduor and Oinas-Kukkonen}{Oduor and
  Oinas-Kukkonen}{2017}]%
        {oduor2017commitment}
\bibfield{author}{\bibinfo{person}{Michael Oduor} {and} \bibinfo{person}{Harri
  Oinas-Kukkonen}.} \bibinfo{year}{2017}\natexlab{}.
\newblock \showarticletitle{Commitment devices as behavior change support
  systems: a study of users' perceived competence and continuance intention}.
  In \bibinfo{booktitle}{{\em International Conference on Persuasive
  Technology}}. Springer, \bibinfo{pages}{201--213}.
\newblock


\bibitem[\protect\citeauthoryear{Olson, Grudin, and Horvitz}{Olson
  et~al\mbox{.}}{2005}]%
        {olson2005study}
\bibfield{author}{\bibinfo{person}{Judith~S Olson}, \bibinfo{person}{Jonathan
  Grudin}, {and} \bibinfo{person}{Eric Horvitz}.}
  \bibinfo{year}{2005}\natexlab{}.
\newblock \showarticletitle{A study of preferences for sharing and privacy}. In
  \bibinfo{booktitle}{{\em CHI'05 extended abstracts on Human factors in
  computing systems}}. \bibinfo{pages}{1985--1988}.
\newblock


\bibitem[\protect\citeauthoryear{Oomen and Leenes}{Oomen and Leenes}{2008}]%
        {oomen2008privacy}
\bibfield{author}{\bibinfo{person}{Isabelle Oomen} {and}
  \bibinfo{person}{Ronald Leenes}.} \bibinfo{year}{2008}\natexlab{}.
\newblock \showarticletitle{Privacy risk perceptions and privacy protection
  strategies}.
\newblock In \bibinfo{booktitle}{{\em Policies and research in identity
  management}}. \bibinfo{publisher}{Springer}, \bibinfo{pages}{121--138}.
\newblock


\bibitem[\protect\citeauthoryear{Park}{Park}{2015}]%
        {park2015men}
\bibfield{author}{\bibinfo{person}{Yong~Jin Park}.}
  \bibinfo{year}{2015}\natexlab{}.
\newblock \showarticletitle{Do men and women differ in privacy? Gendered
  privacy and (in) equality in the Internet}.
\newblock \bibinfo{journal}{{\em Computers in Human Behavior\/}}
  \bibinfo{volume}{50} (\bibinfo{year}{2015}), \bibinfo{pages}{252--258}.
\newblock


\bibitem[\protect\citeauthoryear{Parliament}{Parliament}{2018}]%
        {parliament2018data}
\bibfield{author}{\bibinfo{person}{UK Parliament}.}
  \bibinfo{year}{2018}\natexlab{}.
\newblock \showarticletitle{Data Protection Act 2018}.
\newblock \bibinfo{journal}{{\em URL https://services. parliament.
  uk/bills/2017-19/dataprotection. html\/}} (\bibinfo{year}{2018}).
\newblock


\bibitem[\protect\citeauthoryear{Peer, Brandimarte, Samat, and Acquisti}{Peer
  et~al\mbox{.}}{2017}]%
        {peer2017beyond}
\bibfield{author}{\bibinfo{person}{Eyal Peer}, \bibinfo{person}{Laura
  Brandimarte}, \bibinfo{person}{Sonam Samat}, {and}
  \bibinfo{person}{Alessandro Acquisti}.} \bibinfo{year}{2017}\natexlab{}.
\newblock \showarticletitle{Beyond the Turk: Alternative platforms for
  crowdsourcing behavioral research}.
\newblock \bibinfo{journal}{{\em Journal of Experimental Social Psychology\/}}
  \bibinfo{volume}{70} (\bibinfo{year}{2017}), \bibinfo{pages}{153--163}.
\newblock


\bibitem[\protect\citeauthoryear{Piper}{Piper}{2019}]%
        {piper2019data}
\bibfield{author}{\bibinfo{person}{DLA Piper}.}
  \bibinfo{year}{2019}\natexlab{}.
\newblock \bibinfo{title}{Data protection Laws of the world. 2019}.
\newblock   (\bibinfo{year}{2019}).
\newblock


\bibitem[\protect\citeauthoryear{Polit and Beck}{Polit and Beck}{2010}]%
        {polit2010generalization}
\bibfield{author}{\bibinfo{person}{Denise~F Polit} {and}
  \bibinfo{person}{Cheryl~Tatano Beck}.} \bibinfo{year}{2010}\natexlab{}.
\newblock \showarticletitle{Generalization in quantitative and qualitative
  research: Myths and strategies}.
\newblock \bibinfo{journal}{{\em International journal of nursing studies\/}}
  \bibinfo{volume}{47}, \bibinfo{number}{11} (\bibinfo{year}{2010}),
  \bibinfo{pages}{1451--1458}.
\newblock


\bibitem[\protect\citeauthoryear{Preibusch}{Preibusch}{2013}]%
        {preibusch2013guide}
\bibfield{author}{\bibinfo{person}{S{\"o}ren Preibusch}.}
  \bibinfo{year}{2013}\natexlab{}.
\newblock \showarticletitle{Guide to measuring privacy concern: Review of
  survey and observational instruments}.
\newblock \bibinfo{journal}{{\em International Journal of Human-Computer
  Studies\/}} \bibinfo{volume}{71}, \bibinfo{number}{12}
  (\bibinfo{year}{2013}), \bibinfo{pages}{1133--1143}.
\newblock


\bibitem[\protect\citeauthoryear{Privacy-International}{Privacy-International}{2019}]%
        {privacyint2019nobodys}
\bibfield{author}{\bibinfo{person}{Privacy-International}.}
  \bibinfo{year}{2019}\natexlab{}.
\newblock \bibinfo{title}{No Body's Business But Mine: How Menstruation Apps
  Are Sharing Your Data}.
\newblock   (\bibinfo{year}{2019}).
\newblock
\showURL{%
\url{https://privacyinternational.org/long-read/3196/no-bodys-business-mine-how-menstruation-apps-are-sharing-your-data}}


\bibitem[\protect\citeauthoryear{PWC}{PWC}{2016}]%
        {pwc2016data}
\bibfield{author}{\bibinfo{person}{PWC}.} \bibinfo{year}{2016}\natexlab{}.
\newblock \bibinfo{title}{Data breach notification: 10 ways GDPR differs from
  the US model}.
\newblock   (\bibinfo{year}{2016}).
\newblock
\showURL{%
\url{https://www.pwc.com/us/en/services/consulting/cybersecurity/library/broader-perspectives/gdpr-differences.html}}


\bibitem[\protect\citeauthoryear{Rauniar, Rawski, Yang, and Johnson}{Rauniar
  et~al\mbox{.}}{2014}]%
        {rauniar2014technology}
\bibfield{author}{\bibinfo{person}{Rupak Rauniar}, \bibinfo{person}{Greg
  Rawski}, \bibinfo{person}{Jei Yang}, {and} \bibinfo{person}{Ben Johnson}.}
  \bibinfo{year}{2014}\natexlab{}.
\newblock \showarticletitle{Technology acceptance model (TAM) and social media
  usage: an empirical study on Facebook}.
\newblock \bibinfo{journal}{{\em Journal of Enterprise Information
  Management\/}} (\bibinfo{year}{2014}).
\newblock


\bibitem[\protect\citeauthoryear{Redmiles, Zhu, Kross, Kuchhal, Dumitras, and
  Mazurek}{Redmiles et~al\mbox{.}}{2018}]%
        {redmiles2018asking}
\bibfield{author}{\bibinfo{person}{Elissa~M Redmiles}, \bibinfo{person}{Ziyun
  Zhu}, \bibinfo{person}{Sean Kross}, \bibinfo{person}{Dhruv Kuchhal},
  \bibinfo{person}{Tudor Dumitras}, {and} \bibinfo{person}{Michelle~L
  Mazurek}.} \bibinfo{year}{2018}\natexlab{}.
\newblock \showarticletitle{Asking for a friend: Evaluating response biases in
  security user studies}. In \bibinfo{booktitle}{{\em Proceedings of the 2018
  ACM SIGSAC Conference on Computer and Communications Security}}.
  \bibinfo{pages}{1238--1255}.
\newblock


\bibitem[\protect\citeauthoryear{Reed, Spiro, and Butts}{Reed
  et~al\mbox{.}}{2016}]%
        {reed2016thumbs}
\bibfield{author}{\bibinfo{person}{Philip~J Reed}, \bibinfo{person}{Emma~S
  Spiro}, {and} \bibinfo{person}{Carter~T Butts}.}
  \bibinfo{year}{2016}\natexlab{}.
\newblock \showarticletitle{Thumbs up for privacy?: Differences in online
  self-disclosure behavior across national cultures}.
\newblock \bibinfo{journal}{{\em Social science research\/}}
  \bibinfo{volume}{59} (\bibinfo{year}{2016}), \bibinfo{pages}{155--170}.
\newblock


\bibitem[\protect\citeauthoryear{Renaud, Volkamer, and Renkema-Padmos}{Renaud
  et~al\mbox{.}}{2014}]%
        {renaud2014doesn}
\bibfield{author}{\bibinfo{person}{Karen Renaud}, \bibinfo{person}{Melanie
  Volkamer}, {and} \bibinfo{person}{Arne Renkema-Padmos}.}
  \bibinfo{year}{2014}\natexlab{}.
\newblock \showarticletitle{Why doesn't Jane protect her privacy?}. In
  \bibinfo{booktitle}{{\em International Symposium on Privacy Enhancing
  Technologies Symposium}}. Springer, \bibinfo{pages}{244--262}.
\newblock


\bibitem[\protect\citeauthoryear{Roca and Gagn{\'e}}{Roca and
  Gagn{\'e}}{2008}]%
        {roca2008understanding}
\bibfield{author}{\bibinfo{person}{Juan~Carlos Roca} {and}
  \bibinfo{person}{Maryl{\`e}ne Gagn{\'e}}.} \bibinfo{year}{2008}\natexlab{}.
\newblock \showarticletitle{Understanding e-learning continuance intention in
  the workplace: A self-determination theory perspective}.
\newblock \bibinfo{journal}{{\em Computers in human behavior\/}}
  \bibinfo{volume}{24}, \bibinfo{number}{4} (\bibinfo{year}{2008}),
  \bibinfo{pages}{1585--1604}.
\newblock


\bibitem[\protect\citeauthoryear{Roca, Garc{\'\i}a, and De~La~Vega}{Roca
  et~al\mbox{.}}{2009}]%
        {roca2009importance}
\bibfield{author}{\bibinfo{person}{Juan~Carlos Roca},
  \bibinfo{person}{Juan~Jos{\'e} Garc{\'\i}a}, {and}
  \bibinfo{person}{Juan~Jos{\'e} De~La~Vega}.} \bibinfo{year}{2009}\natexlab{}.
\newblock \showarticletitle{The importance of perceived trust, security and
  privacy in online trading systems}.
\newblock \bibinfo{journal}{{\em Information Management \& Computer
  Security\/}} (\bibinfo{year}{2009}).
\newblock


\bibitem[\protect\citeauthoryear{Romesburg}{Romesburg}{2004}]%
        {romesburg2004cluster}
\bibfield{author}{\bibinfo{person}{Charles Romesburg}.}
  \bibinfo{year}{2004}\natexlab{}.
\newblock \bibinfo{booktitle}{{\em Cluster analysis for researchers}}.
\newblock \bibinfo{publisher}{Lulu. com}.
\newblock


\bibitem[\protect\citeauthoryear{Rousseeuw}{Rousseeuw}{1987}]%
        {rousseeuw1987silhouettes}
\bibfield{author}{\bibinfo{person}{Peter~J Rousseeuw}.}
  \bibinfo{year}{1987}\natexlab{}.
\newblock \showarticletitle{Silhouettes: a graphical aid to the interpretation
  and validation of cluster analysis}.
\newblock \bibinfo{journal}{{\em Journal of computational and applied
  mathematics\/}}  \bibinfo{volume}{20} (\bibinfo{year}{1987}),
  \bibinfo{pages}{53--65}.
\newblock


\bibitem[\protect\citeauthoryear{Schomakers, Lidynia, M{\"u}llmann, and
  Ziefle}{Schomakers et~al\mbox{.}}{2019}]%
        {schomakers2019internet}
\bibfield{author}{\bibinfo{person}{Eva-Maria Schomakers},
  \bibinfo{person}{Chantal Lidynia}, \bibinfo{person}{Dirk M{\"u}llmann}, {and}
  \bibinfo{person}{Martina Ziefle}.} \bibinfo{year}{2019}\natexlab{}.
\newblock \showarticletitle{Internet users' perceptions of information
  sensitivity--insights from germany}.
\newblock \bibinfo{journal}{{\em International Journal of Information
  Management\/}}  \bibinfo{volume}{46} (\bibinfo{year}{2019}),
  \bibinfo{pages}{142--150}.
\newblock


\bibitem[\protect\citeauthoryear{Shirazi and Volkamer}{Shirazi and
  Volkamer}{2014}]%
        {shirazi2014deters}
\bibfield{author}{\bibinfo{person}{Fatemeh Shirazi} {and}
  \bibinfo{person}{Melanie Volkamer}.} \bibinfo{year}{2014}\natexlab{}.
\newblock \showarticletitle{What deters Jane from preventing identification and
  tracking on the Web?}. In \bibinfo{booktitle}{{\em Proceedings of the 13th
  Workshop on Privacy in the Electronic Society}}. \bibinfo{pages}{107--116}.
\newblock


\bibitem[\protect\citeauthoryear{Soffer and Cohen}{Soffer and Cohen}{2014}]%
        {soffer2014privacy}
\bibfield{author}{\bibinfo{person}{Tal Soffer} {and} \bibinfo{person}{Anat
  Cohen}.} \bibinfo{year}{2014}\natexlab{}.
\newblock \showarticletitle{Privacy perception of adolescents in a digital
  world}.
\newblock \bibinfo{journal}{{\em Bulletin of Science, Technology \& Society\/}}
  \bibinfo{volume}{34}, \bibinfo{number}{5-6} (\bibinfo{year}{2014}),
  \bibinfo{pages}{145--158}.
\newblock


\bibitem[\protect\citeauthoryear{Spiekermann, Grossklags, and
  Berendt}{Spiekermann et~al\mbox{.}}{2001}]%
        {SpiGro2001}
\bibfield{author}{\bibinfo{person}{Sarah Spiekermann}, \bibinfo{person}{Jens
  Grossklags}, {and} \bibinfo{person}{Bettina Berendt}.}
  \bibinfo{year}{2001}\natexlab{}.
\newblock \showarticletitle{E-privacy in 2nd generation E-commerce: privacy
  preferences versus actual behavior}. In \bibinfo{booktitle}{{\em Proceedings
  of the 3rd ACM conference on Electronic Commerce}}. ACM,
  \bibinfo{pages}{38--47}.
\newblock


\bibitem[\protect\citeauthoryear{Talukder and Quazi}{Talukder and
  Quazi}{2011}]%
        {talukder2011impact}
\bibfield{author}{\bibinfo{person}{Majharul Talukder} {and}
  \bibinfo{person}{Ali Quazi}.} \bibinfo{year}{2011}\natexlab{}.
\newblock \showarticletitle{The impact of social influence on individuals'
  adoption of innovation}.
\newblock \bibinfo{journal}{{\em Journal of Organizational Computing and
  Electronic Commerce\/}} \bibinfo{volume}{21}, \bibinfo{number}{2}
  (\bibinfo{year}{2011}), \bibinfo{pages}{111--135}.
\newblock


\bibitem[\protect\citeauthoryear{Thomson, Yuki, and Ito}{Thomson
  et~al\mbox{.}}{2015}]%
        {thomson2015socio}
\bibfield{author}{\bibinfo{person}{Robert Thomson}, \bibinfo{person}{Masaki
  Yuki}, {and} \bibinfo{person}{Naoya Ito}.} \bibinfo{year}{2015}\natexlab{}.
\newblock \showarticletitle{A socio-ecological approach to national differences
  in online privacy concern: The role of relational mobility and trust}.
\newblock \bibinfo{journal}{{\em Computers in Human Behavior\/}}
  \bibinfo{volume}{51} (\bibinfo{year}{2015}), \bibinfo{pages}{285--292}.
\newblock


\bibitem[\protect\citeauthoryear{Tifferet}{Tifferet}{2019}]%
        {tifferet2019gender}
\bibfield{author}{\bibinfo{person}{Sigal Tifferet}.}
  \bibinfo{year}{2019}\natexlab{}.
\newblock \showarticletitle{Gender differences in privacy tendencies on social
  network sites: a meta-analysis}.
\newblock \bibinfo{journal}{{\em Computers in Human Behavior\/}}
  \bibinfo{volume}{93} (\bibinfo{year}{2019}), \bibinfo{pages}{1--12}.
\newblock


\bibitem[\protect\citeauthoryear{Tsai, Egelman, Cranor, and Acquisti}{Tsai
  et~al\mbox{.}}{2011}]%
        {tsai2011effect}
\bibfield{author}{\bibinfo{person}{Janice~Y Tsai}, \bibinfo{person}{Serge
  Egelman}, \bibinfo{person}{Lorrie Cranor}, {and} \bibinfo{person}{Alessandro
  Acquisti}.} \bibinfo{year}{2011}\natexlab{}.
\newblock \showarticletitle{The effect of online privacy information on
  purchasing behavior: An experimental study}.
\newblock \bibinfo{journal}{{\em Information systems research\/}}
  \bibinfo{volume}{22}, \bibinfo{number}{2} (\bibinfo{year}{2011}),
  \bibinfo{pages}{254--268}.
\newblock


\bibitem[\protect\citeauthoryear{Utz and Kramer}{Utz and Kramer}{2009}]%
        {utz2009privacy}
\bibfield{author}{\bibinfo{person}{Sonja Utz} {and} \bibinfo{person}{Nicole
  Kramer}.} \bibinfo{year}{2009}\natexlab{}.
\newblock \showarticletitle{The privacy paradox on social network sites
  revisited: The role of individual characteristics and group norms}.
\newblock \bibinfo{journal}{{\em Cyberpsychology: Journal of Psychosocial
  Research on Cyberspace\/}} \bibinfo{volume}{3}, \bibinfo{number}{2}
  (\bibinfo{year}{2009}), \bibinfo{pages}{2}.
\newblock


\bibitem[\protect\citeauthoryear{Vannoy and Palvia}{Vannoy and Palvia}{2010}]%
        {vannoy2010social}
\bibfield{author}{\bibinfo{person}{Sandra~A Vannoy} {and}
  \bibinfo{person}{Prashant Palvia}.} \bibinfo{year}{2010}\natexlab{}.
\newblock \showarticletitle{The social influence model of technology adoption}.
\newblock \bibinfo{journal}{{\it Commun. ACM}} \bibinfo{volume}{53},
  \bibinfo{number}{6} (\bibinfo{year}{2010}), \bibinfo{pages}{149--153}.
\newblock


\bibitem[\protect\citeauthoryear{Venkatesh, Morris, Davis, and Davis}{Venkatesh
  et~al\mbox{.}}{2003}]%
        {venkatesh2003user}
\bibfield{author}{\bibinfo{person}{Viswanath Venkatesh},
  \bibinfo{person}{Michael~G Morris}, \bibinfo{person}{Gordon~B Davis}, {and}
  \bibinfo{person}{Fred~D Davis}.} \bibinfo{year}{2003}\natexlab{}.
\newblock \showarticletitle{User acceptance of information technology: Toward a
  unified view}.
\newblock \bibinfo{journal}{{\em MIS quarterly\/}} (\bibinfo{year}{2003}),
  \bibinfo{pages}{425--478}.
\newblock


\bibitem[\protect\citeauthoryear{Vijayasarathy}{Vijayasarathy}{2004}]%
        {vijayasarathy2004predicting}
\bibfield{author}{\bibinfo{person}{Leo~R Vijayasarathy}.}
  \bibinfo{year}{2004}\natexlab{}.
\newblock \showarticletitle{Predicting consumer intentions to use on-line
  shopping: the case for an augmented technology acceptance model}.
\newblock \bibinfo{journal}{{\em Information \& management\/}}
  \bibinfo{volume}{41}, \bibinfo{number}{6} (\bibinfo{year}{2004}),
  \bibinfo{pages}{747--762}.
\newblock


\bibitem[\protect\citeauthoryear{Voigt and Von~dem Bussche}{Voigt and Von~dem
  Bussche}{2017}]%
        {voigt2017eu}
\bibfield{author}{\bibinfo{person}{Paul Voigt} {and} \bibinfo{person}{Axel
  Von~dem Bussche}.} \bibinfo{year}{2017}\natexlab{}.
\newblock \showarticletitle{The eu general data protection regulation (gdpr)}.
\newblock \bibinfo{journal}{{\em A Practical Guide, 1st Ed., Cham: Springer
  International Publishing\/}} (\bibinfo{year}{2017}).
\newblock


\bibitem[\protect\citeauthoryear{Wang, Norcie, Komanduri, Acquisti, Leon, and
  Cranor}{Wang et~al\mbox{.}}{2011}]%
        {wang2011regretted}
\bibfield{author}{\bibinfo{person}{Yang Wang}, \bibinfo{person}{Gregory
  Norcie}, \bibinfo{person}{Saranga Komanduri}, \bibinfo{person}{Alessandro
  Acquisti}, \bibinfo{person}{Pedro~Giovanni Leon}, {and}
  \bibinfo{person}{Lorrie~Faith Cranor}.} \bibinfo{year}{2011}\natexlab{}.
\newblock \showarticletitle{" I regretted the minute I pressed share" a
  qualitative study of regrets on Facebook}. In \bibinfo{booktitle}{{\em
  Proceedings of the seventh symposium on usable privacy and security}}.
  \bibinfo{pages}{1--16}.
\newblock


\bibitem[\protect\citeauthoryear{Williams, Freedman, and Deci}{Williams
  et~al\mbox{.}}{1998}]%
        {williams1998supporting}
\bibfield{author}{\bibinfo{person}{Geoffrey Williams}, \bibinfo{person}{Zachary
  Freedman}, {and} \bibinfo{person}{Edward Deci}.}
  \bibinfo{year}{1998}\natexlab{}.
\newblock \showarticletitle{Supporting autonomy to motivate glucose control in
  patients with diabetes}.
\newblock \bibinfo{journal}{{\em Diabetes\/}} \bibinfo{volume}{47},
  \bibinfo{number}{1S} (\bibinfo{year}{1998}).
\newblock


\bibitem[\protect\citeauthoryear{Williams and Deci}{Williams and Deci}{1996}]%
        {williams1996internalization}
\bibfield{author}{\bibinfo{person}{Geoffrey~C Williams} {and}
  \bibinfo{person}{Edward~L Deci}.} \bibinfo{year}{1996}\natexlab{}.
\newblock \showarticletitle{Internalization of biopsychosocial values by
  medical students: a test of self-determination theory.}
\newblock \bibinfo{journal}{{\em Journal of personality and social
  psychology\/}} \bibinfo{volume}{70}, \bibinfo{number}{4}
  (\bibinfo{year}{1996}), \bibinfo{pages}{767}.
\newblock


\bibitem[\protect\citeauthoryear{Wold, Esbensen, and Geladi}{Wold
  et~al\mbox{.}}{1987}]%
        {wold1987principal}
\bibfield{author}{\bibinfo{person}{Svante Wold}, \bibinfo{person}{Kim
  Esbensen}, {and} \bibinfo{person}{Paul Geladi}.}
  \bibinfo{year}{1987}\natexlab{}.
\newblock \showarticletitle{Principal component analysis}.
\newblock \bibinfo{journal}{{\em Chemometrics and intelligent laboratory
  systems\/}} \bibinfo{volume}{2}, \bibinfo{number}{1-3}
  (\bibinfo{year}{1987}), \bibinfo{pages}{37--52}.
\newblock


\bibitem[\protect\citeauthoryear{Wright and Javid}{Wright and Javid}{2019}]%
        {wright2019online}
\bibfield{author}{\bibinfo{person}{Jeremy Wright} {and} \bibinfo{person}{Sajid
  Javid}.} \bibinfo{year}{2019}\natexlab{}.
\newblock \showarticletitle{HM Gov Online harms white paper. April 2019}.
\newblock  (\bibinfo{year}{2019}).
\newblock


\end{thebibliography}

\newpage
\onecolumn
\appendix
\section{Study 2 Questionnaire}
\label{sec:study2_questionnaire}
The 43 privacy methods used in the questionnaire in Study 2 were named by participants themselves in Study 1 and compiled into 43 distinct methods.

The instruction was ``From the list below, please rate the tools/methods in terms of whether you have used them \emph{very often} before versus \emph{not used at all or very rarely}, to achieve the purpose of privacy online."

\begin{table*}[h]
\centering
\caption{The list of methods as provided in a tabular form in the questionnaire \& listed in random order for each participant.}
\label{tab:study2_questionnaire}
\footnotesize
%\resizebox{\textwidth}{!}{
\begin{tabular}{lcc} %crrrrlrrrrr}
\toprule
\textbf{Tools/Methods}&\textbf{very often}& \textbf{not used at all or very rarely}\\
\midrule
pseudonyms&$\square$&$\square$\\
paypal instead of online banking& $\square$&$\square$\\
anonymous profile names or emails& $\square$&$\square$\\
not store information online &$\square$&$\square$ \\
Erasery& $\square$&$\square$\\
clear information or history& $\square$&$\square$\\
private browsing or incognito mode& $\square$&$\square$ \\
DuckDuckGo& $\square$&$\square$\\
clear, disallow or limit cookies &$\square$&$\square$\\
Ghostery& $\square$&$\square$\\
not accessing suspicious websites, careful of websites& $\square$&$\square$ \\
anti-tracking extension& $\square$&$\square$\\
switch off location tracking& $\square$&$\square$ \\
Adblock&$\square$&$\square$ \\
anti-malware & $\square$&$\square$\\
anti-spyware& $\square$&$\square$ \\
firewall&$\square$&$\square$ \\
NoScript&$\square$&$\square$\\
VPN&$\square$&$\square$ \\
HTTPS &$\square$&$\square$\\
TOR&$\square$&$\square$ \\
use a proxy&$\square$&$\square$ \\
IPHider&$\square$&$\square$ \\
virtual machines &$\square$&$\square$\\
encryption, encrypt communication or data&$\square$&$\square$ \\
password manager &$\square$&$\square$\\
not reuse passwords &$\square$&$\square$\\
not save password on webpage&$\square$&$\square$ \\
not give personal information, limit sharing of personal information or give minimal personal information &$\square$&$\square$\\
give fake information&$\square$&$\square$\\
set privacy settings or controls &$\square$&$\square$\\
not access accounts in a public place, on public networks or shared computers&$\square$&$\square$\\
read terms of service or business practice &$\square$&$\square$\\
limit use of social network accounts, such as Facebook or others &$\square$&$\square$\\
not use Facebook at all &$\square$&$\square$\\
have several email accounts or have bogus email account for unimportant use &$\square$&$\square$\\
request data collection about you &$\square$&$\square$\\
switch off camera on devices&$\square$&$\square$ \\
not subscribe to newsletters or untick boxes for newsletters&$\square$&$\square$ \\
opt out of data collection or not consent to data collection&$\square$&$\square$ \\
not engage online&$\square$&$\square$ \\
research before engaging online or research before signing up to stuff&$\square$&$\square$ \\
Kaspersky&$\square$&$\square$\\
&\\
Name other privacy methods if you use them: \\
\bottomrule
\end{tabular}
\end{table*}

\newpage
\section{Study 2 Dataset}
\label{sec:study2_dataset}
\begin{table*}[h]
\centering
\caption{Dataset of Privacy Methods \& Count of Usage per Country.}
\label{tab:dataset}
\footnotesize
%\resizebox{\textwidth}{!}{
\begin{tabular}{llrrrr} %crrrrlrrrrr}
\toprule
\textbf{\#}&\textbf{Tools/Methods}&\textbf{US}& \textbf{UK} & \textbf{DE} & \textbf{Total} \\
\midrule
1& not accessing suspicious websites, careful of websites& 265&	251&	250 &766 \\
2&not give personal information or limit sharing &259	&242&	256 &757\\
3&set privacy settings or controls &262&	238 &	257 &757\\
4& not subscribe to newsletters or untick boxes for newsletters&234	&219	 &250 & 703\\
5& Adblock&235	&197 &	262 &694 \\
6& clear information or history& 231 &	235	& 220 &686\\
7& anti-malware &238&217&224  & 679\\
8&firewall&211&	218	&250  &679\\
9& paypal instead of online banking& 201	&226 &	250 & 677\\
10& have several email accounts &214&	183&	261 &658\\
11&clear, disallow or limit cookies &225	&198 &	219 &642\\
12&opt out of data collection or not consent to data collection&216	&202 &	207 &625\\
13&HTTPS &209&	172&	237&618\\
14& not access accounts in a public place/ shared computers &224&210	&171&605\\
15&anti-spyware& 221	&190&	181 &592 \\
16&switch off camera on devices&206	&172	 &209&587 \\
17&private browsing or incognito mode& 194 &	179	& 212&585 \\
18&limit use of social network accounts &189	&186	&209&584\\
19&switch off location tracking&183	 &174	&201 &558\\
20&anonymous profile names or emails& 175 &	134	& 228 &537\\
21&pseudonyms&143	&111& 	229 & 483\\
22&not reuse passwords &162	&148	 &158&468\\
23&not store information online &167&144&154&465\\
24&not save password on webpage&147&	147	&149 &443\\
25&give fake information&128 &102	& 181 & 411\\
26&read terms of service or business practice &154&	147&	100 &401\\
27& password manager &152	&124 &	116 &392\\
28& not use Facebook at all  &131&99	&160&390\\
29&VPN&99&	86	&157 &342\\
30& encryption, encrypt communication or data&117&	103	&119 &339\\
31&anti-tracking extension& 93&	64	&126&283\\
32&not engage online&99	& 97	 &69&265 \\
33&use a proxy&67 &	65&	110&242 \\
34&DuckDuckGo& 79&	39	&77&195\\
35&request data collection about you &54	&38	&81&173\\
36&NoScript&39&	21&	84 &144\\
37&IPHider&42	&35&61 & 138 \\
38&research before engaging online &30 &35& 63 &128\\
39&Kaspersky &30 &	35	& 63&128\\
40&virtual machines &37 &	21&	67 &125\\
41&Tor&32	&23&	67 &122\\
42&Ghostery& 31	&21 &	55 &107\\
43&Erasery& 27&	30&	37&94\\
\bottomrule \\
\end{tabular}\\
\footnotesize{\emph{Notes: This dataset is for the N = 907 sample, and sorted from most used to least used method. \\
This dataset was used in the Cluster and Correspondence Analyses.}}
\end{table*}

\newpage
\section{Correspondence Analysis}
\label{sec:2nd_correspondence_plot}
\begin{figure*}[h]
\centering
\includegraphics[height = 10cm, width=\textwidth]{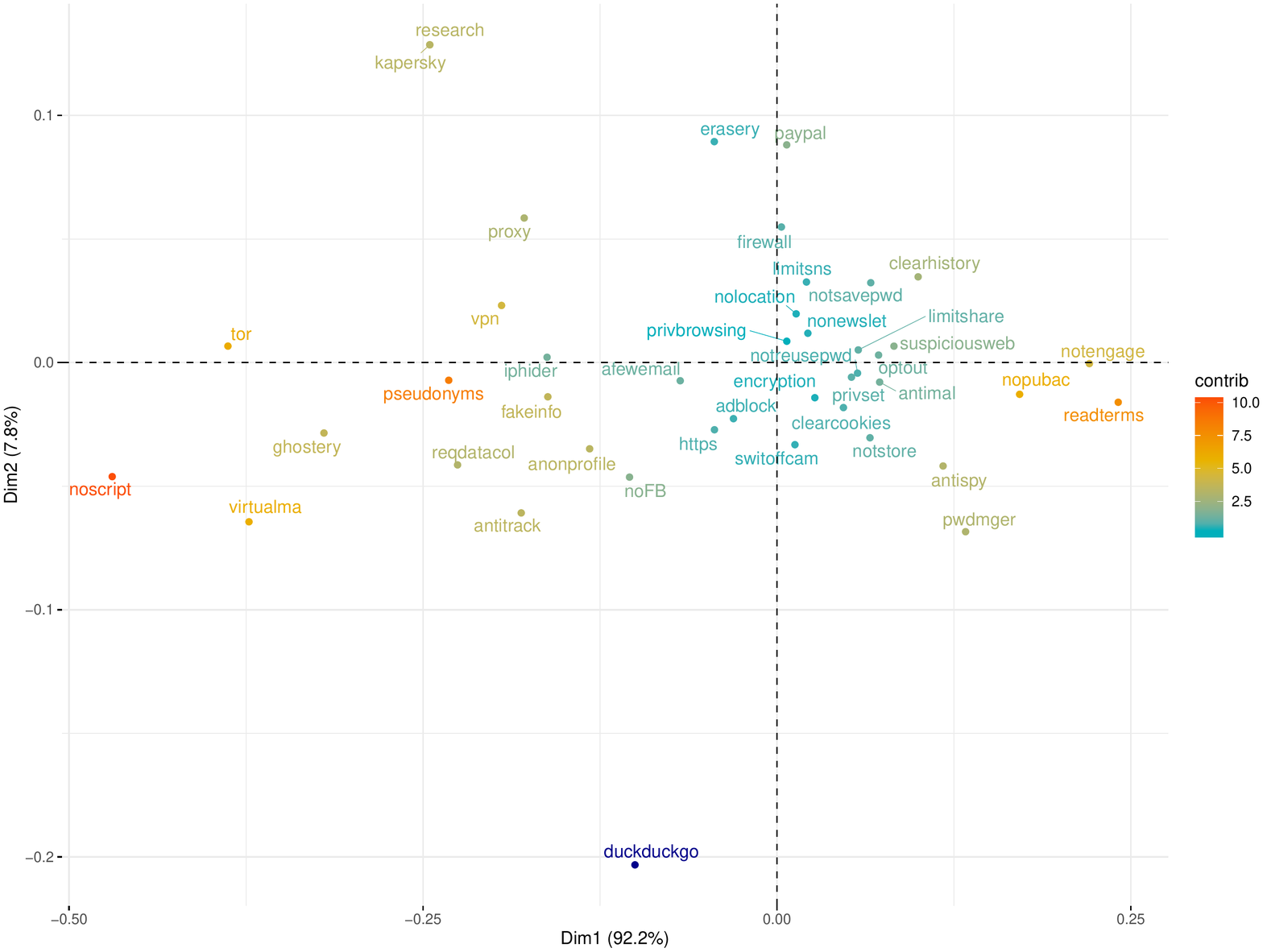}
\caption{Contribution of Privacy Methods to Dimension 1.}%
\label{fig:contribplot}
%\end{subfigure}%
%\vspace{2cm}
\end{figure*}

\newpage
\section{Study 3 Lists Questionnaire}
\label{sec:app_study3_choice}
The instructions provided to participants to choose between two lists of PETs and the follow-up questions are as follow: \\

We provide two lists of privacy technologies that can be used to protect privacy online.
Please take your time to read through both lists.

\textbf{List A contains}: Erasery, Ghostery, Virtual Machine, Tor, NoScript, IPHider, Kaspersky, DuckDuckGo, proxy, anti-tracking extension, VPN and encryption.

\textbf{List B contains}: switch off location tracking, private browsing, HTTPS, anti-spyware, opt-out (of data collection), clear cookies, anti-malware, clear history, Paypal, firewall, Adblock, privacy settings, pseudonyms and anonymous profile.\\

Which of the two lists contains the privacy methods that \textbf{you most often use} to protection your privacy online?

List A\\
\indent List B \\

Please explain why you most often use methods from your selected list.
Please explain why you very rarely (or not at all) use methods from the other list.
[Response in 50 to 100 words]\\

What would support you to use methods from the \textbf{other} list (that is, the one that you did not select)?
In particular, what would help and/or encourage you?\\

\section{Study 3 Perception Questionnaire Scale Items}
\label{sec:study3_scale}
\begin{table*}[h]
\centering
\caption{Scale Items \& Internal Consistency. \\
We provide Cronbach $\alpha$ for Adv. \& Oth.PETs where the scale was set as pairwise comparison between PET types.} %, restricted to significance level $\alpha = .01$}
\label{tab:c_alpha_table}
\footnotesize
\resizebox{\textwidth}{!}{\begin{tabular}{lllll}
\toprule
\textbf{Scale} & \textbf{Items} & \multicolumn{3}{c}{\textbf{Cronbach $\alpha$}} \\
\cline{3-5}\\
&&Adv.PETs && Oth.PETs\\
\midrule
\multirow{4}{*}{Perceived Privacy Competency}& I feel confident in my ability to manage my privacy online && \multirow{4}{*}{.963}\\
& I am capable of protecting my privacy online now\\
& I am able to protect my privacy online now\\
& I feel able to meet the challenge of protecting my privacy online\\
&\\

\multirow{4}{*}{Awareness of PETs}& I have heard of technologies in this list for privacy protection online & \multirow{4}{*}{.932}&& \multirow{4}{*}{.837}\\
&I have encountered the technologies in this list for privacy protection online before \\
&I know how to find technologies in this list to protect my privacy online \\
&I am familiar with the technologies in this list to protect my privacy online \\
&\\

\multirow{4}{*}{Perceived Usefulness of PETs}&Using privacy technologies in this list improves my privacy protection & \multirow{4}{*}{.879}&& \multirow{4}{*}{.842} \\ %Using privacy technologies in this list improves the performance of my privacy protection
& Using privacy technologies in this list increases my level of privacy\\
& Using privacy technologies in this list enhances the effectiveness of my privacy protection \\
& I find privacy technologies in this list to be useful in protecting my privacy \\
%Using privacy technologies in this list improves the performance of my privacy protection.
&\\

\multirow{4}{*}{Perceived Ease of Use of PETs}
&My interaction with privacy technologies in this list is clear and understandable & \multirow{4}{*}{.868}&& \multirow{4}{*}{.798}\\
& Interacting with privacy technologies in this list does not require a lot of mental effort\\
& I find privacy technologies in this list to be easy to use\\
&I find it easy to get privacy technologies in this list to do what I want them to do\\
&\\

\multirow{3}{*}{Social Influence} 
&People who are important to me think that I should use privacy technologies in this list & \multirow{4}{*}{.891}&& \multirow{4}{*}{.881}\\
&People who influence my behaviour think that I should use privacy technologies in this list \\
& People whose opinion I value prefer that I use privacy technologies in this list\\
&\\

\multirow{3}{*}{Behavior Intention} 
&I intend to use privacy technologies in this list in the future & \multirow{4}{*}{.886}&& \multirow{4}{*}{.813}\\
&I will always try to use privacy technologies in this list in my daily life\\
&I plan to use privacy technologies in this list frequently \\
%&\\

%\multirow{3}{*}{Innovativeness}&&& \multirow{4}{*}{.841}&\\
%& If I heard of a new technology, I would look for ways to experiment with it\\
%& Among my peers, I am usually the first to explore new technologies\\
%&I like to experiment with new technologies\\

\bottomrule
\end{tabular}
}
\end{table*}

\newpage
\section{Study 3 Codebook}
\label{sec:study3_codebook}
\begin{table*}[h] 
\centering
\caption{The 47 codes in the final codebook for Study 3} %, restricted to significance level $\alpha = .01$}
\label{tab:codebook}
\footnotesize
\resizebox{\textwidth}{!}{\begin{tabular}{llll}
\toprule
\multicolumn{2}{l}{\textbf{Q1: Reason for choosing one PET type}} & \multicolumn{2}{l}{\textbf{Q2: Support or encouragement}} \\
\textbf{Code} & \textbf{Content} & \textbf{Code} & \textbf{Content}\\
\midrule
\multicolumn{2}{c}{\textbf{Effectiveness of PETs}} & \multicolumn{2}{c}{\textbf{Privacy need}} \\
EFF01 & effective for the privacy I need, adequate, sufficient & PNE01 & if I had a bigger privacy need, extreme privacy \\
EFF02 & used it and it works - so continue to use & PNE02 & needed for different specific privacy protections such as tracking prevention\\
EFF03-other & everyday, regular use & PNE03 & provides more privacy than the one I use\\
EFF03-other & trust, reliable, best method \\
& & \multicolumn{2}{c}{\textbf{Information or training to enhance understanding}} \\

\multicolumn{2}{c}{\textbf{Awareness of PETs}} & INF01 & if I know what they are/what they do \\
AWA1 & familiar with, recognise, heard of, know of & INF02 & Information on how to use or training or education \\
AWA2 & not familiar with, not recognise, not heard of, don't know of & INF03 & Information channel such as advertising, social media, tutorial, app\\
AWA3 & aware of though advertisements & INF04-other &benefits of using, why use\\
AWA4-other & advised, recommended & INF04-other & tech knowledge\\
&\\

\multicolumn{2}{c}{\textbf{Skills needed}} & \multicolumn{2}{c}{\textbf{Usability related supports}} \\
 SKI01 & casual, non-tech user &USS1& convenient  \\
 SKI02 & training needed, need to understand or know more, technical user &USS2& easy to use, simple, clear\\
 SKI03-other & someone to show me & USS3 &easy to install, setup, not many config\\
 &&USS4&lightweight, not many things, easy integration\\
 
 \multicolumn{2}{c}{\textbf{Usability \& cost characteristics of PET}} & USS5& easy access\\
 USR1 & readily or easily available, builtin, presented within service\\
 USR2 & not easily or not readily available  &  \multicolumn{2}{c}{\textbf{Social influence or social support}} \\
 USR3 & easy to use, easy to install, not much effort needed & SOC01 & social influence - if someone I know use it, or someone I trust recommends it\\
 USR4 & complicated to use or complicated to setup &SOC02 & social support - someone teaches me, someone installs it\\
 USR5 & convenient features &SOC03-other & neutral recommend, review, professional reputable company, accreditation\\
 USR6 & cost - available free && \\
 USR7 &cost - expensive or have to pay for & \multicolumn{2}{c}{\textbf{Miscellaneous}}  \\
 USR8-other & accessibility, easy access & INC01 & if I knew they did not pose a security threat, if I trust them\\
 USR8-other & amount of work vs benefit, having to download/install & INC02 & free of cost, affordable\\
 USR8-other & annoyance, discomfort, break things & INC03 & already use both lists\\
 USR8-other & integrated in service & INC04-other & tradeoff with browsing or computer performance \\
 &\\
 
\multicolumn{2}{c}{\textbf{Privacy needed, experience}} \\
PRI01 & no need for privacy \\
PRI02 & privacy vs hamper online experience \\
PRI03 & provide protection, keeps me privacy/safe, increase my privacy \\
PRI05 & use PETs from both lists \\
PRI06-other & over the top, extreme privacy\\
\bottomrule
\end{tabular}
}
\end{table*}

\end{document}